\documentclass{aa}

\usepackage{natbib}
\usepackage{amssymb,verbatim}
\usepackage{graphicx}
\usepackage{appendix}
\usepackage{diagbox}
\usepackage{subfig}
\usepackage{color}
\newcommand{\siml}{\raise -2.truept\hbox{\rlap{\hbox{$\sim$}}\raise5.truept
\hbox{$<$}\ }}
\newcommand{\simg}{\raise -2.truept\hbox{\rlap{\hbox{$\sim$}}\raise5.truept
\hbox{$>$}\ }}

\begin{document}

\titlerunning{Iron in X-COP}

\title{Iron in X-COP: tracing enrichment in cluster outskirts with high accuracy abundance profiles}
\authorrunning{Ghizzardi, Molendi, van der Burg et al}
\author{Simona Ghizzardi\inst{1}, Silvano Molendi\inst{1}, Remco van der Burg\inst{2}, Sabrina De Grandi\inst{3}, Iacopo Bartalucci\inst{1}, Fabio Gastaldello\inst{1},  Mariachiara Rossetti\inst{1}, Veronica Biffi\inst{4,5,6}, Stefano Borgani\inst{7,8,4,9}, Dominique Eckert\inst{10}, Stefano Ettori\inst{11,12}, Massimo Gaspari\inst{13}, Vittorio Ghirardini\inst{14}, Elena Rasia\inst{7,4}}

\institute{
INAF - Istituto di Astrofisica Spaziale e Fisica Cosmica di Milano, Via A. Corti 12, 20133 Milano, Italy\\
\email{simona.ghizzardi@inaf.it}
\and
European Southern Observatory
Karl-Schwarzschild Str. 2, 85748
Garching bei Muenchen, Germany
\and
INAF - Osservatorio Astronomico di Brera, via E. Bianchi 46, 23807 Merate (LC), Italy
\and
Institute of Fundamental Physics of the Universe, via Beirut 2, 34151, Grignano, Trieste, Italy
\and
Harvard-Smithsonian Center for Astrophysics, 60 Garden St., Cambridge, MA, 02138, USA
\and
Universitäts-Sternwarte München, Scheinerstr. 1, 81679, München, Germany
\and
Dipartimento di Fisica, Sezione di Astronomia, Università di Trieste, via Tiepolo 11, 34143, Trieste, Italy
\and 
INAF-Osservatorio Astronomico Trieste, via Tiepolo 11, 34123, Trieste, Italy
\and 
INFN, Instituto Nazionale di Fisica Nucleare, Trieste, Italy
\and 
Department of Astronomy, University of Geneva, ch. d'Ecogia 16, 1290 Versoix, Switzerland
\and
INAF—Osservatorio di Astrofisica e Scienza dello Spazio di Bologna, via Piero Gobetti 93/3, I-40129 Bologna, Italia
\and 
INFN, Sezione di Bologna, viale Berti Pichat 6/2, I-40127 Bologna, Italy
\and 
Department of Astrophysical Sciences, Princeton University, 4 Ivy Lane, Princeton, NJ 08544, USA
\and 
Harvard-Smithsonian Center for Astrophysics, 60 Garden Street, Cambridge, MA 02138, USA
}

\abstract{We present the first metal abundance profiles for a representative sample of massive clusters.  Our measures extend to $R_{500}$ and are corrected for a systematic error plaguing previous outskirt estimates. Our profiles flatten out at large radii, admittedly not a new result, however the radial range and representative nature of our sample extends its import well beyond previous findings. We find no evidence of segregation between cool-core and non-cool-core systems beyond $\sim 0.3 R_{500}$, implying that, as was found for thermodynamic  properties \citep{Ghirardini:2019}, the physical state of the core does not affect global cluster properties.  
Our mean abundance within $R_{500}$ shows a very modest scatter, $< $15\%, suggesting the enrichment process must be quite similar in all these  massive systems.  This is a new finding and has significant implications on  feedback processes.  Together with results  from  thermodynamic properties presented in a previous X-COP paper,  it affords a coherent picture where feedback effects do not vary significantly from one system to another.
By combing ICM with stellar measurements we have found  the amount of Fe diffused in the ICM  to be about ten times higher than that locked in stars. 
Although our estimates suggest, with some strength, that the measured iron mass in clusters is well in excess of the predicted one, systematic errors prevent us from making a definitive statement. Further advancements will only be possible when systematic uncertainties, principally those associated to stellar masses, both within and beyond $R_{500}$, can be reduced. 
}

\keywords{X-rays: galaxies: clusters - Galaxies: clusters: general -  Galaxies: clusters: intracluster medium - cosmology: large-scale structure}
\maketitle

\section{Introduction}
\label{sec:intro}

Fall into the largest gravitational wells in the Universe, i.e. clusters of galaxies, must lead to the heating of gas to very high temperatures; what is by no means forgone is that the same gas be enriched in heavy elements. Indeed, while the former property of the Intra-Cluster Medium (ICM) results from simple gravitational collapse, the latter requires that gas be processed in stars and re-ejected into the  ICM in the form of heavy elements. 
Radiative cooling and feedback processes can of course leave their imprint on thermodynamic properties of the ICM, however, beyond core regions, they do so in the form of modest modifications over substantial gravitational heating \citep{Pratt:2010}, conversely, metals in the hot gas phase trace exclusively feedback mechanisms. 

Metals can in principle be a powerful probe, their quantity and distribution in the ICM can be used to provide important clues on the nature of feedback processes \citep[see][ and refs. therein]{Biffi:2018}. However, thus far, abundances have been of limited use, mainly because of the lack of dedicated observational programs of cluster samples out to large radii. An adequate radial range is needed both to measure the distribution and assess the total amount of Fe in the ICM and, ultimately, to constrain feedback processes occurring at high redshift during and perhaps even before the proto-cluster formation phase. 
Unfortunately, extending X-ray measurements out to large radii is quite difficult, particularly so for abundances. As discussed in \citet{Molendi:2016}, reliable measurements have been carried out to $\sim 0.6 R_{500}$, sampling about 1/3 of the total gas mass in clusters and only on archival samples with no guarantee  of representativity \citep{Leccardi_metal:2008, Mernier:2017, Lovisari:2019}.  Moreover measurements are known to be plagued by a variety of systematic issues \citep{Buote:2000,Leccardi:2007, Leccardi_metal:2008, Leccardi_temp:2008}.  

A full census of metals in clusters requires that the stellar component also be estimated. This is not a trivial measurement in itself, moreover it requires that assumptions be made about: the Initial Mass Function (IMF), the stellar population synthesis models, stellar formation histories etc. \citep[see][for a detailed discussion]{Behroozi:2010}.
Thus, claims that the amount of Fe in the ICM is in excess of what can be produced by the stars in the cluster \citep{RA14,Lowenstein13} should be regarded with some caution. 

In this paper we derive abundance profiles for the  X-COP\footnote{The XMM Cluster Outskirts Project (X-COP) is  an XMM-Newton Very Large Program dedicated to Cluster outskirts. It has been extracted from the Planck PSZ1 catalog by making a high cut in SNR and placing further constraints on: minimum apparent size, redshift range and maximum galactic N$_{\rm H}$.} sample \citep{Eckert_XCOP:2017}. Our measurements are unprecedented for 4 reasons: 1) they are carried out on a representative  \citep{Eckert_XCOP:2017}, albeit small, sample; 2) they are made on massive systems ($M_{500} > 3.5 \cdot 10^{14} M_{\odot}$), i.e. on systems better approximating the ideal "closed box";  3) they extend, for virtually all systems, to $R_{500}$ and 4) they have been corrected for a systematic error which has affected most past Fe abundance measurements.

The paper is organized as follows. In Sect. \ref{sec:datax} we describe our {\it XMM-Newton} data set,  data reduction and analysis, including modifications to the spectral analysis, specifically developed for Fe abundance measurements, the most important being the exclusion of the L-shell region from spectral fitting. 
In Sect. \ref{sec:fe_icm} we present results from the X-ray analysis, we show the X-COP Fe abundance profile and compare it with those measured from other samples such as the ones reported in \citep{Leccardi_metal:2008, Lovisari:2019, Mernier:2017}. We also compute deprojected Fe profiles,
Fe mass profiles and investigate scaling relations between the Fe mass and other cluster observables.
In Sect. \ref{sec:datao} we present stellar mass profiles for 7 of our 12 systems and investigate scaling relations between stellar mass and X-ray observables. In Sect. \ref{sec:comb}, we combine X-ray and optical measurements to take a census of iron in clusters and compare the total estimated Fe mass with the one expected from supernovae (hereafter SN).    
In Sect. \ref{sec:disc} we interpret and discuss our results.

Throughout the paper, we assume a $\Lambda$ cold dark matter cosmology with $H_0=70 \mathrm{~km ~ s}^{-1} \mathrm{Mpc}^{-1}$, $\Omega_\mathrm{m}=0.3$, $\Omega_\Lambda=0.7$ and $E(z) = \sqrt{\Omega_\mathrm{m}(1+z)^3+\Omega_\Lambda}$ for the evolution of the Hubble parameter. 
For each cluster we use as reference radii $R_{500}$ and $R_{200}$, defined at the over-densities of $\Delta = 500$ and 200, respectively, with respect to the critical value $\rho_c = 3H^2_0 {\frac{E(z)^2}{8\pi G}}$, and computed using the X-COP hydrostatic mass profiles in \citet{Ettori:XCOP2019}.
All the quoted errors hereafter are at the $1\sigma$ confidence level.

\section{X-ray data analysis}
\label{sec:datax}

\emph{XMM-Newton} observations, X-ray data analysis pipeline and sample are described in detail in \citet{Ghirardini:2018,Ghirardini:2019}.
In the following we summarize the main steps of this analysis.

\subsection{Data reduction}
\label{sec:dataxana}

All data are reduced using XMMSAS v13.5 and the Extended Source Analysis Software \citep[ESAS;][]{Snowden:2008}. The first basic steps produce calibrated events files for each observation (with \emph{emchain} and \emph{epchain}; the pn chain is run twice to create events files also for pn out-of-time events). We filter out time period affected by soft proton flares with \emph{mos-filter} and \emph{pn-filter}. We then estimate the  contamination of residual soft protons to the spectrum by comparing IN and OUT count rates, where IN are the count rates of the MOS measured in the 7.5-11.8 keV energy band from regions inside the Field of View (FoV), and OUT outside the FoV, in the unexposed corners of the MOS detectors \citep{Deluca_molendi_inout:2004,Leccardi_temp:2008}. 
We run the  XMMSAS tool \emph{ewavelet} to detect point sources within the FoV and correct the resulting point source list for the spatial dependence of the fraction of the cosmic X-ray background (CXB) that is resolved by the instrument. This correction consists in excising only the sources with a measured count rate greater than a certain count rate threshold, determined by comparing our count rate distribution with the LogN-LogS distribution of CXB sources, down to which our source detection is complete \citep[details in][]{Ghirardini:2018}. We leave the fainter sources to enforce a constant flux threshold across the FoV and avoid biasing local measurements of the CXB intensity.

\subsection{Spectral analysis in X-COP}
\label{sec:dataxspec}
We extract spectra in concentric annuli around the X-ray peak up to $\sim R_{500}$, estimated from the hydrostatic mass \citep[see][]{Ettori:XCOP2019}, with the ESAS routines \emph{mos-spectra} and \emph{pn-spectra}. Subsequently we use filter-wheel-closed data to estimate the high-energy particle background contribution with \emph{mos-back} and \emph{pn-back}. Products of these tools are also appropriate response matrices and effective areas files for extended sources. The output spectra are re-binned with a minimum of 5 counts per bin to ensure stable fitting results, and the data below 0.5 keV are discarded to avoid EPIC calibration uncertainties in this energy range.  

\begin{figure}
\centerline{\includegraphics[angle=0,width=9.8cm]{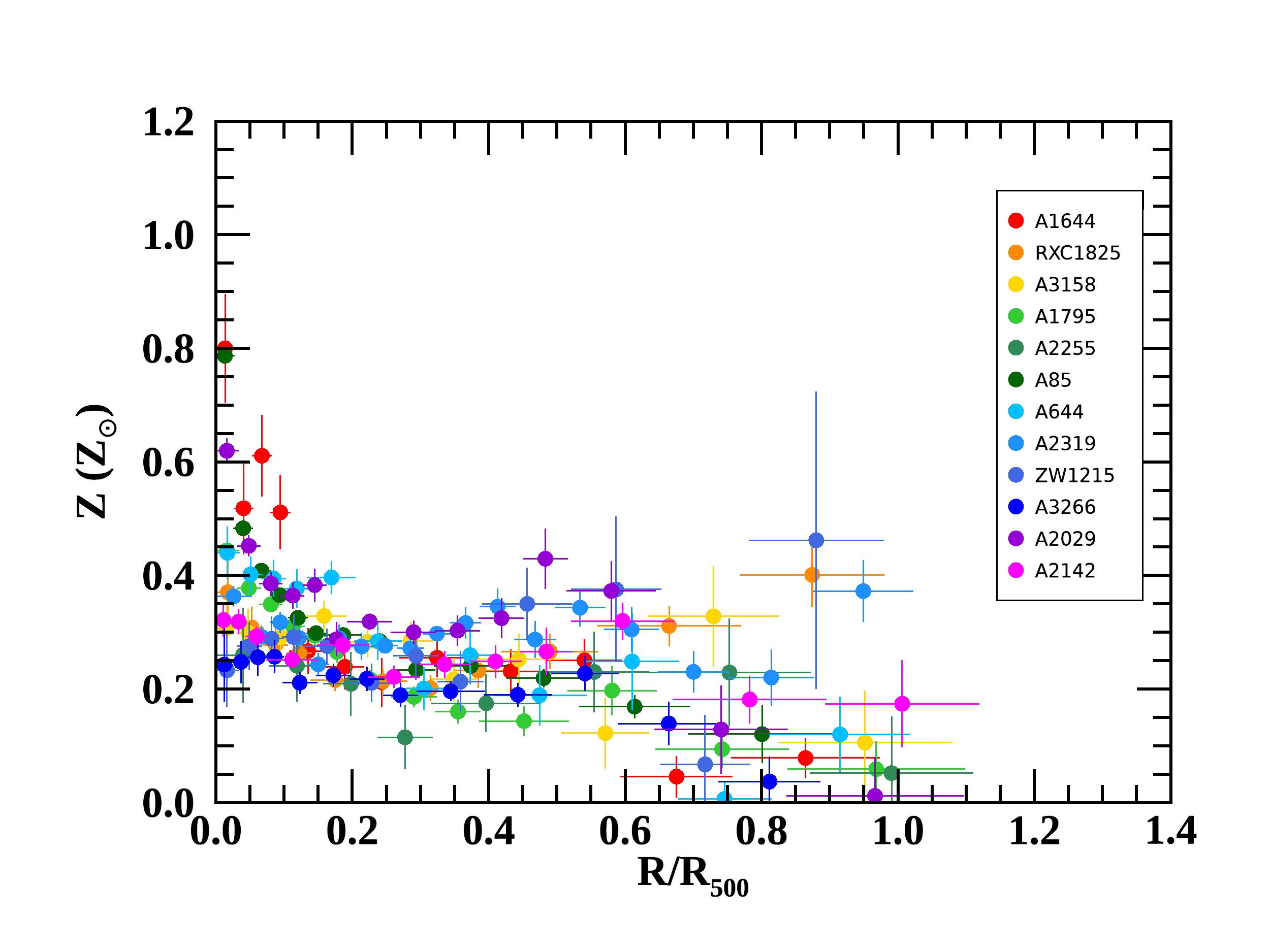}}
\caption{Iron abundance profiles as a function of R/R$_{500}$, derived with standard X-COP analysis (see Sect. \ref{sec:dataxspec}).  Clusters in the inset are ordered by their total mass starting from the least massive, i.e. A1644. Abundances reported here and throughout the X-ray sections of this paper are in solar units as defined in \citet{AG:1989}.}
\label{fig:zfeproj0}
\end{figure}

We use {\tt XSPEC} v12 \citep{Arnaud_XSPEC:1996} and {\tt ATOMDB} v3.0.7 to fit the spectra and determine the plasma best-fit parameters according to the Cash-statistics. Our strategy to estimate the physical quantities from the observed spectra consists in modeling all the individual background components and the source spectra. The background components,  described in detail in \citet{Ghirardini:2018}, are: high-energy particle background, sky background and residual soft protons. We model the source emission of each region with an absorbed single temperature APEC model with temperature, emission measure and metal abundance free to vary. We note that, in the current version of  {\tt XSPEC}, the metal abundance parameter in APEC has a lower hard boundary at 0, this has significant implications which will be discussed in Sect.\ref{sec:negative}. The solar abundance table is set to \citet{AG:1989}. MOS and pn spectra of each region are fitted jointly, as are spectra of different observations of the same regions. 
The energy range considered in the fitting is 0.5-12 keV, we ignore energy ranges where bright and time variable fluorescence lines are present, i.e. 1.2-1.9 keV for the two MOS and 1.2-1.7 keV and 7.0-9.2 keV for the pn.  Gas temperature, density, entropy and pressure profiles for the whole X-COP sample are presented in \citet{Ghirardini:2019},  Fe abundance profiles are plotted in Fig. \ref{fig:zfeproj0}. Hence forward we will refer to this analysis as the "standard" analysis.

\begin{figure}
\centerline{\includegraphics[angle=0,width=9.8cm]{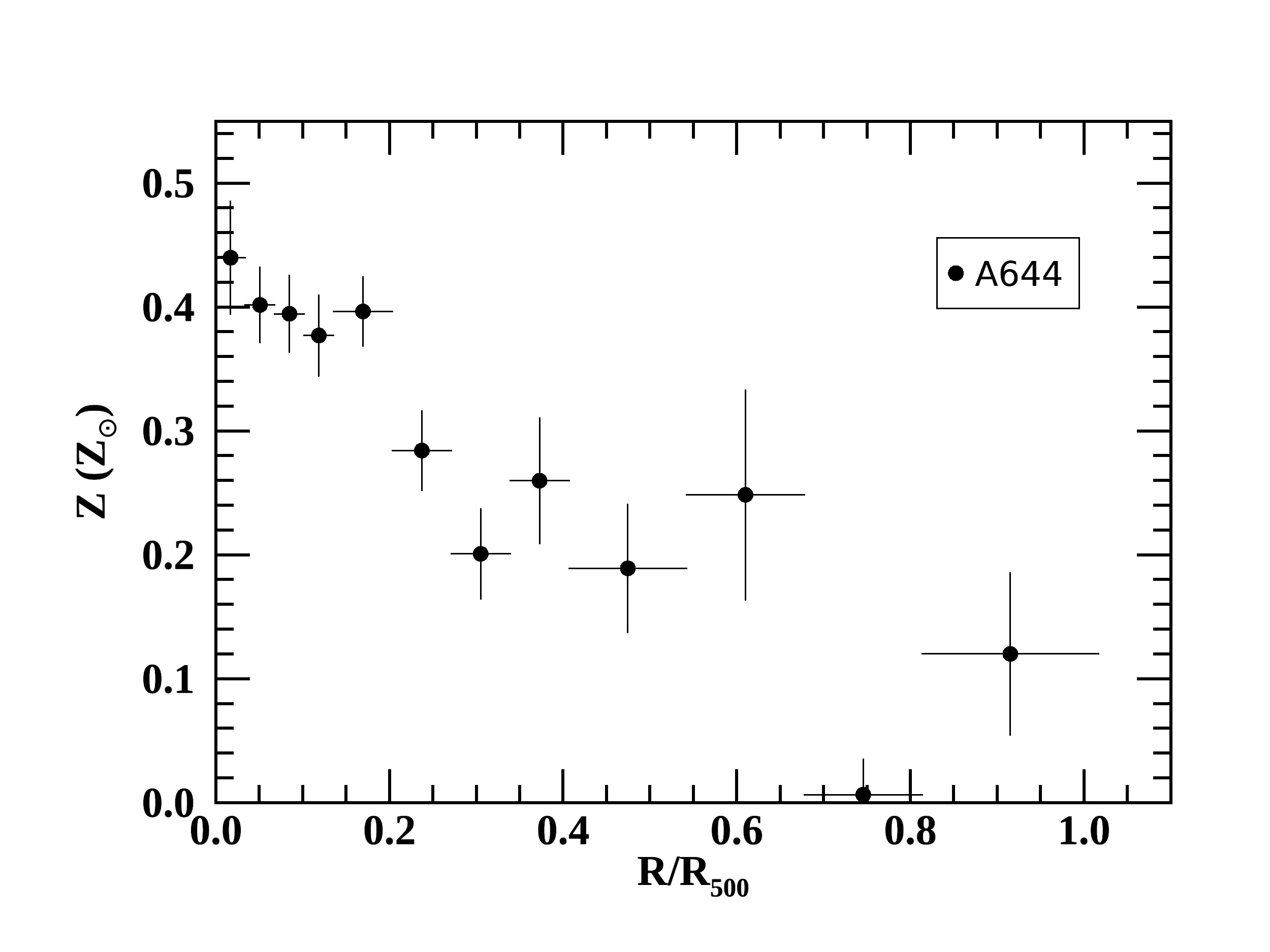}}
\caption{Iron abundance profile for A644 as a function of R/R$_{500}$, derived with standard X-COP analysis (see Sect. \ref{sec:dataxspec}).}
\label{fig:zfeA644}
\end{figure}

\subsection{Modifications to X-COP spectral analysis}
\label{sec:modif_spec_anal}
In this subsection we shall look into modifications of the X-COP spectral analysis specifically
designed to improve abundance measurements. 

\subsubsection{The vanishing metals}
\label{sec:externalbins}

Inspection of Fig. \ref{fig:zfeproj0} shows that, in several instances, the measured abundance drops to a value close to zero, in a specific bin, only to increase again in the following. What is even more puzzling is that the computed error on the near zero measure is very small, an example of this is provided by A644, see Fig. \ref{fig:zfeA644}. Similar jumps in abundance profiles have been found by other authors. In \citet{Mernier:2017} the abundance profiles of A2029,  A133 and A2597, see their Appendix A,
feature a prominent drop in metallicity; incidentally, for A2029, which is also in our sample, we measure a very similar jump. In \citet{Lovisari:2019}, no individual profiles are reported, however, their Fig. 4 shows that for $R> 0.7 R_{500}$ profiles tend to separate out with a lower branch located close to 0.
We investigated the nature of these jumps through dedicated simulations.

\begin{figure}
\centerline{\includegraphics[angle=-90,width=8.8cm]{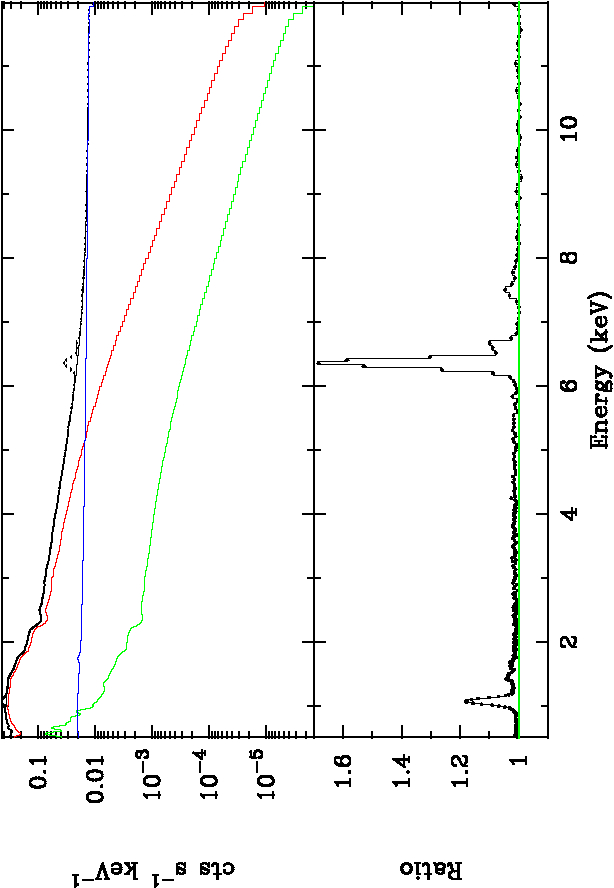}}
\caption{Simulated ICM spectrum from an intermediate radius. In the top panel we show the breakdown of the total spectrum (black) in  source (red), sky foreground and background  (green) and instrumental background (blue) components. In the bottom panel, to highlight the role of line emission, we show the ratio of simulated data to model with the abundance parameter in the source model set to 0 (see Sect. \ref{sec:externalbins} for details).}
\label{fig:sim1}
\end{figure}

\begin{figure}
\centerline{\includegraphics[angle=-90,width=8.8cm]{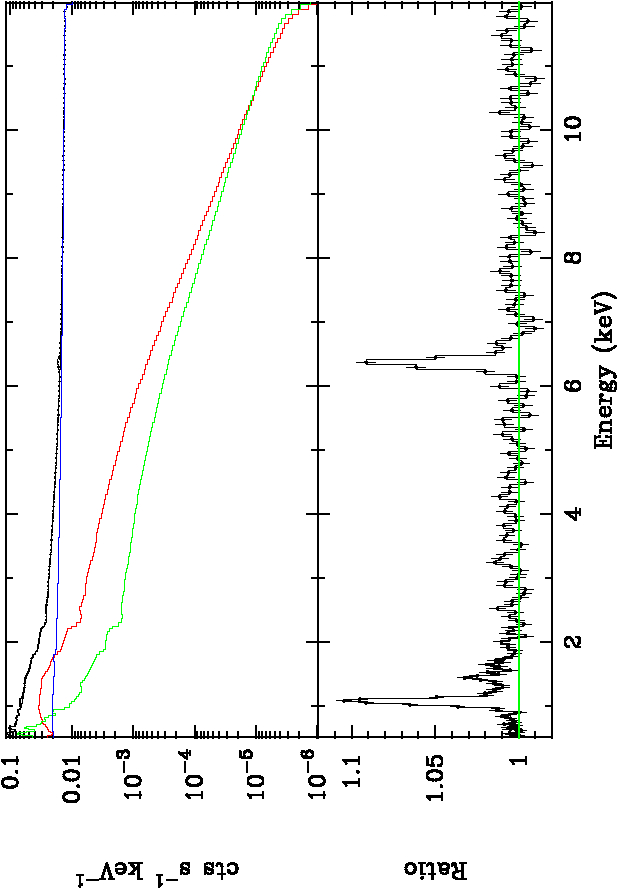}}
\caption{Same as for Fig. \ref{fig:sim1} with the one difference that the surface brightness of the ICM emission has been reduced by a factor of 10 to be representative of cluster outskirts.}
\label{fig:sim2}
\end{figure}

In Fig. \ref{fig:sim1} we show a simulation comprising 3 main components: a source (red line), modelled as  a thermal spectrum (APEC) with a temperature of 5 keV, an abundance of 0.3 $Z_\odot$ and a surface brightness of $2.5\times 10^{-14} {\rm \, erg \, cm^{-2} s^{-1}}$ in the 0.5-2.0 keV band, typical of intermediate radii; a sky component (green line), comprising both X-ray foregrounds and backgrounds and last, but by no means least, an instrumental background component (blue line). 
In the bottom panel of Fig. \ref{fig:sim1} we show the simulated data over model ratio, where the abundance parameter of the APEC component has been set to 0. We clearly see that the abundance measure is driven by the K${\alpha}$ line at 6.7 keV. In Fig. \ref{fig:sim2} we show the same simulation with one difference: the source surface brightness has been dialed down by a factor of 10  to $2.5\times 10^{-15} {\rm \, erg \, cm^{-2} s^{-1}}$ in the 0.5-2.0 keV band. In this new simulation the instrumental background dominates above $\sim$ 2 keV and the sky foregrounds below $\sim 0.7$ keV. An intriguing consequence, highlighted in the bottom panel of Fig. \ref{fig:sim2}, is that the intensity of the Fe L-shell emission with respect to the total continuum is comparable to that of the Fe K$\alpha$ line and that both lines contribute to the abundance measurement.

From these exercises, we infer that, as we move out to larger radii and the surface brightness of the ICM becomes progressively smaller, the L-shell emission, or more precisely  what the fitting algorithm attributes to L-shell emission (see Sect. \ref{sec:l-shell} for details), plays an increasing role in the measurement of the abundance.

\begin{figure*}
\centerline{\includegraphics[angle=0,width=18cm]{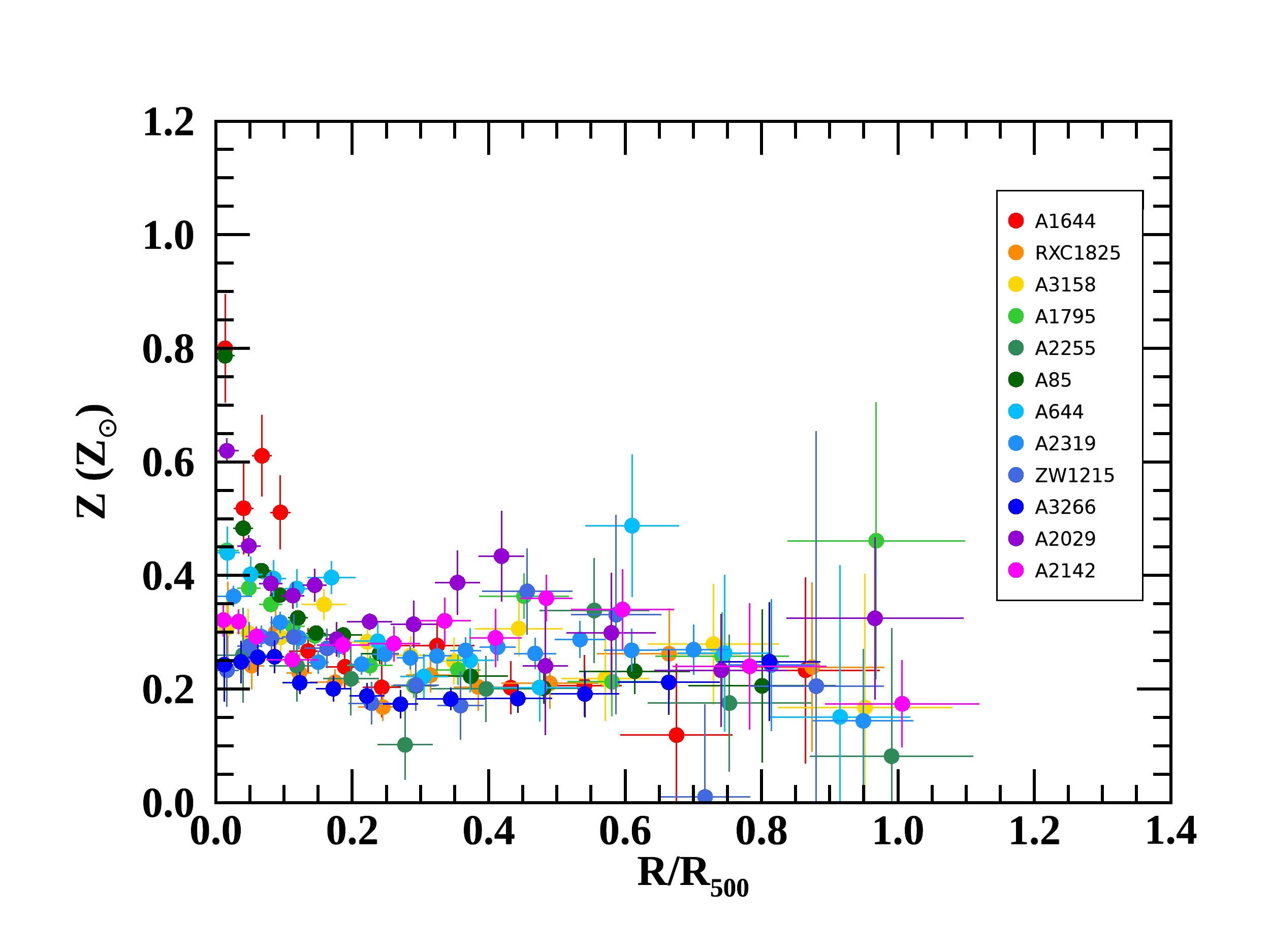}}
\caption{Iron abundance profiles as a function of R/R$_{500}$. Abundance are measured solely through  the Fe K$\alpha$ line  (see Sec. \ref{sec:externalbins} for details). Clusters in the inset are ordered as in Fig.\ref{fig:zfeproj0}.}
\label{fig:zfeproj}
\end{figure*}

\begin{table*}
\centering
\caption{Average abundance $\bar Z$ (and total scatter $\sigma$) in regions: $ R/R_{500} > 0.3$,  $0.3< R/R_{500} < 0.6$, and $R/R_{500} > 0.6$ obtained through standard and modified spectral analysis.} 
\renewcommand\arraystretch{1.5}
\begin{tabular}{|c|c|c|c|c|}
\hline
& \multicolumn{2}{|c|}{Standard analysis}  &  \multicolumn{2}{|c|}{Modified analysis}  \\
\hline
&  ${\bar{Z}}$ & $\sigma$  & $ {\bar{Z}}$ & $\sigma$   \\
\hline
& \multicolumn{4}{|c|}{ (/Z$_\odot$)} \\

\hline
\hline
 $ R/R_{500} > 0.3$ &  0.259 $\pm$  0.003  &  0.067 $\pm$ 0.006 &   0.244  $\pm$ 0.005  & 0.052 $\pm$ 0.005 \\
\hline
 $0.3 <  R/R_{500} < 0.6$ &   0.272 $\pm$  0.003 & 0.046 $\pm$ 0.005 &  0.244 $\pm$  0.006 & 0.052 $\pm$ 0.006\\
\hline
 $R/R_{500} > 0.6$ &   0.154 $\pm$  0.008   & 0.103 $\pm$ 0.014 &  0.242 $\pm$  0.016 & 0.056 $\pm$ 0.008 \\
\hline
\end{tabular}
\renewcommand\arraystretch{1}

\label{tab:intsc_zproj_cfr}
\end{table*}

In light of these findings, we decided to test whether the L-shell emission is responsible for the sudden jumps  and drops in abundance seen in Figs. \ref{fig:zfeproj0} and \ref{fig:zfeA644}. We refitted our spectra excluding the 0.9-1.3 keV energy range, where the emission is observed. Our revised profiles are plotted in Fig. \ref{fig:zfeproj}. As we can see: 1) most measurements close to 0 are shifted to higher values; 2) the errors associated to the new measurements are larger and 3) several measurements located at relatively high abundances  have shifted down. The overall result is that, in the outskirts, the abundance profiles of the different clusters are now less scattered and indicative of a flat metal abundance distribution.

To quantify the changes obtained by fitting  only the Fe K$\alpha$ line, we compare mean abundances derived using the standard fitting method (that we label as "standard" procedure) with those derived by excluding the 0.9-1.3 keV energy range (that we refer to as "modified" procedure). 
In the comparison we exclude the central regions where the source surface brightness is high and spectral results are insensitive to systematics associated to the background. Namely, we restrict the test to the radial range $R/R_{500} > 0.3 $,  which we further split into two sub-intervals:  $0.3< R/R_{500} < 0.6$, and $R/R_{500} > 0.6$.  Results are reported in 
Table \ref{tab:intsc_zproj_cfr}, where mean abundances and total scatters\footnote{In this paper we report total scatters even if formally intrinsic scatters may be more appropriate. The reason for this is that we suspect that, here and elsewhere, our statistical errors might be somewhat overestimated; if this is indeed the case, estimates of the intrinsic scatter will be biased low and incorrect conclusions about  properties of the samples we are investigating might be drawn. By making use of the total scatter as an upper limits to the intrinsic scatter we adopt a conservative approach that prevents us from making excessive claims.}, with their statistical errors, are shown, both for the whole $R/R_{500} > 0.3 $ radial range and for the two sub-intervals. The average metal abundance measured through the standard analysis drops significantly from $Z = 0.272 \pm 0.003 Z_\odot$ for $0.3< R/R_{500} < 0.6$ to $Z = 0.154 \pm 0.008 Z_\odot$ for $R > 0.6 R_{500}$. 
By applying the modified method, the average abundance remains unchanged from $Z = 0.244 \pm 0.006 Z_\odot$ for $0.3< R/R_{500} < 0.6$ to $Z = 0.242 \pm 0.016 Z_\odot$ for $R > 0.6 R_{500}$. 
 In Table \ref{tab:intsc_zproj_cfr} we report the scatter of the data about the average value in the same radial ranges. When fitting with the standard method, the scatter in the outskirts, $R > 0.6 R_{500} $, is high,  $ \sigma = 0.103 \pm 0.014$ (i.e. $\sim 67 \% $), with a notable increase with respect the $0.3< R/R_{500} < 0.6$ radial range. The adoption of the modified recipe reduces the scatter to $\sim 20 \%$  over the whole range $R/R_{500} > 0.3 $, with little variations from $0.3< R/R_{500} < 0.6$ to $R > 0.6 R_{500}$. 
 
\begin{figure}
\centerline{\includegraphics[angle=0,width=9.8cm]{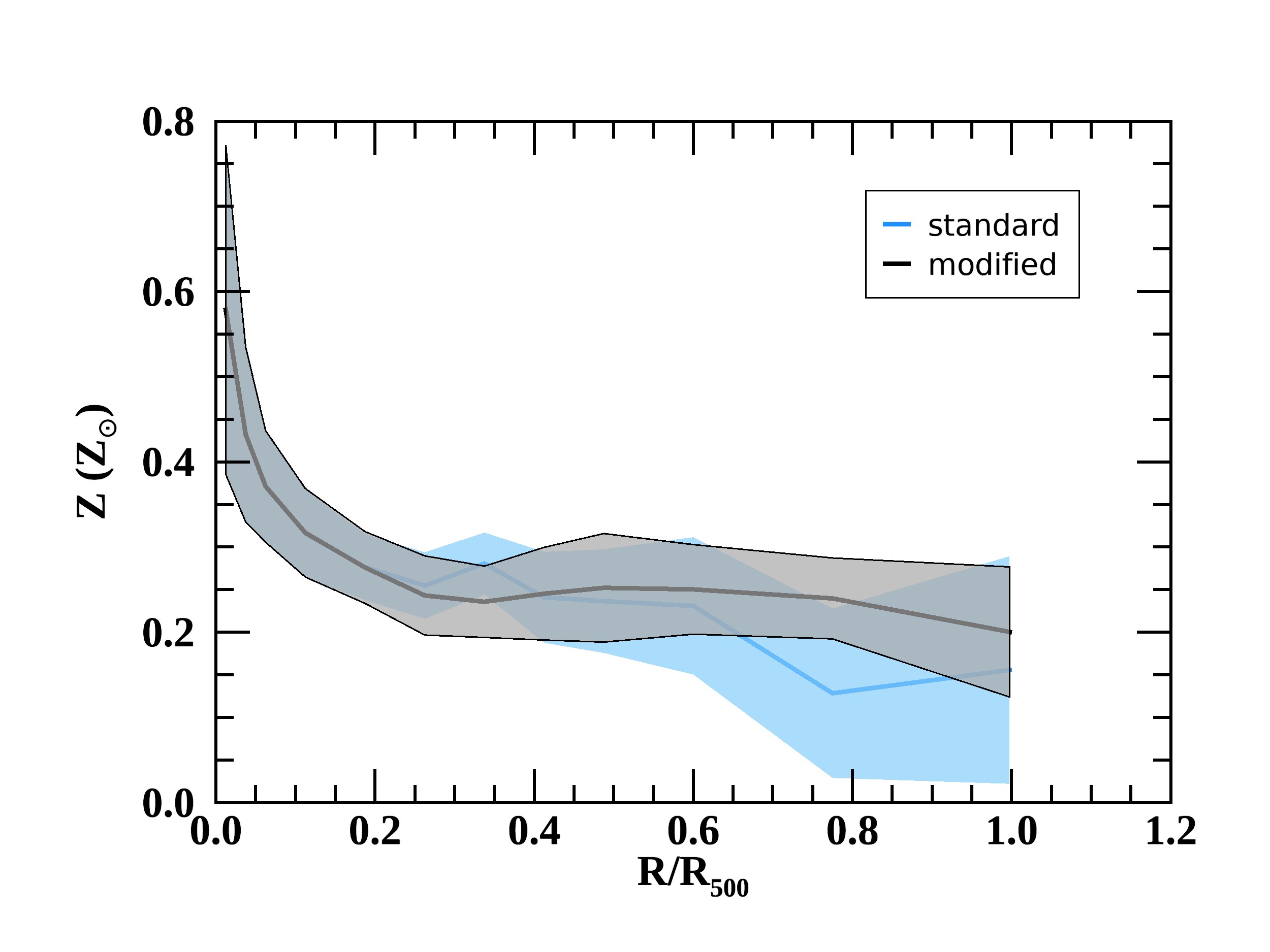}}
\caption{Average metallicity profile (thick line) obtained through the standard (light blue) and the modified (gray) spectral analysis; the 1$\sigma$-scatter is shown as a shaded region.}
\label{fig:stand-mod}
\end{figure}

The global change in the profiles induced by this new method can be appreciated in Fig. \ref{fig:stand-mod}, where we plot the average metallicity profile (derived following the approach detailed in Sect. \ref{sec:projz}) for our sample  using the standard spectral analysis (blue area) and the modified spectral analysis (grey area).  The new fitting procedure provides a flatter and more uniform profile up to $\sim R_{500}$, with a substantially reduced scatter. 
As a side note, we point out that the two profiles, which are indistinguishable at small radii, slowly separate as we move to larger radii, which is precisely what we expect if the difference is due to the increasing importance of the background in the spectral analysis.

\begin{figure}
\centerline{\includegraphics[angle=0,width=9.8cm]{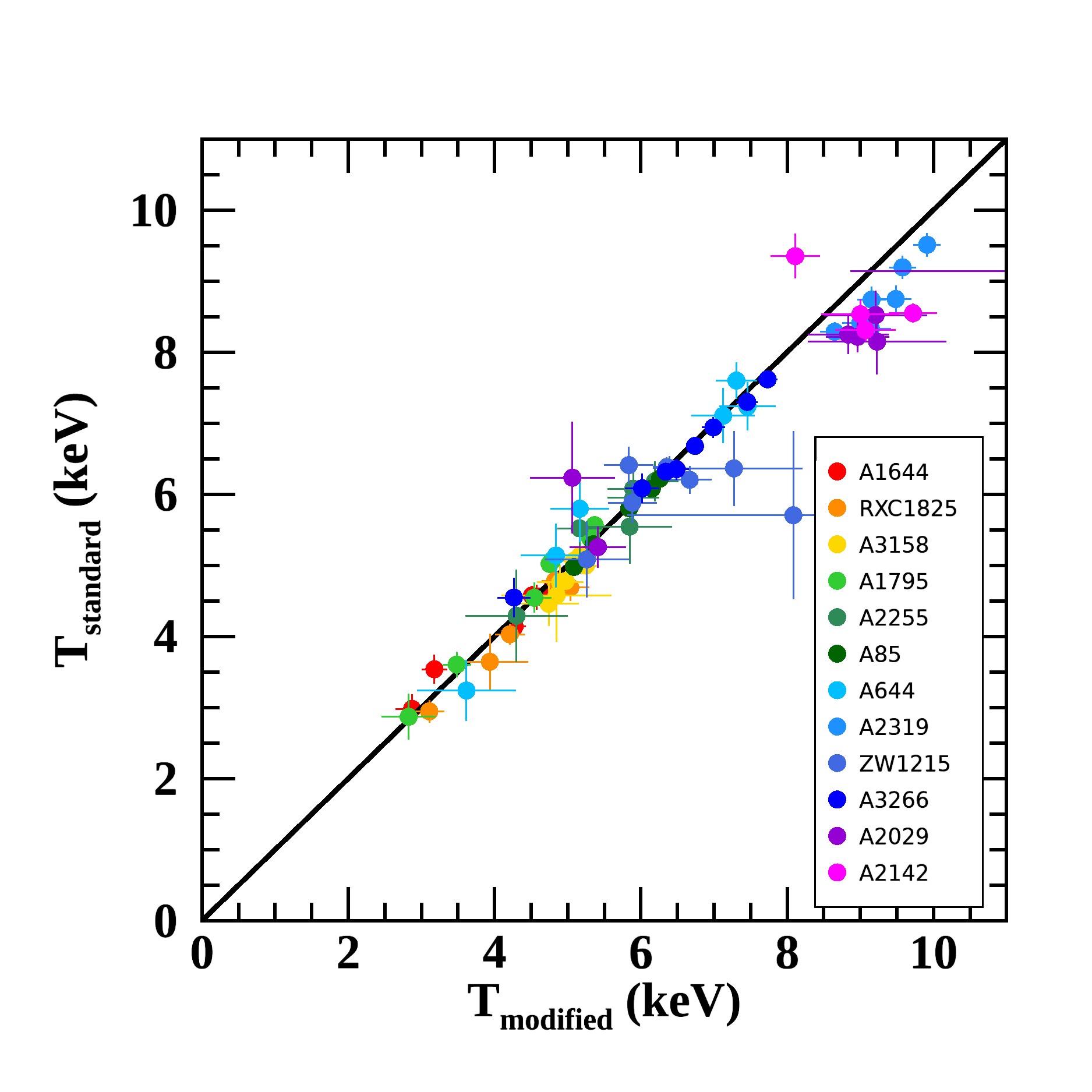}}
\caption{Comparison between the temperatures for the sample derived through the standard method and the modified method. Only points beyond $0.3 R_{500}$ are shown. Clusters in the inset are ordered as in Fig.\ref{fig:zfeproj0}.}
\label{fig:cfr_temp}
\end{figure}

The modified procedure does not affect temperature estimates. In Fig. \ref{fig:cfr_temp} we compare the temperatures obtained with the modified recipe with those derived through the standard spectral analysis. Only points beyond $0.3 R_{500}$ are plotted. The agreement is good, the small differences observed in the high temperature range are associated to known calibration issues \citep[e.g.][]{Nevalainen:2010} between the EPIC soft and hard energy bands. These are, quite likely, the same calibration issues responsible for the underestimation of metal abundances from the L-shell (see Sect. \ref{sec:l-shell} for details).

\subsubsection{The L-shell issue}
\label{sec:l-shell}

We could simply adopt our new abundance measurements by pointing out that they are self consistent Fe K$\alpha$ line measurements and that the effects introduced by Fe L-shell emission contributions do not concern us. However we prefer to take a different point of view, namely that Fe L-shell emission could, at least in principle, provide useful measurements and if they do not we can, at the very least, speculate as to what the problem might be. An important clue comes from the equivalent width of the L-shell emission, which, for a high temperature plasma ($kT > 4 \,{\rm keV } $), is very modest, of the order of few tens of eV. The reason it becomes important and in some instances dominant, as shown by our analysis, is that the equivalent width of the Fe K$\alpha$ line with respect to the dominant continuum, i.e. the instrumental background, rapidly drops as a function of radius.
So, the question of why the Fe L measurements are unreliable and Fe K$\alpha$ are not can be recast into a more meaningful form: why can the  Fe K$\alpha$ measurements be extended to very low equivalent width while the Fe L ones cannot?

Let us start by examining the  Fe K$\alpha$ measurements, these are made with respect to a well determined continuum dominated by one highly reproducible component, namely the instrumental background, moreover, at the relevant energies, the spectral resolution is high and the measurement of the intensity of the line is limited to a narrow energy range. Under these circumstances, the recovered abundance will be characterized by large statistical and small systematic errors or, as some might say, it will be an accurate but not a precise measurement.
Let us now consider the  Fe L measurements, these are made in a region of the spectrum where more than one continuum component contributes, it is also where effective areas peak and calibration issues will impact most significantly on Maximum-Likelihood estimations.
Under these circumstances, there is a strong possibility that the fitting algorithm will use the Fe L-shell emission to "fix" local residua associated either to an insufficient modeling of the continuum components or to the limited calibration of the instrument. 
In other words, L-shell measures will have small statistical errors and, often, large undetected systematic errors leading to precise but inaccurate measurements.

Finally, before closing our discussion, we  will consider the use of 
Fe L-shell measurements of hot (e.g. $>$4 keV) plasmas from an ideal instrument characterized by negligible calibration errors on effective areas and a highly reproducible background.  This is not entirely idle speculation as ESA's next large X-ray mission ATHENA \citep{Nandra_Athena:2013} is being designed with very stringent calibration requirements in mind.
Under the standard assumption of Collisional Ionization Equilibrium (CIE), for temperatures larger than $\sim$ 4 keV, the fraction of Fe in the form of FeXXIV is less then 10\% of the total, the bulk being in FeXXV. This immediately implies that, to achieve systematic uncertainties below 10\% on any Fe L-shell based abundance measure, the CIE will have to be understood to better than the 1\% level. Note also that this is a necessary condition but by no means a sufficient one as several other issues enter into the computation of emission lines. Putting to the side atomic transitions, we  note that the distribution between different excitation states of FeXXIV ions,  which, like CIE, is critically dependent on collisions, needs to be know to a high precision.  All these considerations suggest that a robust measure of Fe abundance should be based on transitions occurring in the dominant Fe ion, which, for temperatures in the 2-10 keV range, are Fe K-shell transitions in FeXXV. Use of less abundant ions, such as FeXXIV, requires a detailed understanding of the equilibrium properties of plasmas  which may become available with the advent of high spectral resolution measurements such as those afforded by XRISM \citep{Tashiro_XRISM:2018,GT:2018} and the ATHENA XIFU \citep{Barret_XIFU:2013, Cucchetti:2019}. 

\subsubsection{Going negative}
\label{sec:negative}

As pointed out in Sect.\ref{sec:dataxspec}, the metal abundance parameter in APEC has a hard lower limit at 0. Such an approach may seem reasonable at a first glance, however it can lead to  biased measurements. In the case at hand, for large radii where abundances are small and statistical errors large, enforcing a hard limit at 0 for the metallicity can  lead to biased results. This is discussed in Appendix A of  \citet{Leccardi_metal:2008} and is treated in greater detail in Appendix \ref{appendix:bias} of this paper.  The obvious solution to this potential bias is to allow abundances to assume negative values. Unfortunately the current implementation of the APEC model within XSPEC does not allow this and we had to resort to a  more indirect approach.
We refitted our spectra substituting the APEC component with MEKAL, which does allow for negative values. We compared results between runs where the metallicity was forced to be positive and runs where it was allowed to assume negative values. In the latter case we found  no instance where the best fitting value was smaller than zero; we did however find a few measures where the 1$\sigma$ confidence interval extended to negative abundances.  For these measurements, in the case of forced positive 
abundance fits, the confidence regions were of course cutoff at zero, however the  errors from the show par command, which are the square roots of the diagonal elements of the covariance matrix, were the same  measured when abundances were allowed to go negative and very close to the error determined by subtracting the best fit value from the upper bound. Thus, at least for the present sample, limiting abundance fits to positive values does not appear to introduce significant biases. We caution our readers that this is by no means a general result and that, for another sample,  result could be quite different. We refer to App.\ref{appendix:bias} for a more detailed discussion of this issue. Of course the definitive solution will be to change the XSPEC code to allow negative values for APEC abundances. 

\section{Iron in the ICM}
\label{sec:fe_icm}

\subsection{Abundance profiles}
\label{sec:projz}
As discussed in detail in Sect. \ref{sec:externalbins} we adopt measures of the Fe abundance entirely based  on the Fe K$\alpha$ line. Ours are the first profiles extending out to $\sim R_{500}$ for a representative sample of massive systems,  for more details see Sect. \ref{sec:intro} and \citet{Eckert_XCOP:2017}. As expected, they feature a significant spread, spanning from 0.23 to 0.80 $Z_\odot$, in the core and in the circum-core regions, where low core entropy systems have more prominent abundance peaks than high core entropy systems \citep{Leccardi:2010}. 
Profiles flatten out beyond $R > 0.3 \, R_{500}$ to values comprised between $0.15 - 0.35 \, Z_{\odot}$, with an averaged value of $Z = 0.244 \pm 0.005 Z_\odot$ and remain remarkably flat out to $R_{500}$. 

\begin{figure}
\centerline{\includegraphics[angle=0,width=9.8cm]{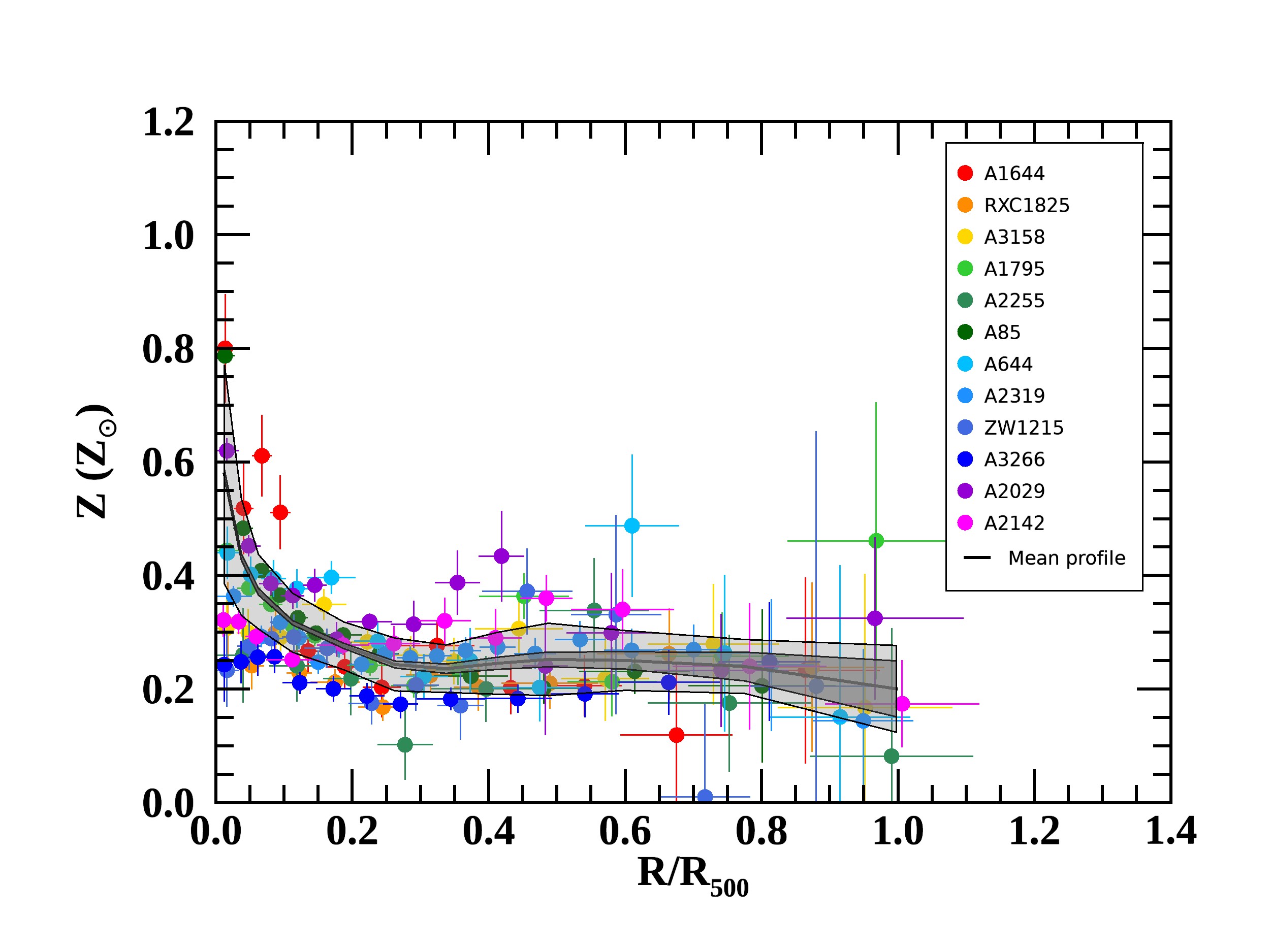}}
\caption{Same as Fig. \ref{fig:zfeproj}  with the average profile (thick line) overlaid. The dark and light shaded areas indicate respectively the statistical error and the total scatter. Clusters in the inset are ordered as in Fig.\ref{fig:zfeproj0}.}
\label{fig:zprof_ave}
\end{figure}

In addition, the distribution of metal abundance  appears uniform across the sample, with little variation from cluster to cluster at radii larger than 0.3$R_{500}$ (total scatter $\sim 20 \%$, see Table \ref{tab:intsc_zproj_cfr}).

\begin{table*}
\centering
\caption{Mean abundance ($\bar Z$) profile for the full sample, the cool-core sub-sample and the non-cool-core sub-sample; $\epsilon$ and $\sigma$ are respectively the statistical error and the total scatter (see text). For the full sample we also report the median abundance profile $Z_{\rm median}$, along with the 1$\sigma$ statistical error, that will be discussed in Appendix B.}
\renewcommand\arraystretch{1.3}
\begin{tabular}{|c|c|c|c|c|c|c|c|c|c|c|}
\hline
& \multicolumn{4}{|c|}{Full Sample} & \multicolumn{3}{|c|}{Cool-Core SubSample} & \multicolumn{3}{|c|}{Non-Cool-Core SubSample} \\
\hline
Radius &  $\bar{Z}$  & $\epsilon$ & $\sigma$ & $Z_{\rm median}$ & $\bar{Z}$ & $\epsilon$ & $\sigma$  & $\bar{Z}$  & $\epsilon$  & $\sigma$ \\
\hline
 (/$R_{500}$) & \multicolumn{10}{|c|}{(/Z$_\odot$)}  \\
\hline
\hline
0.000 - 0.025 & 0.578 & 0.008 & 0.193 & 0.440 $\pm$ 0.024 & 0.648 & 0.009 & 0.158 & 0.330 & 0.017 & 0.048 \\
0.025 - 0.050 & 0.432 & 0.006 & 0.103 & 0.363 $\pm$ 0.013 &  0.475 & 0.007 & 0.086 & 0.325 & 0.011 & 0.048 \\
0.050 - 0.075 & 0.371 & 0.006 & 0.066 & 0.293 $\pm$ 0.014 & 0.408 & 0.007 & 0.040 & 0.297 & 0.010 & 0.038 \\
0.075 - 0.150 & 0.317 & 0.004 & 0.052 & 0.299 $\pm$ 0.008 &   0.342 & 0.005 & 0.037 & 0.276 & 0.006 & 0.046 \\
0.150 - 0.225 & 0.276 & 0.005 & 0.042 & 0.275 $\pm$ 0.010 & 0.291 & 0.006 & 0.024 & 0.258 & 0.007 & 0.051 \\
0.225 - 0.300 & 0.243 & 0.006 & 0.046 & 0.241 $\pm$ 0.011 & 0.259 & 0.009 & 0.044 & 0.231 & 0.008 & 0.045 \\
0.300 - 0.375 & 0.236 & 0.008 & 0.042 & 0.249 $\pm$ 0.013 & 0.236 & 0.014 & 0.044 & 0.236 & 0.010 & 0.041 \\
0.375 - 0.450 & 0.245 & 0.011 & 0.054 & 0.274 $\pm$ 0.022 & 0.252 & 0.019 & 0.069 & 0.242 & 0.013 & 0.046 \\
0.450 - 0.525 & 0.252 & 0.013 & 0.064 & 0.263 $\pm$ 0.034 & 0.237 & 0.023 & 0.065 & 0.259 & 0.016 & 0.062 \\
0.525 - 0.675 & 0.250 & 0.015 & 0.053 & 0.268 $\pm$  0.023 & 0.222 & 0.027 & 0.028 & 0.263 & 0.018 & 0.056 \\
0.675 - 0.875 & 0.240 & 0.025 & 0.047 & 0.240 $\pm$ 0.038 & 0.231 & 0.048 & 0.041 & 0.243 & 0.030 & 0.049 \\
0.875 - 1.120 & 0.200 & 0.049 & 0.076 & 0.174 $\pm$ 0.083 & 0.316 & 0.108 & 0.081 & 0.170 & 0.055 & 0.035 \\

\hline
\end{tabular}
\renewcommand\arraystretch{1}
\label{tab:mean_prof}
\end{table*}

In Fig. \ref{fig:zprof_ave} we overlay the average profile on the individual ones.
The bins adopted for the average profile are listed in the first column of Table \ref{tab:mean_prof}. To derive the average profile, we followed the approach described in \citet{Leccardi_temp:2008}:
in each bin, the average metallicity has been computed by performing a weighted average on values of the different clusters whose original bins are (even partially) included in the bin of the average profile.
More precisely, we assign a weight $w_{i,j,k}$ to the $j$-th bin of the $i$-th cluster,  which measures the fraction of the  $j$-th bin included in $k$-th bin of the average profile; $j$ bins which are totally included in a $k$ bin are assigned a weight of 1. 
These weights are combined with the usual statistical weights $1/\sigma_{i,j}^2$ associated to the abundance measures.

Thus, the average metallicity in the $k$-th bin is given by:

\begin{equation}
    \bar{Z}_k ={ { \sum\limits_{i=1}^{n}{  \sum\limits_{j=1}^{m_i}{ w_{i,j,k} { { Z_{i,j} } \over { \sigma^2_{i,j} } } } } } \over {  \sum\limits_{i=1}^{n}{  \sum\limits_{j=1}^{m_i}{ w_{i,j,k} { { 1 } \over { \sigma^2_{i,j} } } } } }   } \, ,
\end{equation}

\noindent
where $Z_{i,j}$ and $\sigma_{i,j}$ are, respectively, the metal abundance and the error measured in the $j$-th bin of the $i$-th cluster. Index $i$ runs from 1 to the number of clusters in the sample, $n$, while $j$ runs from 1 to the number of bins of the profile of the $i$-th cluster, $m_i$.

Similarly, statistical errors $\epsilon_k$ are given by
\begin{equation}
    \epsilon_k = {1 \over \sqrt{ \sum\limits_{i=1}^{n}{  \sum\limits_{j=1}^{m_i}{ w_{i,j,k} { { 1 } \over { \sigma^2_{i,j} } } } } } } \, ,
\end{equation}
        
\noindent
and the total scatter $\sigma_k$ is obtained as: 

\begin{equation}
    \sigma_k = { \sqrt{ \sum\limits_{i=1}^{n}{  \sum\limits_{j=1}^{m_i}{ w_{i,j,k} { { \left(Z_{i,j} -\bar{Z}_k \right)^2  } \over { \sigma^2_{i,j} } } } } } \over \sqrt{ \sum\limits_{i=1}^{n}{  \sum\limits_{j=1}^{m_i}{ w_{i,j,k} { { 1 } \over { \sigma^2_{i,j} } } } } } } \, .
\end{equation}

\noindent
The average profile $\bar{Z}$ is plotted in Fig. \ref{fig:zprof_ave} and values are reported in Table \ref{tab:mean_prof} along with the statistical errors $\epsilon$ and the total scatters $\sigma$ (columns 2-4).

The mean abundances beyond $0.3 R_{500}$ range from 0.2 to 0.25 Z$_\odot$, remaining extremely flat all the way out to $R_{500}$. Measurements have very small statistical uncertainties, smaller than 6\% within $0.675 R_{500}$ and  still smaller than 25\% in the two last bins covering the $[0.675-1] R_{500}$ range.
Excluding the central bins where the scatter of the metallicity is dominated by the diversity between cool-core/non-cool-core systems, the total scatter stays below  $0.065 Z_\odot$ ($\lesssim 25\%$) on almost all the cluster volume, except for the last bin, where, due to the large statistical errors, the total scatter reaches $0.076 Z_\odot$ ($\sim 38\%$).

\begin{figure}
\centerline{\includegraphics[angle=0,width=9.8cm]{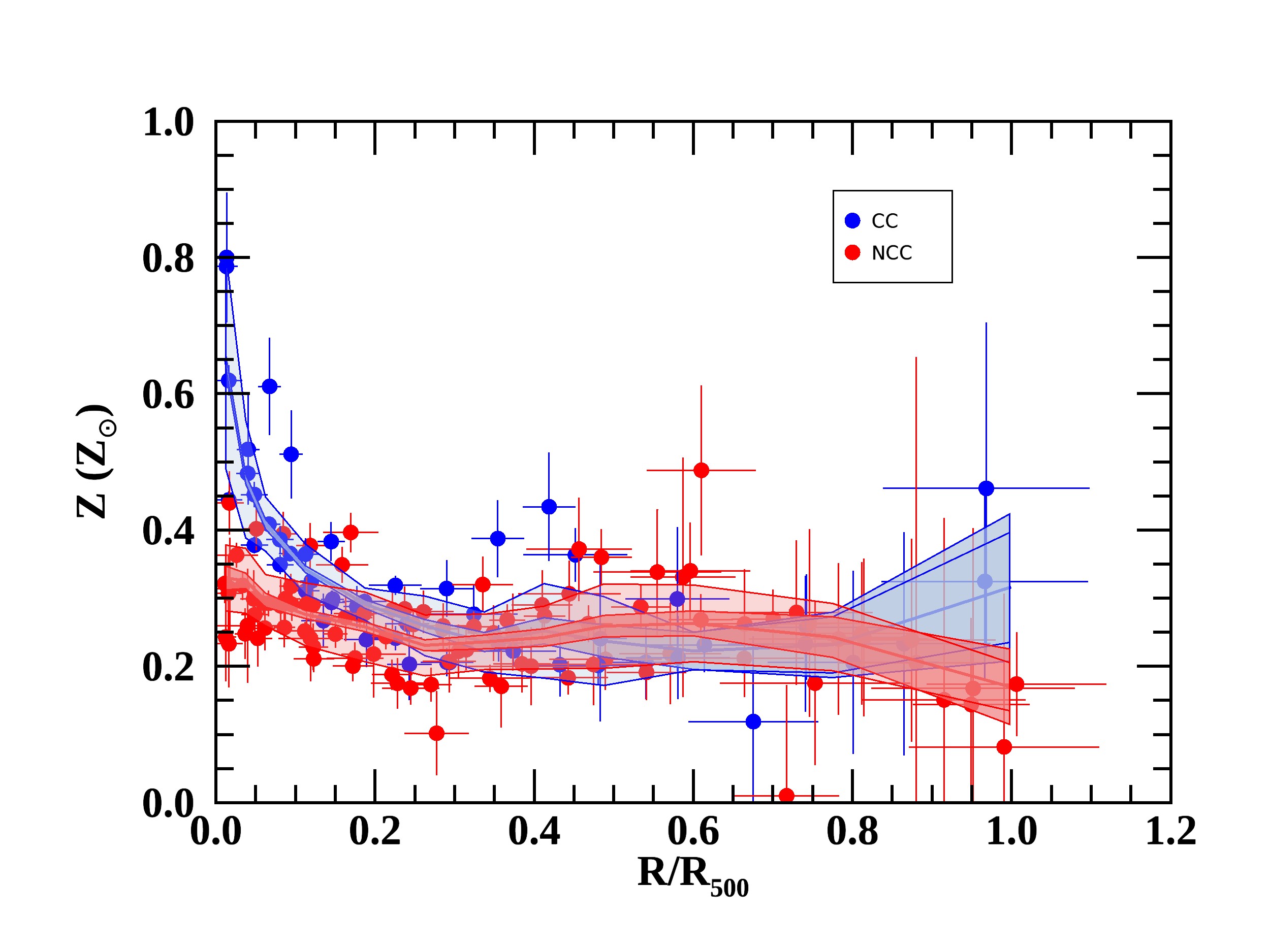}}
\caption{Abundance profiles as a function of R/R$_{500}$ for cool-core (blue) and non-cool-core (red) subsamples. Average profiles for both subsamples are overlaid. The dark and light shaded areas indicate respectively the statistical error and the total scatter.}
\label{fig:zprof_CC_NCC}
\end{figure}

To investigate any dependence of the metal abundance on the dynamical state of the clusters, we divided our sample into two subsamples, namely cool-core and non-cool-core systems. To separate clusters into these two classes, we adopt as an indicator the central entropy $K_0$ provided by  \citet{Cavagnolo:2009}. We consider as cool-cores all the clusters with central entropy $K_0 < 30 \, {\rm keV \, cm}^2$. Our sample includes 4 cool-core systems and 8 non-cool-core systems.
In Fig. \ref{fig:zprof_CC_NCC} we report mean profiles separately for the two subsamples, cool cores are plotted in blue, while non-cool cores are shown in red. Values for the mean profiles are reported in Table \ref{tab:mean_prof} along with the statistical errors $\epsilon$ and the total scatter $\sigma$ (columns 6-11). 
As expected, the two average profiles differ in the innermost bins. On the contrary, they do not reveal any significant discrepancy beyond $ 0.3 R_{500}$.
The overall mean metal abundance at $R > 0.3 R_{500}$ is $Z = 0.254^{+0.018}_{-0.013} Z_\odot$ with a total scatter $\sigma = 0.058 \pm +0.009 Z_\odot$ ($\sim 23\%$) for the cool-core subsample and $Z = 0.249^{+ 0.008}_{- 0.008} Z_\odot$ with a total scatter $\sigma = 0.051 \pm 0.005 Z_\odot$ ($\sim 20\%$) for the non-cool-core subsample, showing an excellent agreement between the two subsamples.  

From Figs. \ref{fig:zfeproj} to \ref{fig:zprof_CC_NCC}  we derive three key results: 1) the metal abundance profile is remarkably flat for $R > 0.3 R_{500}$; 2) the profile beyond $0.3 R_{500}$ shows a modest scatter implying that, whatever the enrichment mechanism may be, it must provide a uniform level of  metal abundance for all the clusters over  most of their volume; 3) cool-core and non-cool-core clusters have similar metal distributions beyond $0.3 R_{500}$. If, on the one hand, differences in the metal content at the center of clusters can be ascribed to the early contribution of stars currently residing in the BCGs invariably found in low-entropy systems \citetext{see \citealp{Degrandi:2004} and \citealp{DeGrandi:2014} for a discussion on this point}, on the other, the lack of differences in the rest of the cluster volume implies the enrichment mechanisms at work in those regions are not related to phenomena which are typical of only one of the two classes, such as relatively recent merging events, which are signatures of disturbed systems \citep[see also][]{Urdampilleta:2019,Mernier_review:2018}.

\subsection{Systematic uncertainties}
\label{sec:systx}
In Sect. \ref{sec:externalbins} we have identified and corrected  a major source of 
systematic errors on measurements in low surface brightness regions located at large radii. Here we investigate further sources 
of systematic errors through a complementary approach i.e. we turn to hydro-dynamic simulations.  

An important assumption, made in all estimates of the Fe abundance, is that spectra can be fit with single temperature and abundance  models, although it is well known that some degree of 
multi temperature and multi abundance has to be present \citep[e.g.][]{Molendi:2016} and that this can have an impact on measurements \citep[e.g.][]{Buote:2000}. 
Simulations have been employed to address this point by several authors. 
\citet{Rasia:2008} produced spectra starting from the  distribution of temperature and 
abundance in their simulations, convolved them with instrumental response functions and 
then fitted them with single temperature and abundance models. 
They found that, for temperatures above $\sim$3 keV, where the abundance measure is 
dominated by the K$\alpha$ line, the measured metallicity is within  5\% 
of the emission weighted (hereafter EW) abundance derived directly from the simulation. 
Since all our spectra feature temperatures in excess of 
3 keV (see Fig. \ref{fig:cfr_temp}) and the vast majority of 4 keV, we can conclude that any bias in the  measurements will be very small and confined to those spectra with temperatures smaller than 4 keV. 

An important aside, which we will pick up again in Sect. \ref{sec:disc}, is that only for the external regions of  massive clusters, where K$\alpha$ emission is sufficiently strong, can we proceed with a measure of the metallicity that is not plagued by the many biases, e.g. \citet{Rasia:2008}, and plasma code uncertainties \citep{Mernier_review:2018} associated to L-shell emission.

Having argued that K$\alpha$ based measurements of the abundance are affected
by very modest biases with respect to the EW abundance, 
we now quantify a possible systematic error due to the assumption  that the spectrally measured abundances, 
which are emission-weighted, are equivalent to mass-weighted (hereafter MW) ones.  
Differences between EW and MW estimates arise when multi-temperature plasma is present, as is the case in  outer regions where, on top of declining temperature profiles, clumps of denser and colder material might be present.
Given the correlation between entropy and metallicity, found both in observational data \citep[e.g.][]{Leccardi:2010}
and in cosmological simulations \citep{Biffi:2017}, these clumps of high-density, 
low-entropy gas could bias high the EW abundance estimate. 

\begin{figure}
\centerline{\includegraphics[angle=-90,width=8.8cm]{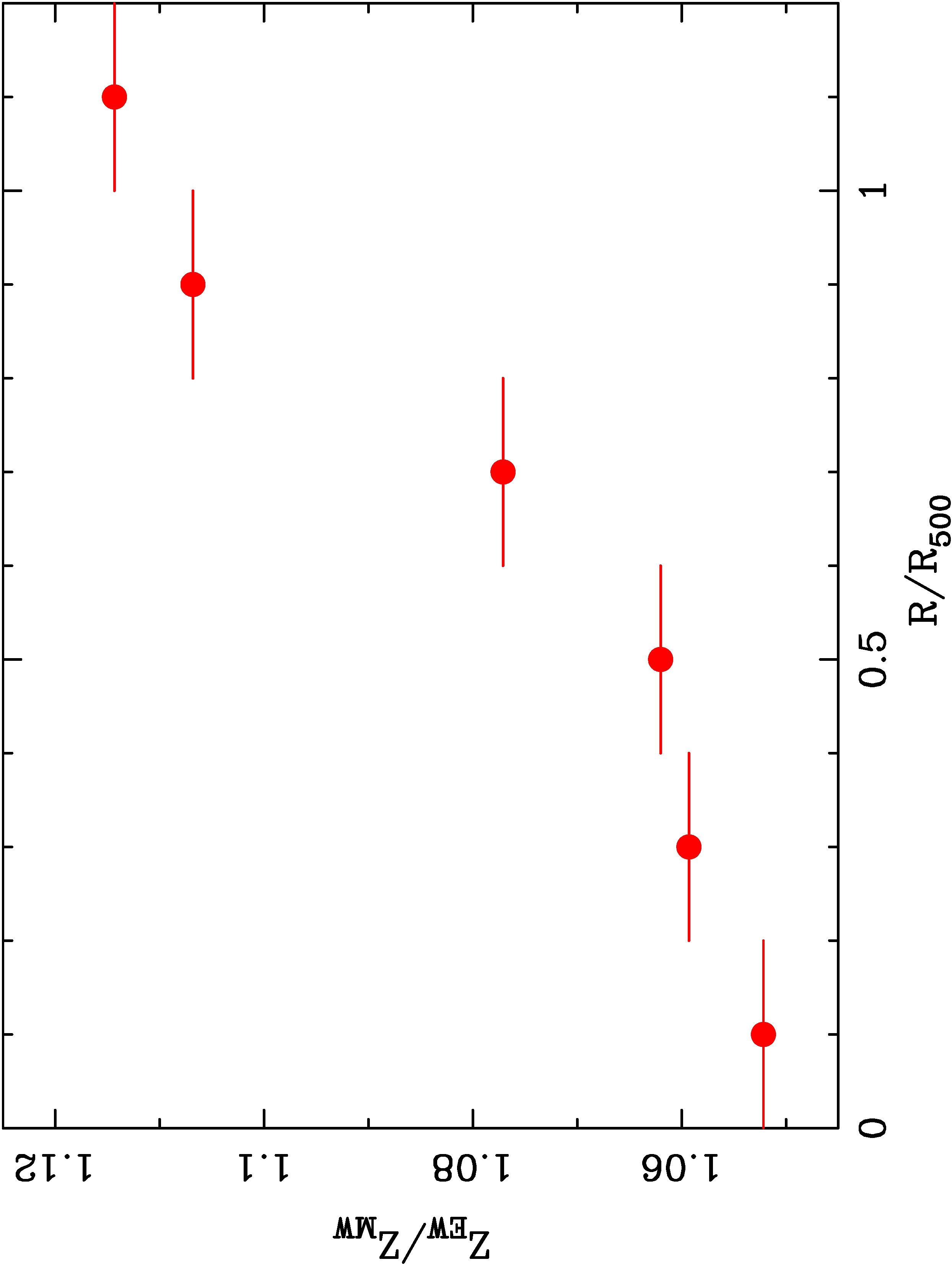}}
\caption{Ratio of the EW to MW abundance profiles as a function of $R/R_{500}$ averaged over  29 
simulated clusters, see Sect. \ref{sec:systx} for details.}
\label{fig:zratio}
\end{figure}

The complex conditions described above cannot be captured by simple simulations such as those presented in Sect. \ref{sec:modif_spec_anal}.
We therefore investigate this aspect through a set of clusters extracted from hydrodynamical cosmological simulations. 
The general properties of this suite, derived from a modified version of GADGET-3, are described in \citet{Rasia:2015}. 
The history of the ICM enrichment and the origin of the outskirts metallicity in present-day clusters are  discussed respectively 
in \citet{Biffi:2017} and \citet{Biffi:2018},
where details of the sub-grid models linked to stellar formation, stellar evolution and chemical production, are also provided. 
We consider 29 massive clusters (25 with temperatures within $R_{500}$ in the range $4-10$ keV and 4 with temperatures around $2-3$ keV) 
and produce both EW and MW maps of Fe with the  Smac code \citep{Dolag:2005}. The maps are generated in pixels of about 4 kpc on the side and span in projection along the line of sight 10 Mpc, thereby including the contribution of gas particles  in the cluster surroundings. 
We divided each map in concentric annuli centered on the emission 
peak with bounding radii $0-0.2 R_{500}$, $0.2-0.4 R_{500}$, $0.4-0.6 R_{500}$,
$0.6-0.8 R_{500}$, $0.8-1.0 R_{500}$ and $1.0-1.2 R_{500}$ and averaged the EW and MW abundances over 
these regions.
Next, we computed the EW over MW ratio radial profile for each system and, 
finally averaged it over all systems.  
As expected, see Fig. \ref{fig:zratio}, the bias increases, but only moderately, as we move from center to periphery, it lies roughly between 5\% and 10\% over the radial range of interest $\sim 0.4-1.0 R_{500}$.

Finally we combine the two effects: for the bias on the estimate of the EW abundance from the single phase spectral modeling we take the 5\% value estimated by \citet{Rasia:2008}, for the EW to MW bias, using results presented in Fig. \ref{fig:zratio}, we take 10\%.
 We combine these two measures and assume an overall 15\% systematic error on abundances. Note that this is a conservative estimate as the two effects go in opposite directions and should  tend to cancel each other out, rather than build up. 

Our systematic error is smaller than typical measurement errors for 
individual systems, implying that, on single clusters, we are dominated by statistical uncertainties. Conversely, as shown in Fig. \ref{fig:zprof_ave} and Tables \ref{tab:intsc_zproj_cfr} and \ref{tab:mean_prof}, our systematic error is larger than the statistical error on the average profile, 
suggesting that, despite the relatively small size of our sample, further improvements in the measure of the mean cluster abundance in the outskirts will come from advancements in analysis methods rather than from an increase in  sample size.

\subsection{Comparison with previous measurements}

\begin{figure}
\centerline{\includegraphics[angle=0,width=9.8cm]{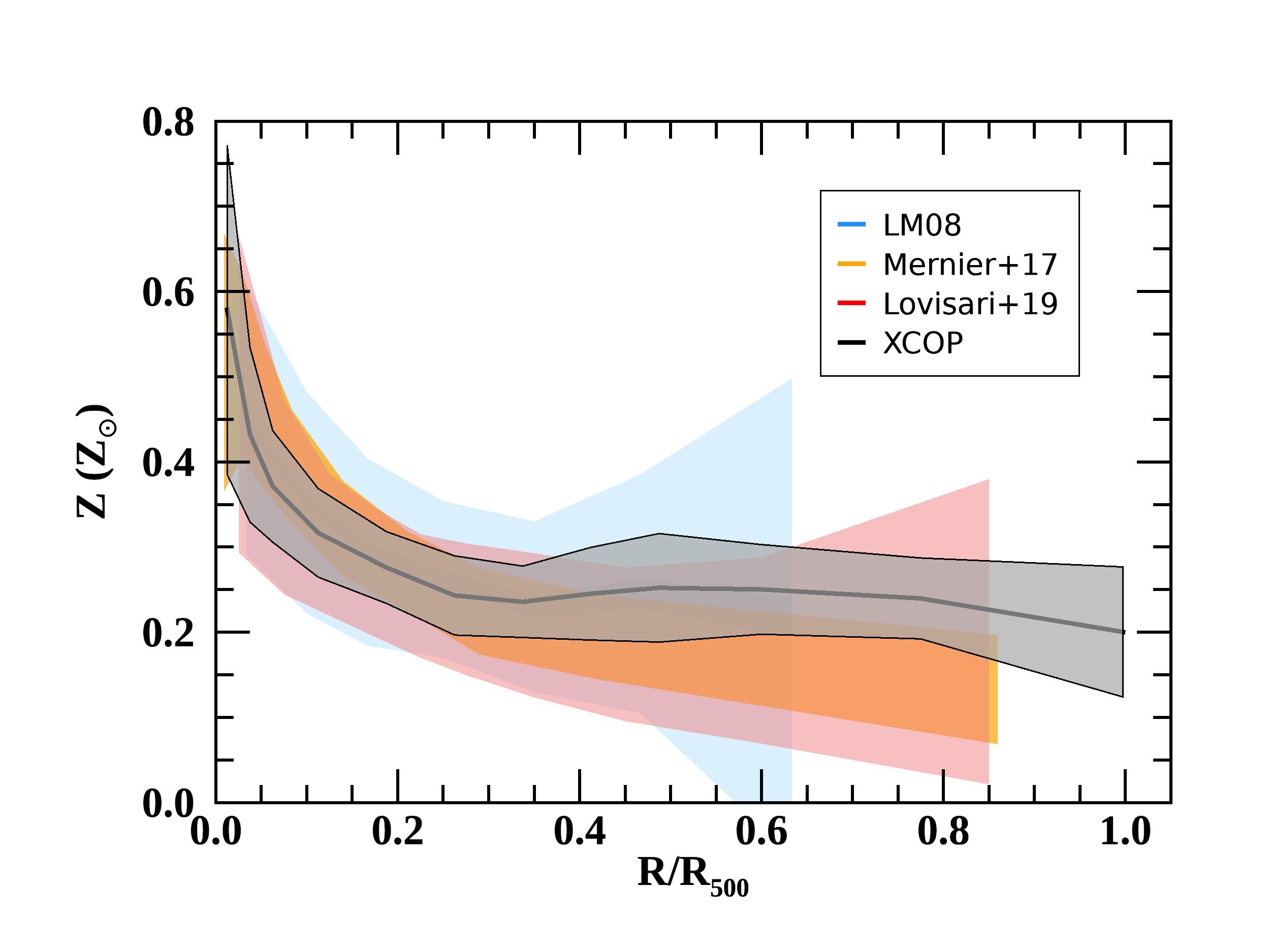}}
\caption{Abundance profiles as a function of R/R$_{500}$: comparison between different measurements.}
\label{fig:zprof_cfr}
\end{figure}

In Fig. \ref{fig:zprof_cfr} we compare our abundance profile with those obtained by other authors. The profile reported in \citet{Leccardi_metal:2008} extends to about $0.6 R_{500}$ and in its outermost bins are characterized by a very large scatter.  It is worth pointing out that part of this scatter reflects background systematics, which at the time were identified, but only partially understood. Our profile fits comfortably within the rather weak bound posed by the \citet{Leccardi_metal:2008} measures. 

The CHEERS sample \citep{Mernier:2017} was constructed to investigate cluster cores rather than outskirts, it comprises mostly nearby systems and only for a handful of systems can measures be extended beyond intermediate radii. Starting from $\sim 0.3R_{500}$ their mean profile features a slow decline. While some overlap between the shaded regions indicating the scatter of the CHEERS and X-COP profiles is present all the way out to the largest radii, mean values start to differ significantly beyond $0.45 R_{500}$. The decline seen in this profile might be due to the same artifact we have found in our own sample. Indeed, as can be seen in the Appendix of \citet{Mernier:2017}, 3 of their systems namely: A133, A2029 and A2597 show evidence of a sudden drop in metal abundance at large radii. In the case of A2029, which is also part of our sample, the drop is essentially the same we find from our  "standard" analysis. However it must be recognized that other factors could contribute to shaping the mean abundance profile reported in  \citet{Mernier:2017}, more specifically: 1) the CHEERS sample includes a lot of cooler systems (galaxy groups and elliptical galaxies); 2) it was fitted with SPEX  rather than XSPEC. 

Finally, we compare our measures with those from \citet{Lovisari:2019}. As for the CHEERS sample, the focus of the analysis is on the central regions, namely on the impact of the AGN on the central metal abundance, rather than on the outskirts, and only for a limited number of systems can abundances be measured out to large radii.
Unfortunately, the authors do not provide profiles for individual systems, however, from their Fig. 4, we see that beyond $\sim 0.6 R_{500}$ measures tend to split up with some  around 0 and others at larger values, this could be yet another manifestation of the artifact described in Sect. \ref{sec:externalbins}.

 We do not compare our measurements with those from Suzaku \citep[e.g.][]{Urban:2017} for 2 main reasons:
 1) measures refer to a handful of systems for which only a  limited azimuthal sampling is available; 2) unlike all samples discussed here, the spectral analysis performed for the Suzaku data does not rely on full background modeling, more specifically the instrumental background, which is critical for this analysis, as it dominates at the Fe-K$\alpha$ line energy, is subtracted rather than modeled.
 We note that a similar choice has been made by other authors, indeed neither \citet{Mernier:2017} or \citet{Lovisari:2019} compare their iron abundance measurements with those from Suzaku.
 
In summary, what emerges from this comparison is that our abundance profile is unique in several ways: 1) it is constructed from a representative sample of massive systems;
2) it is based on individual profiles that, thanks to the off-set pointings afforded by the X-COP Very Large Program, extend to $R_{500}$;
3) it features a small statistical scatter all the way to large radii and last but not least
4) it has been corrected for a major systematic error that has impacted on previous measurements.
As we shall see in the next sections, these properties allow us to perform an estimate of the metal content in massive clusters of unprecedented quality.

\subsection{Deprojection and cumulative iron masses.}
\label{sec:depro}

\begin{figure}
\centerline{\includegraphics[angle=0,width=9.8cm]{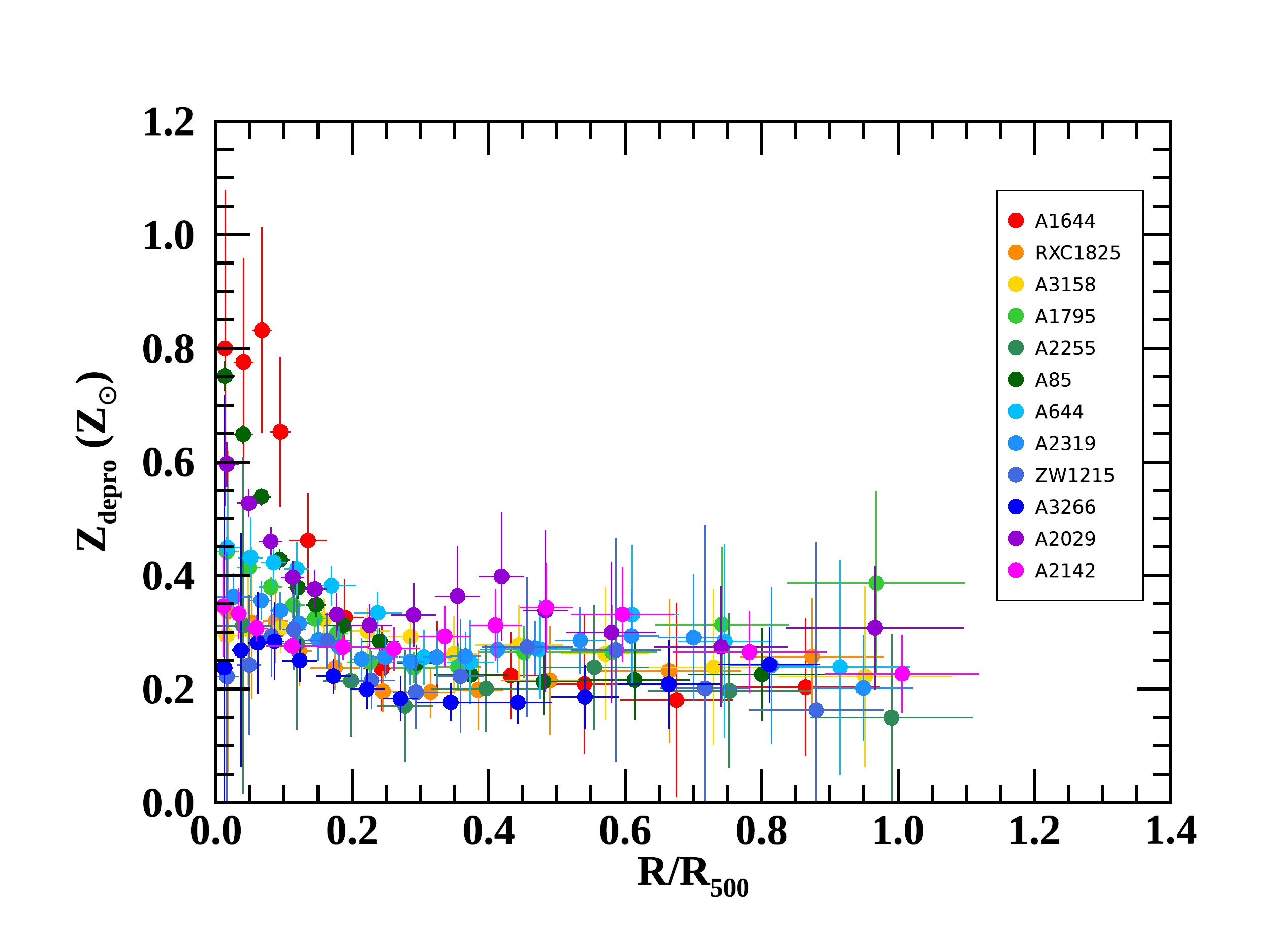}}
\caption{Deprojected iron abundance profile as a function of $R/R_{500}$. Clusters in the inset are ordered as in Fig.\ref{fig:zfeproj0}.}
\label{fig:zfeprof}
\end{figure}

 We define the deprojected iron abundance  as: $Z_{\rm depro}=n_{\rm Fe}/(Z_{n,\odot} n_{\rm H})$, where  $n_{\rm Fe}$ and $n_{\rm H}$ are the iron and hydrogen densities, by number, respectively, and $Z_{n,\odot}$ is the solar iron abundance by  number. Assuming solar abundances reported in \citet{AG:1989}, $Z_{n,\odot} = 4.68 \times 10^{-5} $.
We adopted the standard onion-skin technique to deproject abundances 
\citep{Kriss83,Ettori_depro:2002a}, including a correction factor to account for the emission of the cluster beyond the outermost bin \citep[see][for details]{Ghizzardi_M87:2004, McLaughlin:1999}.
Data have been slightly smoothed, with a boxcar average of 3-points width, before deprojection, to reduce nonphysical fluctuations that would be enhanced by the deprojection process. 
In Fig. \ref{fig:zfeprof}  we show the deprojected abundance profiles for the whole sample.  $Z$ profiles, both projected and deprojected, for individual cluster, are reported in Appendix \ref{sec:app_zprofs}.
The average deprojected metallicity at $R > 0.3 R_{500}$ is   $Z_{\rm depro} = 0.242 \pm 0.008$.

  The iron mass enclosed within a sphere of radius $R$, $M_{\rm Fe}(<R)$ \citep[see][]{Degrandi:2004}, can be expressed as:

\begin{equation}
     M_{\rm Fe} (<R) = 4\pi A_{\rm Fe} m_{\rm H} {Z_{n,\odot}}~ \int_0^R Z_{\rm depro}(r) ~n_{\rm H}(r)~ r^2 dr, 
\label{eq:mfe}
\end{equation} 

\noindent
where $A_{\rm Fe}$ is the atomic weight of iron, and $m_{\rm H}$ is the atomic unit mass.
The hydrogen density $n_{\rm H}$ is derived from the gas density $n_{\rm gas}$ through the usual relation $ n_{\rm gas}= \left(1 + n_e/n_{\rm H} \right)n_{\rm H} = 2.21 n_{\rm H}$, where $n_e$ is the electron density; $n_{\rm gas}$ has been obtained through deprojection as detailed in \citet{Ettori:XCOP2019} and \citet{Eckert_clump:2015}.

\begin{figure}
\centerline{\includegraphics[angle=0,width=9.8cm]{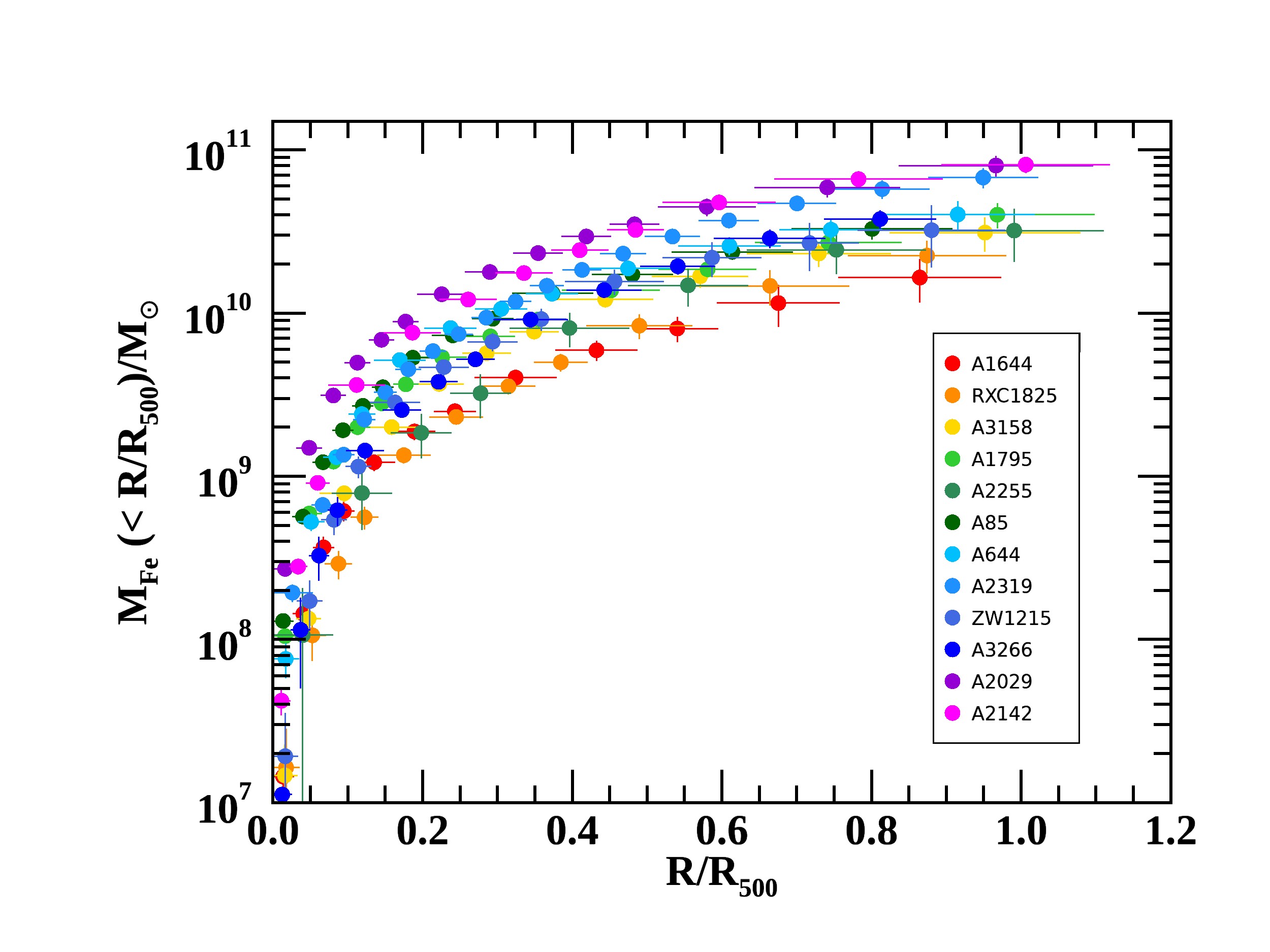}}
\caption{Iron mass profiles as a function of $R/R_{500}$. Clusters in the inset are ordered as in Fig.\ref{fig:zfeproj0}.}
\label{fig:mfecum}
\end{figure}

In Fig. \ref{fig:mfecum} we show all the cumulative iron mass profiles $M_{\rm Fe}(<R)$.  At $R=R_{500}$ the iron mass $M_{\rm Fe,500} \equiv M_{\rm Fe}( < R_{500})$  for our systems ranges roughly between $10^{10}$M$_\odot$ and $10^{11}$M$_\odot$.
$M_{\rm Fe,500}$ values for all clusters in our sample are listed in Table \ref{tab:mfe500-mstar500}. 

\begin{table*}
\centering
\caption{Iron mass $M_{\rm Fe,500}$, gas mass $M_{\rm gas, 500}$, hydrostatic mass $M_{500}$, and stellar mass $M_{\rm star,500}$ enclosed within $R_{500}$ for the X-COP clusters.}
\renewcommand\arraystretch{1.5}
\begin{tabular}{|c|c|c|c|c|}
\hline
Cluster  & $M_{\rm Fe,500}$ & $M_{\rm gas,500}$  & $M_{\rm 500}  ^{(a)}$ &  $M_{\rm star,500}$\\
 & ($10^{10}$M$_\odot)$ &  ($10^{13}$M$_\odot)$ &  ($10^{14}$M$_\odot)$ &   ($10^{12}$M$_\odot)$ \\
\hline
A1644 & $ 2.01^{+0.57}_{-0.62}$ &  $4.77^{+0.09}_{-0.10}$ & $3.48 \pm 0.20$ & - \\
RXC1825 & $ 2.72^{+0.62}_{-0.59}$ & $5.94^{+0.07}_{-0.07}$ & $4.08 \pm 0.13$ & - \\
A3158 & $ 3.28^{+0.81}_{-0.79}$ & $6.73^{+0.08}_{-0.07}$ & $4.26 \pm 0.18$  & - \\
A1795 & $ 4.19^{+0.69}_{-0.75}$ & $6.92^{+0.08}_{-0.09}$ & $4.63 \pm 0.14$ & $ 3.02^{+0.28}_{-0.28} $ \\
A2255 & $ 3.23^{+1.21}_{-1.03}$ & $8.80^{+0.14}_{-0.14}$ & $5.26 \pm 0.34$ & - \\
A85 & $ 4.24^{+0.60}_{-0.58}$ & $9.17^{+0.09}_{-0.08}$ & $5.65 \pm 0.18$ & $ 2.10^{+0.33}_{-0.33} $ \\
A644 & $ 4.40^{+0.89}_{-0.94}$ & $8.05^{+0.10}_{-0.11}$ & $5.66 \pm 0.48$ & $ 3.70^{+0.38}_{-0.38} $ \\
A2319 & $ 7.14^{+0.95}_{-0.90}$ & $14.51 ^{+0.10}_{-0.10}$ & $7.31 \pm 0.28$ & $ 5.11^{+0.46}_{-0.46} $  \\
ZW1215 & $ 3.61^{+1.56}_{-1.77}$ & $8.72^{+0.08}_{-0.08}$ & $7.66 \pm 0.52$ & $ 3.34^{+0.39}_{-0.39} $  \\
A3266 & $ 4.91^{+0.63}_{-0.57}$ & $11.94^{+0.11}_{-0.09}$ & $8.80 \pm 0.57$ & - \\
A2029 & $ 8.33^{+1.13}_{-1.14}$ & $13.34^{+0.16}_{-0.17}$ & $8.82 \pm 0.35$ & $ 6.51^{+0.47}_{-0.47} $ \\
A2142 & $ 8.07^{+0.81}_{-0.84}$ & $14.80^{+0.14}_{-0.13}$ & $8.95 \pm 0.26$ & $ 6.97^{+0.51}_{-0.51} $ \\
\hline
\end{tabular}
\renewcommand\arraystretch{1}
\begin{list}{}{}
\item[Notes:]
   $\mathrm{^{(a)}}$ Hydrostatic mass from \citet{Ettori:XCOP2019}.
\end{list}
\label{tab:mfe500-mstar500}
\end{table*}

\subsection{Mass-weighted abundances and iron-to-gas-mass ratio}

From the deprojected profiles $Z_{\rm depro}$ derived in Sect. \ref{sec:depro}, we can compute mass weighted abundances within a given radius, $Z_{\rm mw} (<R)$. We define $Z_{\rm mw} (<R)$ as 
\begin{equation}
 Z_{\rm mw} (<R) = { { \int_0^R Z_{\rm depro}(r) n_{\rm H}(r) {r^2 dr}} \over{\int_0^R{n_{\rm H}(r) r^2 dr}}}
\, . 
\end{equation}



\begin{figure}
\centerline{\includegraphics[angle=0,width=9.8cm]{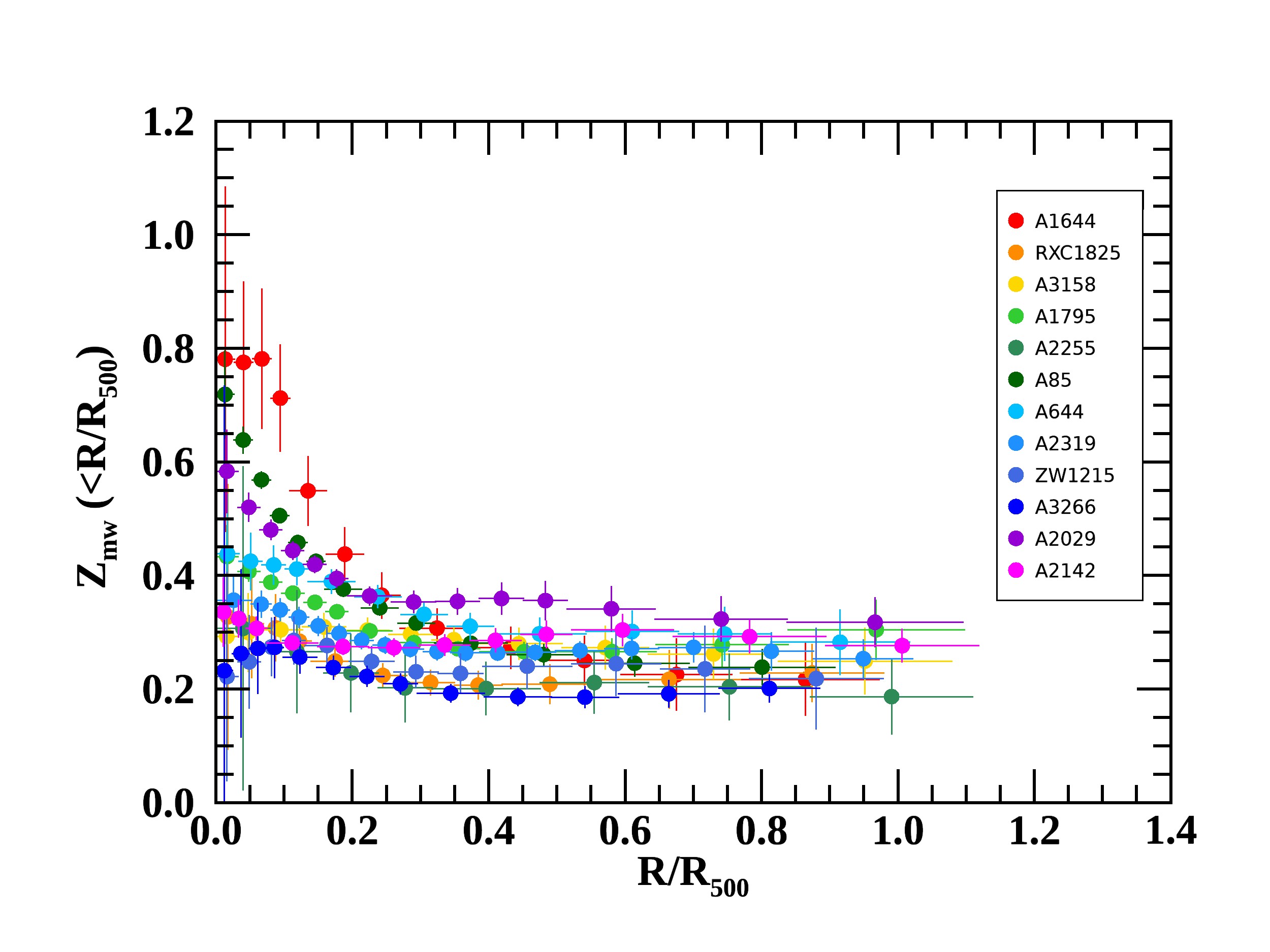}}
\caption{Profiles of mass-weighted iron abundance within a given radius as a function of $R/R_{500}$ for all systems in our sample. Clusters in the inset are ordered as in Fig.\ref{fig:zfeproj0}.}
\label{fig:mfe_su_mgas_resc_to_Zfe}
\end{figure}

\noindent
Mass weighted abundances for all our systems are reported in Fig. \ref{fig:mfe_su_mgas_resc_to_Zfe}. 
 Working with integrated quantities has the advantage that the cumulative functions are more regular than density functions. Consequently the mass-weighted $Z_{\rm mw}(<R)$ profiles appear smoother and more regular than the $Z_{\rm depro}$ profiles.
At $R_{500}$ the mean value of the mass-weighted abundance $Z_{\rm mw,500} =0.247^{+0.013}_{-0.012} Z_\odot$. 
The total scatter of $Z_{\rm mw,500}$ about the mean value is small, $0.037 \pm 0.008 Z_\odot,$ i.e. $\sim 15\%$. This finding has significant implications that will be discussed in Sect. \ref{sec:disc}.

\subsection{Scaling relations}
\label{sec:scal_rel_mfe}

Having derived for the first time robust estimates of Fe masses within $R_{500}$, we investigate scaling relations between $M_{\rm Fe,500}$ and other ICM observables, namely the gas mass $M_{\rm gas,500}$ and the total mass $M_{500}$ enclosed within $R_{500}$; $M_{\rm gas,500}$ is obtained as usual by integrating $n_{\rm gas}$, while the total masses $M_{500}$ have been derived in \citet{Ettori:XCOP2019}, under the assumption of hydrostatic equilibrium. Values for $M_{\rm gas,500}$ and $M_{500}$ for our sample are reported in Table \ref{tab:mfe500-mstar500}.

\begin{table*}

\centering
\caption{Best fit parameters for scaling relations. $\alpha$ and $\beta$ are defined in eqn. \ref{eqn:fe_scal_rel} and \ref{eq:mstar}, $\sigma$ is the intrinsic scatter about the best fit line in the log-log space.} 
\renewcommand\arraystretch{1.5}
\begin{tabular}{|c||c|c|c||c|c|c|}
\hline
& \multicolumn{3}{|c||}{$M_{\rm gas,500}$}  &  \multicolumn{3}{|c|}{$M_{500}$}  \\
\hline \hline
&  $\alpha$ & $\beta$ &  $\sigma (\%)$  &  $\alpha$ & $\beta$ & $ \sigma (\%)$  \\
\hline
$M_{\rm Fe,500}$ & $0.01^{+0.03}_{-0.03}$ & $1.10^{+0.20}_{-0.19} $ & $15.0 \pm 3.2 $ & $0.005^{+0.04}_{-0.04}$ & $1.18^{+0.27}_{-0.26} $ & $ 22.3 \pm 4.7 $\\
\hline
$M_{\rm star,500}$ & $-0.03^{+0.08}_{-0.08}$ & $1.11^{+0.60}_{-0.52} $ & $ 20.7 \pm 5.9 $ &  $-0.07^{+0.08}_{-0.09}$ & $1.39^{+0.68}_{-0.75} $ & $ 22.0 \pm 6.1 $ \\
\hline
\end{tabular}
\renewcommand\arraystretch{1}

\label{tab:scal_rel}
\end{table*}

\begin{figure}
\centerline{\includegraphics[angle=0,width=9.8cm]{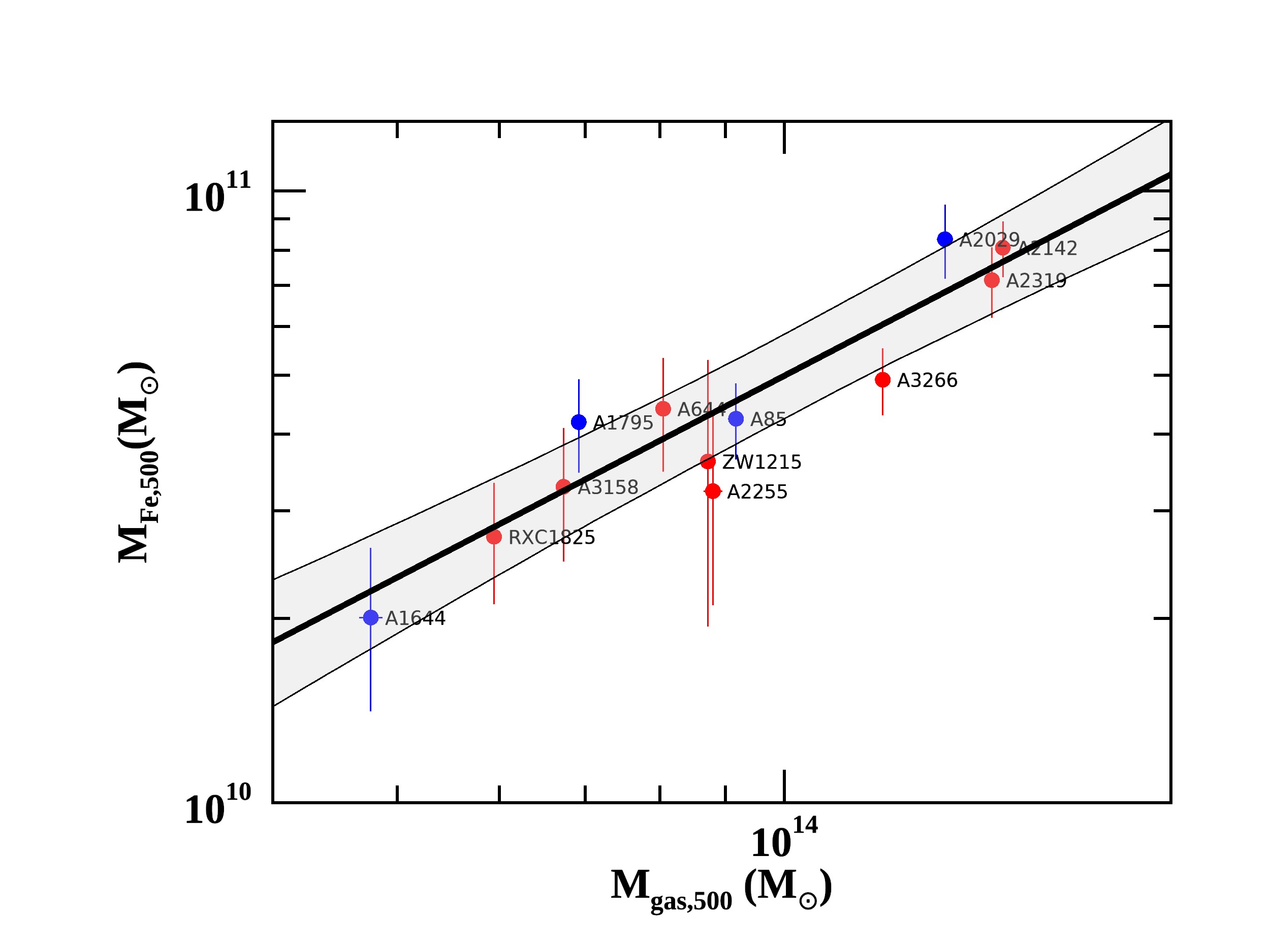}}
\centerline{\includegraphics[angle=0,width=9.8cm]{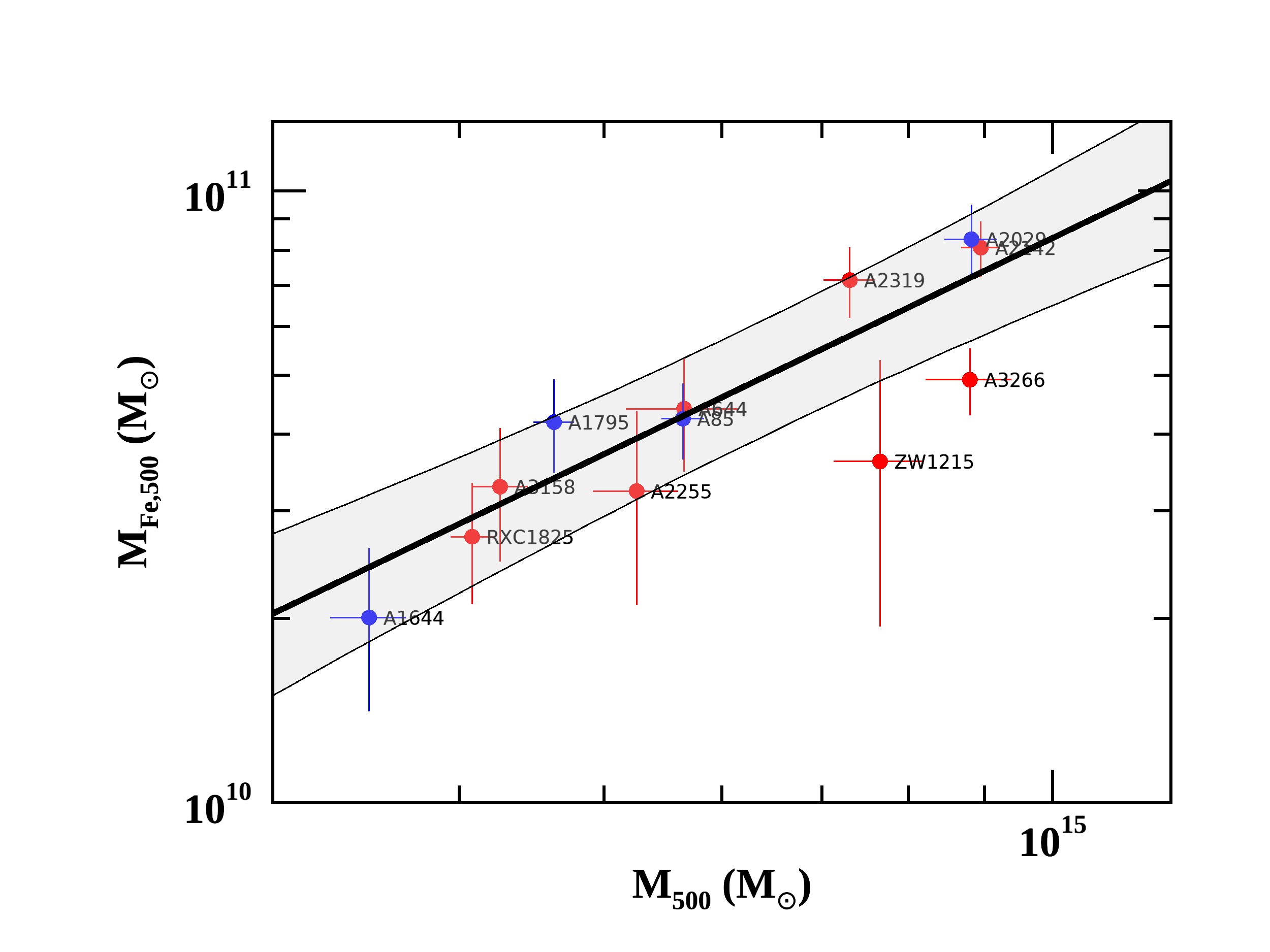}}
\caption{(Top panel) $M_{\rm Fe}$ within $R_{500}$ vs $M_{\rm gas}$ within $R_{500}$; (Bottom panel) $M_{\rm Fe}$ within $R_{500}$ vs $M_{\rm tot}$ within $R_{500}$. Blue and red mark cool-core and -non-cool-core systems, respectively. Black lines are the best fits and shaded areas indicate the 1$\sigma$ confidence regions, including intrinsic scatters.}
\label{fig:mfe_mgas_mtot}
\end{figure}

In Fig. \ref{fig:mfe_mgas_mtot} we show  $M_{\rm Fe,500}$ as a function of $M_{\rm gas,500}$ (top panel) and $M_{500}$ (bottom panel) for all clusters in our  sample. 
In both cases, we observe a very clear correlation without any difference between cool-core and non-cool-core clusters (marked with blue and red circles, respectively).

We model the relations with a standard power law:

\begin{equation}
    {{M_{\rm Fe,500}} \over {\hat{M}_{\rm Fe,500}}}= 10^{\alpha} \left({{M_X \over {\hat{M}_X}}}\right)^\beta  ,
    \label{eqn:fe_scal_rel}
\end{equation}

\noindent
where  $M_X= M_{\rm gas,500}$ or $M_X= M_{500}$. 
We center the relation on the pivot values $\hat{M}_{\rm Fe,500} = 4.24 \times 10^{10} M_\odot $, $\hat{M}_{\rm gas,500} = 8.80 \times 10^{13} M_\odot $, and $\hat{M}_{\rm 500} = 5.66 \times 10^{14} M_\odot$ set at the median of the distributions of $M_{\rm Fe,500}$, $M_{\rm gas,500}$, and $M_{500}$ respectively.

We fit our data, by performing a linear regression analysis, in the log-log space, using the IDL package {\sl linmix\_err.pro} by \citet{Kelly:2007}, based on Bayesian inference, which treats  measurement errors in both variables and allows an intrinsic scatter about the regression line. Though we are aware of the unreliability of intrinsic scatter estimates for our data (see Sect. \ref{sec:externalbins}), we decided to include this quantity into our regression analysis, to account for some intrinsic spread. Conscious of the possible underestimation of this quantity (see Sect. \ref{sec:externalbins}) we will, conservatively, make use of the total scatter as a measure of the dispersion of the data about the best fit value.

The best-fitting values and the total scatter are listed in the first row of Table \ref{tab:scal_rel}, in both cases the slope value is close to 1, meaning that the relation is consistent with being linear.
No segregation between cool-cores and non-cool-cores is found in either relations.
This suggests a lack of any causal connection between the mechanism responsible for the formation of a cool core and the overall enrichment of the ICM.

Focusing on the relation $M_{\rm Fe,500} - M_{\rm gas,500}$, the total scatter of the data about the best fit power-law relation is $\sigma = (15.0 \pm 3.2) \%$. 
It is worth pointing out that the tight relation between $M_{\rm Fe,500}$ and $M_{\rm gas,500}$ is not connected to the flatness of the $Z$ profiles, but to the small scatter in the  average abundance profile. 
If the scatter in the average abundance profile were high, clusters having similar $M_{\rm gas,500} $ could have different $M_{\rm Fe,500}$, leading to a large scatter in the scaling relation.
Scatter for the $M_{\rm Fe,500} - M_{500}$ relation is slightly larger  $\sigma = (22.3 \pm 4.7) \%$ than for the $M_{\rm Fe,500} - M_{\rm gas,500}$ relation.

Scatter values reported in Table \ref{tab:scal_rel} (see also Sect. \ref{sec:scale_mstar}) are very similar and reciprocally consistent within 1-2 $\sigma$. Nonetheless, the scatter for the $M_{\rm Fe,500}$ vs $M_{\rm gas,500}$  relation is the lowest and best constrained  making the correlation between gas and Fe mass the tightest amongst the scaling relations we have investigated. Henceforth, in the admittedly limited mass range covered by X-COP, $M_{\rm gas,500}$ can be used as a proxy for $M_{\rm Fe,500}$ and, the latter can be estimated for any massive system from the former with an uncertainty of $\sim 15\% $. 
Of course, this is just another way of saying that our distribution of  $Z_{\rm mw,500}$ measures features a small scatter.


\section{Optical data}
\label{sec:datao}

The X-COP sample has seven clusters in common with the Multi Epoch Nearby Cluster Survey (MENeaCS). For these, we have deep optical imaging data, taken with the purpose of performing a weak-lensing analysis of the clusters (Herbonnet et al., in prep.) and their constituent galaxies \citep{sifon18}, and to study transient phenomena \citep[particularly intra-cluster supernovae,][]{sand11}. \citet{vdB15} (hereafter vdB15) combined the original g- and r-band imaging with additional u- and i-band imaging and performed a study of the stellar mass content of these clusters. The optical analysis presented in this paper is based on their data set. We provide a brief summary here, for more details we  refer to vdB15.

The basis of the study is g- and r-band imaging data taken with MegaCam at the Canada-France-Hawaii Telescope (CFHT). Imaging data in the u- and i-bands was also acquired using the Wide Field Camera at the Isaac Newton Telescope. Sources are detected in the r-band, and aperture fluxes are measured in each filter stack from PSF-homogenised images using Gaussian weight functions. Typical 5-sigma aperture magnitude limits are 24.3, 24.8, 24.2 and 23.3 in the ugri-filters, respectively. The aperture fluxes form the basis for the spectral energy distribuion (SED) fitting, from which stellar mass to light ratios are estimated for each galaxy. The stellar population libraries from \citet{Bruzual_Charlot:2003} are used to  model the SED. The star formation history is parametrized as $ SFR \propto e^{-t/\tau}$, where the time-scale $\tau$ is allowed to range between 10 Myr and 10 Gyr. We assume a \citet{chabrier03} IMF, solar metallicity, and the \citet{Calzetti:2000} dust extinction law.  Using the total flux measured in the r-band, luminosities are converted into stellar masses. Initially each galaxy is assumed to be part of the cluster (to set the luminosity distance), in a second step a statistical subtraction of fore- and background interlopers is performed using multi-band photometry available in the COSMOS field \citep{muzzin13a}. 

\subsection{Stellar mass profiles}
\label{sec:dataoana}

To accumulate stellar mass radial profiles, we centre clusters on the X-ray centroids, as opposed to the BCGs  in vdB15. The impact of how these profiles are centred is generally small (the centres differ by at most 2\% of $R_{500}$).

\begin{figure}
\centerline{\includegraphics[angle=0,width=9.8cm]{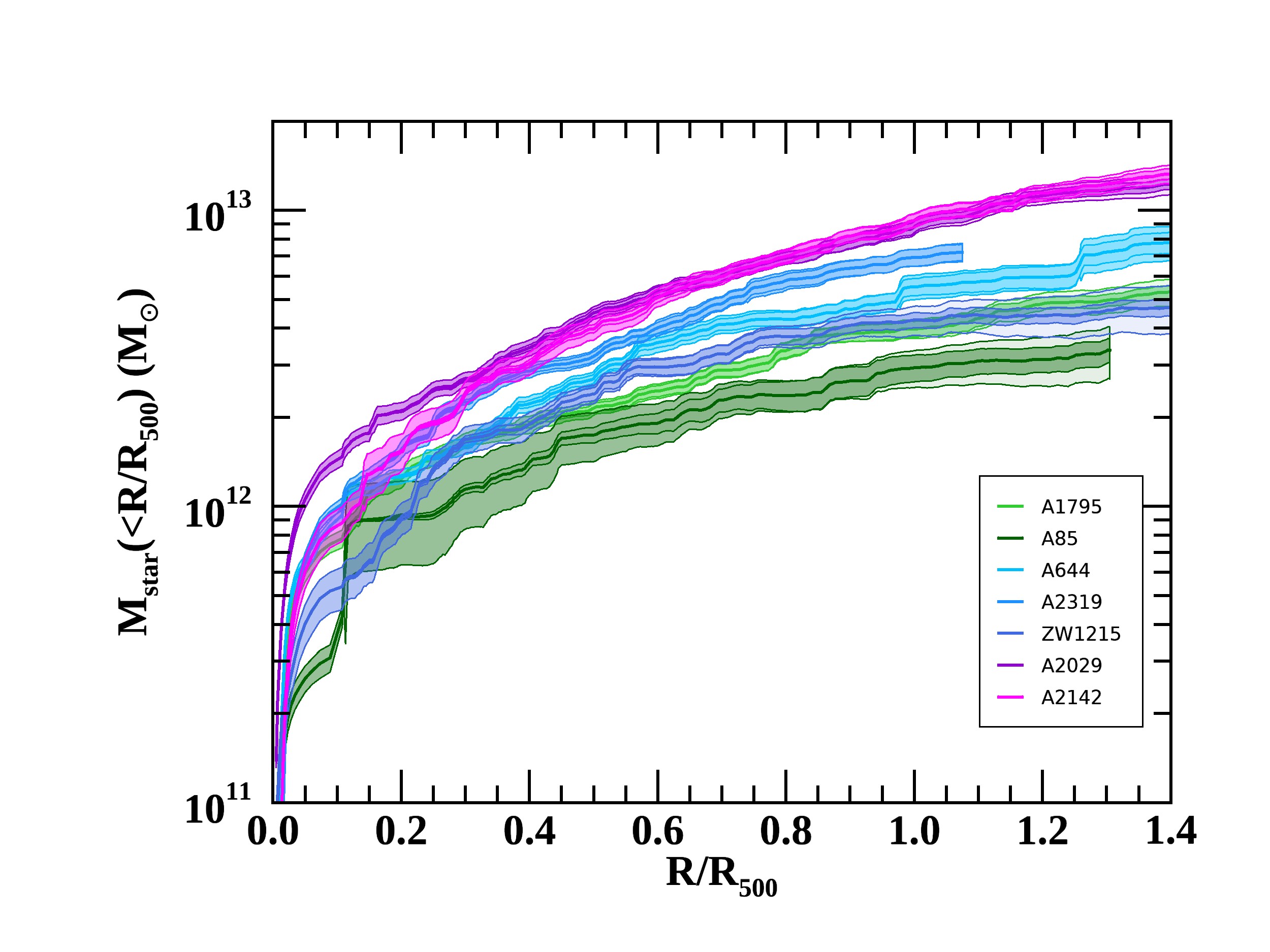}}
\caption{Cumulative stellar mass profiles. Dark shades are statistical uncertainties, faint shades are cosmic variance uncertainties.}
\label{fig:mstarprof}
\end{figure}

For the profiles we consider all galaxies with stellar masses in excess of $10^9\,\mathrm{M_{\odot}}$ \footnote{We note that these galaxies make up the vast majority of the stellar mass component in a typical cluster galaxy population. To showcase this, we integrate the SMF, which was measured in \citet{vdb18} to have a \citet{Schechter:1976} form with low-mass slope of $\alpha \approx -1$ and characteristic mass $M^*=10^{10.8}\,\mathrm{M_{\odot}}$, down to infinitely low masses. If, instead, we only consider galaxies with $M_{\star}>10^{9}\,\mathrm{M_{\odot}}$, this would account for more than 98\% of the stellar mass that is present in the entire population.}. 
We point out that the BCG is included in our mass computation, while the Intra-Cluster Light (hereafter ICL) is not.
We consider, for each galaxy, the statistical uncertainty on the estimated stellar masses, using the measured flux uncertainties. For this we bootstrap the flux measurements within their uncertainties, and perform 100 perturbations for each galaxy. When we combine the stellar masses of all galaxies, we account for these uncertainties on individual stellar masses. 
We account for and subtract the contribution of fore- and background galaxy interlopers as described in vdB15. Briefly, we make use of the COSMOS field and, for consistency, only consider data taken in the ugri-bands. The subtraction procedure introduces a Poisson noise term (statistical uncertainty), but also a systematic uncertainty related to how representative the reference field is of the true cluster field background. Since the COSMOS field is relatively small, there is a substantial uncertainty due to field-to-field (often called “cosmic”) variance. We estimate this uncertainty using \citet{moster11} and tested in vdB15 that the estimated variance is consistent with the scatter obtained when considering the four spatially-independent CFHT Legacy Deep fields used for the background subtraction.
Stellar masses are integrated to derive the cumulative stellar mass profile for each cluster, $M_{\rm star}(<R)$. Profiles for $M_{\rm star}(<R)$ are shown in Fig. \ref{fig:mstarprof}. The statistical uncertainty due to flux errors on individual galaxies and the systematic uncertainty on the profile due to field-to-field variance are shown separately in the cumulative stellar-mass profiles (see Fig. \ref{fig:mstarprof}).
It is worth pointing out that the cumulative stellar masses plotted in Fig. \ref{fig:mstarprof} are derived by integrating within a projected radius. To derive the stellar mass enclosed within a sphere of radius  $R_{500}$, we perform a correction assuming a gNFW for the galaxy distribution (see vdB15), with a concentration parameter $c = 0.72$ and a slope $\alpha=1.64$. We find that 75$\%$ of the mass obtained integrating along the line of sight lies within the sphere with radius $R_{500}$. We therefore multiply the stellar mass estimates by a factor 0.75. The values of $M_{\rm star,500}$ corrected in this manner are listed in the last column of Table \ref{tab:mfe500-mstar500}. 

\subsection{Scaling relations for stellar masses and stellar mass uncertainties} 
\label{sec:scale_mstar}

As previously done for $M_{\rm Fe,500}$, we investigate scaling relations for $M_{\rm star,500}$. 

\begin{figure}
\centerline{\includegraphics[angle=0,width=9.8cm]{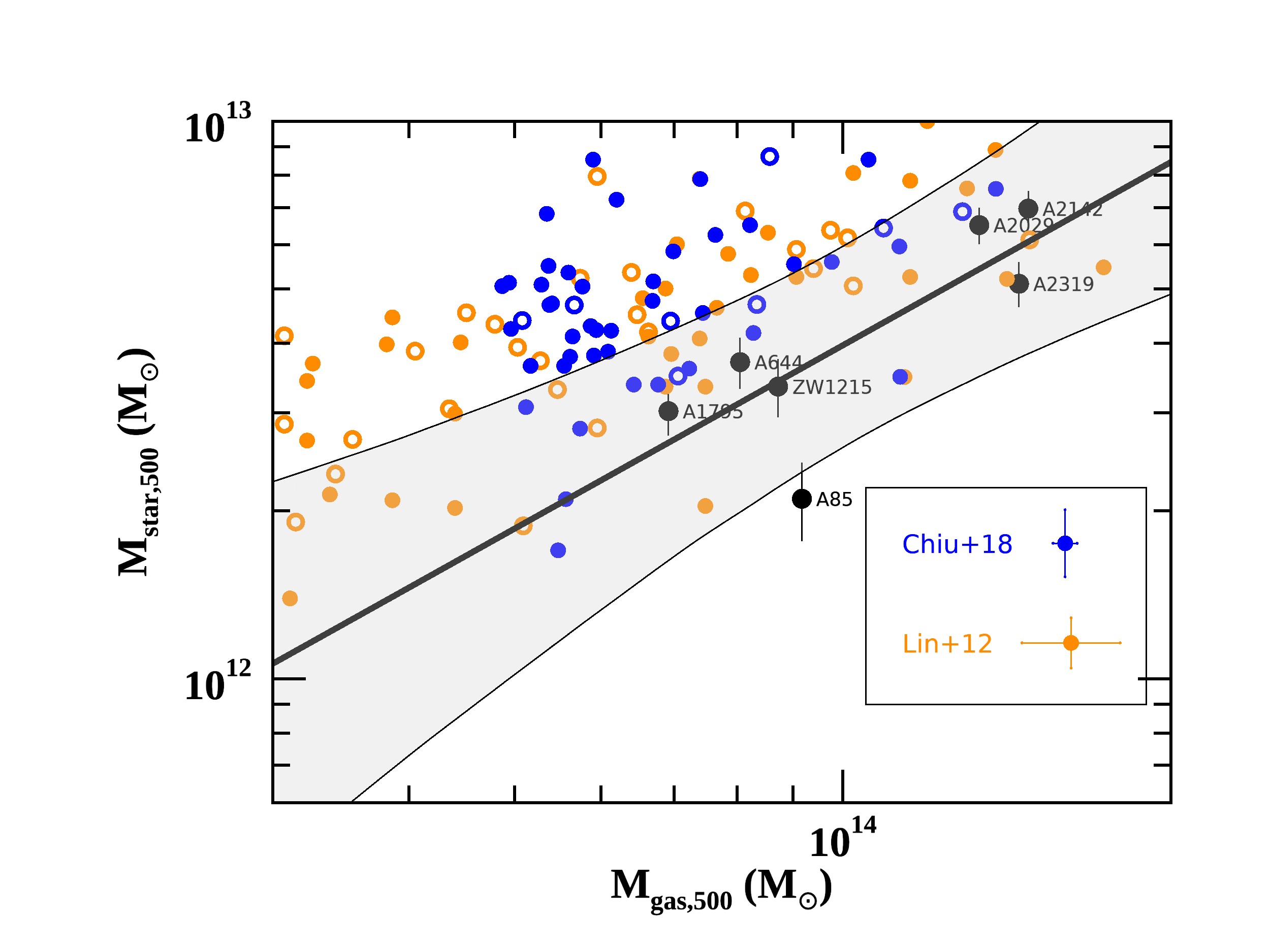}}
\centerline{\includegraphics[angle=0,width=9.8cm]{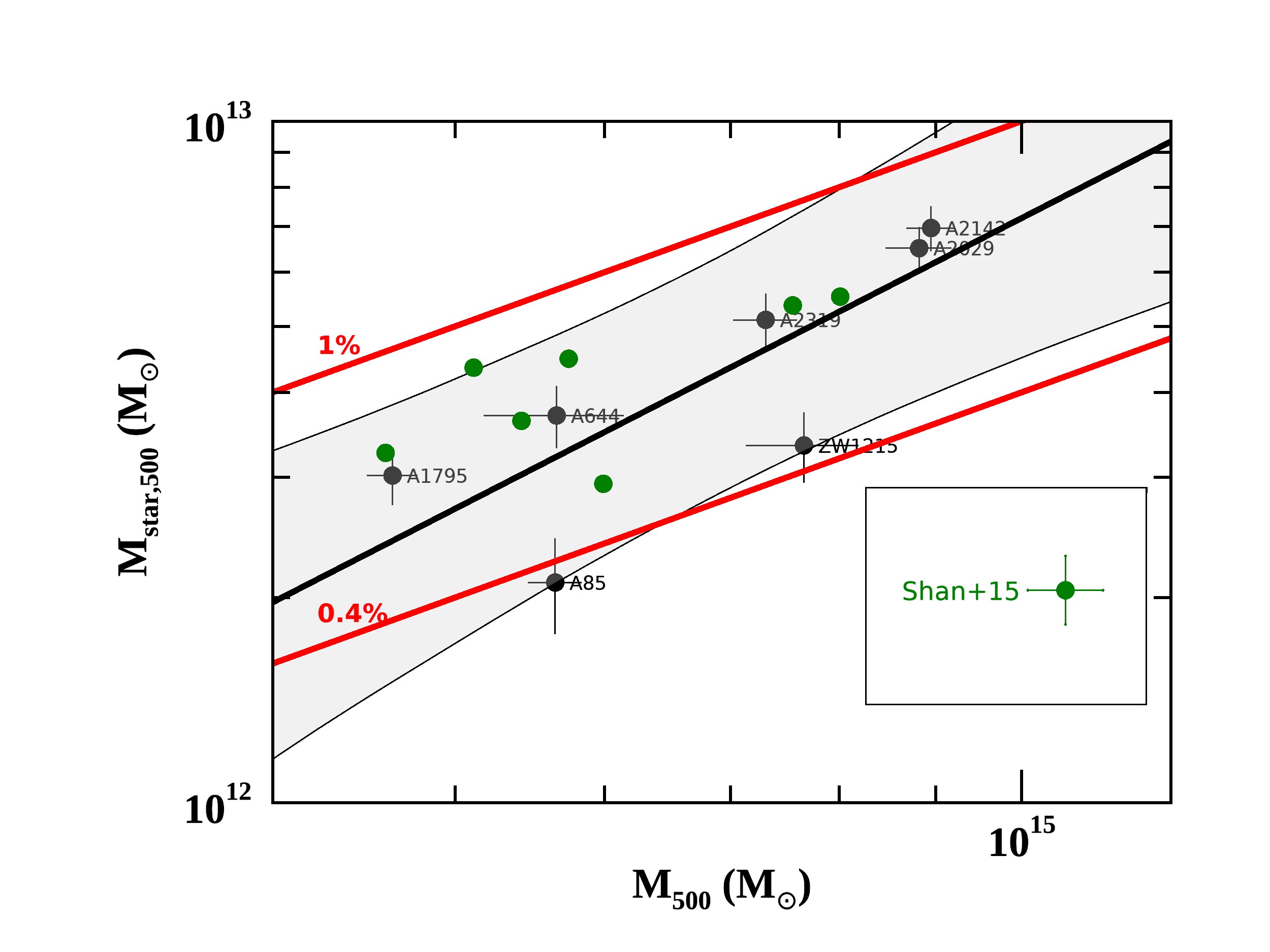}}
\caption{Stellar mass within $R_{500}$ as a function of the gas mass (top panel) and total mass (bottom panel) within $R_{500}$ for the X-COP sample (black circles). Black lines are the best fits and shaded areas indicate the 1$\sigma$ confidence regions, including intrinsic scatters. In the top panel we plot for comparison data from \citet{Chiu:2018} (blue circles) and \citet{Lin:2012} (yellow circles) samples. Low redshifts clusters of both samples are marked with open circles. In the bottom panel we report data by \citet{Shan:2015} (green circles). We omit error bars for these samples to avoid overcrowding the plot: typical error bars are shown in the legend. Red lines in the bottom panel represent two fixed levels (0.4\% and 1\%) of stellar fraction.}
\label{fig:starmass500}
\end{figure}
In Fig. \ref{fig:starmass500} we plot $M_{\rm star,500}$ vs $M_{\rm gas,500}$ (top panel) and $M_{\rm star,500}$ vs $M_{500}$ (bottom panel).
We fit the relations with a standard power law:
\begin{equation}
    {{M_{\rm star,500}} \over {\hat{M}_{\rm star,500}}}= 10^{\alpha} \left({{M_X} \over {\hat{M}_X}}\right)^\beta , 
    \label{eq:mstar}
\end{equation}
\noindent
for both $M_X=M_{\rm gas,500}$ or $M_X=M_{500}$. 
 We adopt the same pivot values for $\hat{M}_{\rm gas,500}$ and $\hat{M}_{500}$ used in Sect. \ref{sec:scal_rel_mfe} and set $\hat{M}_{\rm star,500} = 3.70 \times 10^{12} M_\odot $, at the median of the  $M_{\rm star,500}$ distribution.
Best-fitting values are listed in the second row of Table \ref{tab:scal_rel}. In both cases the slope value is compatible with 1, meaning that the relation is consistent with being linear.
%
Interestingly, the best fit relations for ${M}_{\rm star,500} - {M_{\rm gas,500}}$ and ${M}_{\rm star,500} - M_{500}$ feature  modest scatters, comparable to those for the corresponding relations for $M_{\rm Fe,500}$  plotted in Fig. \ref{fig:mfe_mgas_mtot}. 
These findings suggest that processes underpinning metal enrichment proceed at essentially the same pace in all our objects.

Remarkably, A85 is an outlier in both the relations plotted in Fig. \ref{fig:starmass500}, having a rather low stellar content for its mass. This is in agreement with the findings presented by \citet[see their Fig. 7]{Shan:2015}, where A85 features a low stellar fraction (comparable to ours), but at odds with estimates from \citet{Kravtsov:2018} who find $M_{\rm star,500} \sim 7-8 \times 10^{12} M_\odot$, which is extremely high and would bring A85 to be an outlier on the opposite side of our scaling relations. The reason for this discrepancy is not clear: as we will discuss below, comparison between different stellar measurements are quite delicate and should be treated carefully. The impact of the eventual underestimation of the stellar mass of A85 will be discussed in Sect. \ref{sec:iron-share}.

Comparing stellar mass measurements from different cluster samples requires some caution, since stellar masses are estimated via complicated models, which include many physical processes and are based on different assumptions. 
For example, assuming a \citet{salpeter55}-diet Initial Mass Function (IMF) in deriving stellar masses from luminosities, provides values that exceed by a factor $ \sim 2$  those obtained using a \citet{chabrier03} mass function. Similarly, different choices for stellar population synthesis (SPS),  stellar evolution,  star formation history (SFH), dust attenuation,  stellar mass function etc. are sources of possible bias and can induce significant systematics  on the final stellar mass estimation. Comparison between samples, where different assumptions have been made, can be used to provide an estimate of systematics affecting stellar mass measurements.

 We compare our ${M}_{\rm star,500}$ vs. ${M_{\rm gas,500}}$ measurements with those reported in \citet{Chiu:2018} and \citet{Lin:2012}, hereafter C18 and L12. Their values are shown in Fig. \ref{fig:starmass500}  as blue and yellow symbols respectively. We omit error bars for these two samples to avoid overcrowding the plot. 
C18 present stellar masses for a sample of 91 clusters; we restricted the comparison to clusters having $M_{500}$ in a range similar to ours, namely: $ 4.5 \times 10^{14} - 10^{15} M_\odot $. 
C18 clusters cover a wide range in redshift ($0.25 < z < 1.25$). To highlight possible differences, we show low-redshift systems, $ z < 0.4$, as open circles. The L12 sample includes 94 clusters that span the redshift range $0-0.6$; they assume a \citet{Kroupa:2001} IMF,  consequently stellar masses for their sample (yellow circles in Fig. \ref{fig:starmass500}) have been rescaled by a factor 0.76 (see C18), to bring them to \citet{chabrier03} values.
Low-redshift clusters ($z < 0.1$) are marked with open circles. 
Unfortunately, L12 do not report $M_{500}$ for their systems, so we do not apply any restriction to the masses of their sample. 
The bulk of both samples has stellar masses that are higher than ours, by a factor $1.5-3$, there is however some overlap between our data and theirs. Differences seem to diminish at higher gas (and stellar) masses. Discrepancies are still present when we restrict the comparison to the low-redshift subsamples (open circles), meaning they cannot be ascribed to the high redshift objects, nor be interpreted in any evolutionary framework.
Both C18 and L12 datasets exhibit a significant scatter; to quantify differences between the samples, we fit C18 and L12 using our power law model (Eqn. \ref{eq:mstar}) and compare best fit relations. L12 do not explicitly report error bars for their data, so we assign each point of their sample a typical error bar (see their Fig. 3) of 0.045  dex for $M_{\rm gas,500}$ and for $M_{\rm star,500}$ in logarithmic scale. Best fit values for C18 and L12, along with the total scatters and the assigned values of pivot masses, are reported in Table \ref{tab:chiu-lin-bf}. 
\begin{table*}

\centering
\caption{Parameters for scaling relation $M_{\rm star,500}$  vs $M_{\rm gas,500}$ (top panel in Fig. \ref{fig:starmass500}) for the samples by \citet{Chiu:2018} and \citet{Lin:2012}. $\hat{M}_{\rm gas,500}$ and $\hat{M}_{\rm star,500}$ are the adopted pivot masses; $\alpha$, $\beta$ and $\sigma$ are the best fit parameters as in Table  \ref{tab:scal_rel}.} 
\renewcommand\arraystretch{1.5}
\begin{tabular}{|c||c|c||c|c|c|}
\hline
Sample & \multicolumn{2}{|c||}{pivot masses $(M_\odot)$} & \multicolumn{2}{|c|}{best-fit-values} & intrinsic scatter \\
\hline
& $\hat{M}_{\rm gas,500}$  &  $\hat{M}_{\rm star,500}$ & $\alpha$ & $\beta$ & $\sigma (\%)$  \\
\hline \hline
Chiu+2018 & $6.13 \times 10^{13}$ & $4.69 \times 10^{12}$
& $-0.011 \pm 0.016$ & $ 0.48 \pm 0.12 $ & $32.3 \pm 2.3$ \\
\hline
Lin+2012 &  $4.51 \times 10^{13}$ &   $4.71 \times 10^{12}$ & $-0.014 \pm 0.015 $ & $0.64 \pm 0.04 $ & $36.8 \pm 2.7$  \\
\hline
\end{tabular}
\renewcommand\arraystretch{1}

\label{tab:chiu-lin-bf}
\end{table*}
Note that the best fit for C18 does not match values reported by the authors in their paper, because we included only massive clusters.

Since differences among samples seem to diminish at higher masses,  we choose to evaluate stellar mass discrepancies at three reference gas mass values: $M_{\rm gas,500} = [6.5,8.8,15] \times 10^{13} M_\odot$, which are approximately the minimum, the median and the maximum  $M_{\rm gas,500}$ values for our sample.  At these three reference gas masses, C18 (L12) estimates exceed ours  by  [90\%, 58\%, 12\%] ([78\%, 56\%, 24\%]), respectively. Values at the minimum reference gas mass should be taken with caution both for C18 and L12: data from C18 have been restricted to massive systems, with $M_{500}> 4.5 \times 10^{14} M_\odot$, approximately the mass of A1795, however, as shown in the top panel of Fig. \ref{fig:starmass500},  many C18 points feature $M_{\rm gas,500}$ values that are significantly lower than that found for A1795. Indeed, while measuring gas masses is quite simple and straightforward,  estimates of $M_{500}$ are  subject to bias, depending on the adopted method. The low gas-mass values for many C18 data, instill some doubt that  bias could be present and that we are plotting (and comparing) also clusters whose mass is below the $4.5 \times 10^{14} M_\odot $ threshold. Since the stellar fraction is well-known to increase when $M_{500}$ decreases, including these (possibly) less massive systems could increase the discrepancy with our measurements. Moreover, the L12 sample includes low-mass systems, since we could not apply any selection on mass, and the discrepancy at the minimum reference gas mass may be overestimated also in this case. As a consequence, at the minimum reference gas mass,  the discrepancy should be regarded with some caution, while values at the median and maximum reference gas masses are more robust. Globally, assuming a systematic discrepancy of $50\%-60\%$ between our measurements and those reported in C18 and L12 seems reasonable.


Let us now consider the ${M}_{\rm star,500} - M_{500}$,
in the bottom panel of Fig. \ref{fig:starmass500} we show two lines corresponding to two levels of stellar fraction, namely  $M_{\rm star,500}/M_{500}$ = $1\%$ and $0.4\%$. All our clusters are within this range (except for A85 which is slightly below).
Interestingly, our stellar fractions are in agreement with those reported by \citet{Leau:2012}, and almost one order of magnitude larger than those found by \citep{Girelli:2020}. As pointed out in \citet{Leau:2012}, stellar fractions estimates are subject to substantial systematic errors, similar to those we have estimated for the ${M}_{\rm star,500} - M_{\rm gas,500}$ relation.

\section{Combining iron in the ICM and stars}
\label{sec:comb}

 Having derived estimates of the iron mass in the ICM for all our objects and of stellar masses for a sizeable fraction of them, we will now merge this information to address two fundamental questions. 1) How is iron shared between stars and ICM? 2) Is the iron mass measured in our clusters consistent with expectations based on SN rates?  We will start by addressing the first of these questions.

\subsection{Iron Share}
\label{sec:iron-share}

The iron share $\Im$ is defined as the ratio between iron diffused into the ICM and locked into stars. We shall evaluate $\Im$ within $R_{500}$, i.e. 

\begin{equation}
    \Im_{500} = {M_{\rm Fe,500} \over {M^{\rm star}_{\rm Fe,500}}}\, .
\label{eq:iron-share-def}
\end{equation}

\noindent
 $\Im_{500}$ quantifies how  stars and intracluster plasma share the total cluster iron content. The iron diffused in the ICM,
$M_{\rm Fe,500}$, has been derived from Eq. \ref{eq:mfe},
while $M^{\rm star}_{\rm Fe,500}$ is determined from our stellar mass estimates (see Sect. \ref{sec:dataoana}), assuming on average solar metallicity, i.e. $Z^{\rm star} \simeq  Z_{m,\odot}$ \citep[see][]{RA14}\footnote{Stars have a range of metallicity depending on where and when they were formed, and there is a known relationship between galaxy stellar mass and metallicity. On average the stellar metallicities (weighted by stellar mass) of cluster galaxies should be close to solar and perhaps even super-solar, e.g. \citet{Maoz:2010} and refs. therein. This would, slightly, further raise the total iron mass in clusters, and increase the discrepancy with the expectations from SNe (see Sect. \ref{sec:iron-yield}).}, where $Z_{m,\odot}$ is the solar iron abundance by mass, $Z_{m,\odot} = A_{\rm Fe} Z_{n,\odot} X$, $A_{\rm Fe}$ is the atomic weight of iron (see Eq. 4) and $X$ the hydrogen mass fraction.  Following the standard approach \citep[see][]{RA14,Maoz:2017} we adopt the \citet{Asplund:2009} values so that $Z_{n,\odot}=3.16 \times 10^{-5}$, $X=0.7$ provide $Z_{m,\odot} = 0.00124$. For brevity we will shorten $Z_{m,\odot}$ using the standard notation $Z_\odot$. 
The iron share can then be rewritten as:

\begin{equation}
    \Im_{500} \equiv { {M_{\rm Fe,500}} \over {Z_\odot M_{\rm star,500} }} \, .
\label{eq:iron-share}
\end{equation}

\begin{figure}
\centerline{\includegraphics[angle=0,width=9.8cm]{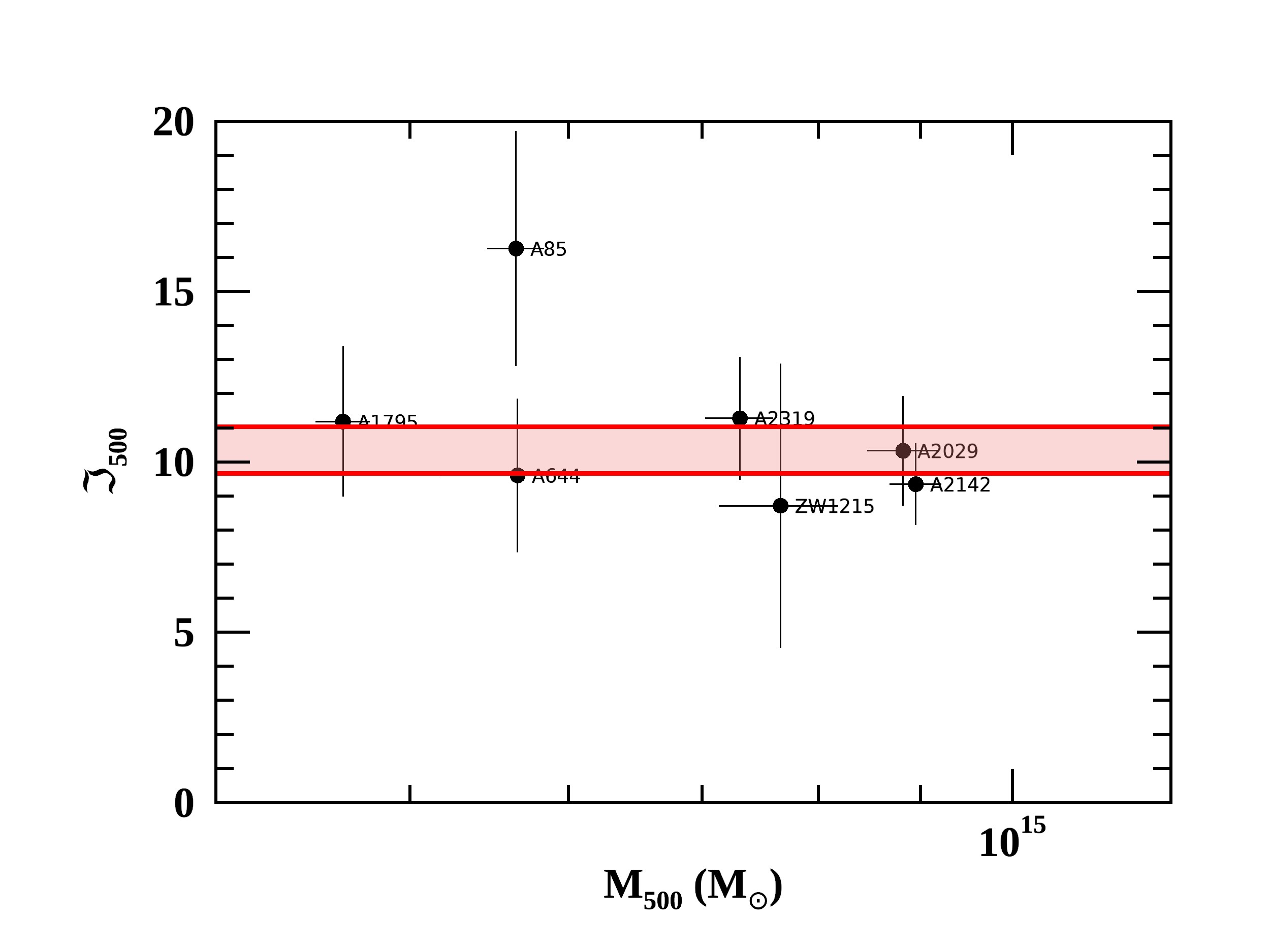}}
\caption{Iron Share for the clusters of our sample. The red band is the 68$\%$ confidence interval about the mean.}
\label{fig:share}
\end{figure}

\noindent
Iron share measurements are plotted in Fig. \ref{fig:share} and listed in Table \ref{tab:iron-share-yfe}.
As we can see, values vary between 8 and 11, reaching a maximum of about 16 for A85. 
The mean value for $\Im_{500}$ is $10.35 \pm 0.69$ 
with a total scatter $\sigma= 1.48$, (i.e. $\sim 14\%$).
In Sect. \ref{sec:scale_mstar} we noted that A85 is particularly poor in stellar mass and that a broad range of values is found in the literature, some consistent with ours \citep{Shan:2015}, others much larger \citep{Kravtsov:2018}. A higher value for the A85 stellar mass could reconcile its iron-share with the mean value of the sample, however if we were to assume the  value reported by \citet{Kravtsov:2018}, the iron-share for A85 would stand out in Fig. \ref{fig:share} on the opposite side of the red band.

Our measurements show that, on average, the iron locked in stars contributes only 10\% to the total iron in clusters. It would therefore seem that feedback effects, independently of where and when they took place, are very efficient, in the sense that  90\% of the iron in clusters, while originating from the stellar component, has managed to escape galaxies and pollute the hot gas. Thus, we can visualize stars as "iron factories" producing metal only to hand it over to the ICM.


 \begin{table}
 \centering
 \caption{Iron shares $\Im_{500}$ and effective iron yields $\mathcal{Y}_{\rm Fe,\odot}$.}
 \renewcommand\arraystretch{1.5}
 \begin{tabular}{|c|c|c|}
 \hline
Cluster & $\Im_{500}$ & $\mathcal{Y}_{\rm Fe, \odot}$ \\
 \hline
 A1795 &  11.18 $\pm$  2.18 &  7.07 $\pm$  1.27  \\
A85 &  16.26 $\pm$  3.43 & 10.01 $\pm$  1.99  \\
A644 &   9.60 $\pm$  2.23 &  6.15 $\pm$  1.29  \\ 
A2319 &  11.28 $\pm$  1.78 &  7.12 $\pm$  1.03  \\ 
ZW1215 &   8.71 $\pm$  4.14 &  5.63 $\pm$  2.40  \\
A2029 &  10.32 $\pm$  1.59 &  6.57 $\pm$  0.92 \\ 
A2142 &   9.34 $\pm$  1.17 &  6.00 $\pm$  0.68  \\ 
 \hline
 \end{tabular}
 \renewcommand\arraystretch{1}
 
 \label{tab:iron-share-yfe}
 \end{table}
 

\subsection{Effective Iron yield}
\label{sec:iron-yield}

The effective iron yield $\mathcal{Y}_{\rm Fe}$ provides the efficiency with which stars produce iron in clusters.
It is defined as the total Fe mass divided by  the mass of the gas that went into stars:

\begin{equation}
    \mathcal{Y}_{\rm Fe} = { {M^{\rm star}_{\rm Fe,500} + M_{\rm Fe,500}} \over  M_{\rm star,500}(0)} ,
\end{equation}

\noindent
where $M_{\rm star,500}(0)$ is the mass of gas that went into stars whose present mass is reduced to $M_{\rm star,500}$ by the mass return from stellar mass loss, i.e. $M_{\rm star,500}(0) = r_o M_{\rm star,500}$, where $r_o$ is the  return factor.
Following \citet{RA14} and \citet{Maraston:2005}, we shall assume $r_o = 1/0.58 $.
By dividing $\mathcal{Y}_{\rm Fe}$  by the Fe solar abundance, $Z_\odot$ \citep{Asplund:2009}, 
we can express our result in solar units, i.e:
\begin{equation}
\mathcal{Y}_{\rm Fe,\odot}  \equiv \mathcal{Y}_{\rm Fe} /Z_\odot  .  
\end{equation}

\begin{figure}
\centerline{\includegraphics[angle=0,width=9.8cm]{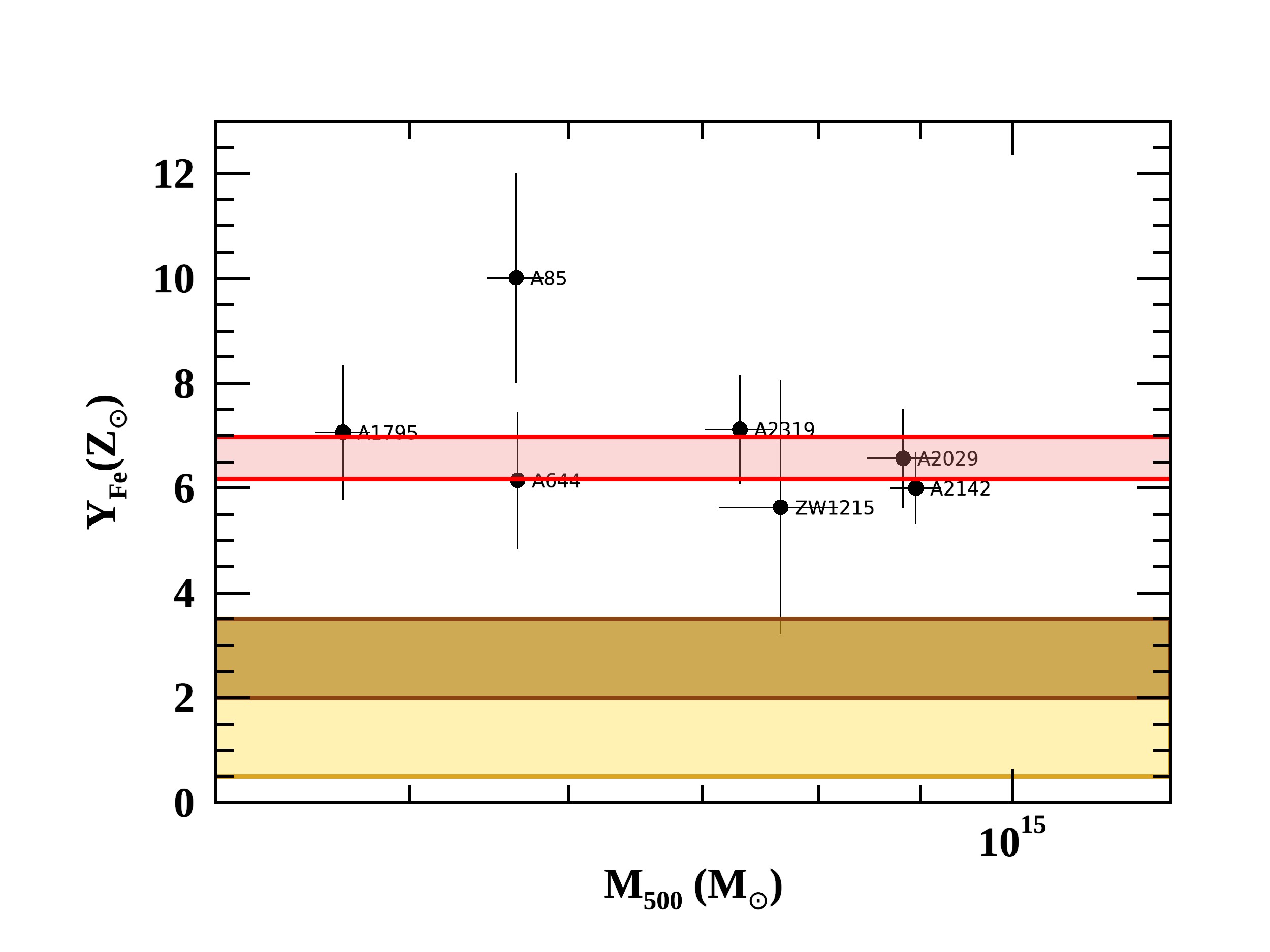}}
\caption{Effective iron yield for the clusters of the sample. The red band is the 68$\%$ confidence interval about the mean. The yellow band shows the expected value computed through the SN yields derived from \citet{Maoz:2017} and \citet{RA14} (see text for details); the brown band represents the expected value derived assuming a higher SNIa explosions' rate in galaxy clusters than in the field, following \citet{Maoz:2017}.}
\label{fig:yfe}
\end{figure}

The effective iron yield, in solar units, for the objects in our sample, is plotted in Fig. \ref{fig:yfe}, values are reported in Table \ref{tab:iron-share-yfe}. The mean value for $\mathcal{Y}_{\rm Fe}$ is $6.58 \pm 0.40$ and the  
total scatter $0.87$ (i.e. $\sim 13\%)$.

Estimates of the expected $\mathcal{Y}_{\rm Fe}$  have been derived by several authors, e.g. \citet{RA14}, \citet{Maoz:2017}, it is computed as the product of the Fe mass produced by a SN explosion, $y$, and the number of SN events produced per unit mass of gas turned into stars, $k$. 
Both contributions from Ia and CC SN are considered as they are of the same order. Thus, $\mathcal{Y}_{\rm Fe}$ can be written as:

\begin{equation}
  \mathcal{Y}_{\rm Fe} = y_{\rm Ia}  \cdot  k_{\rm Ia} + y_{\rm CC}  \cdot k_{\rm CC},  
\label{eq:iron-yield}
\end{equation}
where Ia and CC subscripts refer to the two different SN types. 
For Ia, following \citet{Maoz:2017}, we assume $y_{\rm Ia} = 0.7 \, M_\odot$ and $k_{\rm Ia} = 1.3\times 10^{-3} \,  M_\odot^{-1}$.  \citet{RA14},  see also \citet{Greggio:2011}, suggest that $ k_{\rm Ia}$ could be as high as $2.5\times 10^{-3} \, M_\odot^{-1}$. 
However this high value is based on early measurements of the SN Ia rate in local Sb galaxies requiring some uncertain assumptions about the star formation histories of such galaxies.
For CC SN, following \citet{Maoz:2017}, we assume $ y_{\rm CC} = 0.074 \,  M_\odot $ and $ k_{\rm CC} = 1.0 \times 10^{-2}  \,  M_\odot^{-1}$ . Substituting the above values in Eqn. \ref{eq:iron-yield} and dividing by the solar abundance we get,
\begin{equation}
\mathcal{Y}_{\rm Fe,\odot}= 0.93 \, Z_\odot .
\end{equation}

Given the simplified fashion in which this calculation was carried through, i.e. assuming one average value for Ia and CC yields and SN rates, a rather large uncertainty should be associated with $\mathcal{Y}_{\rm Fe}$.  Although the factor of 2 error suggested in \citet{RA14} might be too generous, see discussion above, we shall nonetheless assume it as part of our conservative approach in interpreting our measurements.

The expected effective Fe yield is shown in Fig. \ref{fig:yfe} as a yellow shaded region. Remarkably, measured values, which range from 6 to 10, are much higher, even when assuming a highly conservative factor of 2 uncertainty in the expected effective Fe yield.

Recently, it has been shown \citetext{see \citealp{Maoz:2017} and \citealp{FM:2018}} that type Ia SN explosions in galaxy clusters are more frequent than in the field. If, following \citet{Maoz:2017}, we assume a SN$_{\rm Ia}$ rate per unit mass  $k_{\rm Ia}  = (5.4\pm 2.3) \times 10^{-3} \, M_\odot^{-1}$  we derive $\mathcal{Y}_{\rm Fe,\odot}= 2.6^{+0.9}_{-0.6} \, Z_\odot $ (gold shaded region in Fig. \ref{fig:yfe}). This revised SN rate brings the expected effective yield closer but still well below the measured ones.

\subsection{Uncertainties on iron share and yield}
\label{sec:sys_comb}
Here we evaluate possible uncertainties in the calculation of the iron share and yield. We consider three distinct sources of errors: 1) those associated to the stellar mass estimates; 2) those associated to ICM mass estimates and 3) those associated to comparing stellar and ICM measurements within a given radius.
A summary of  systematics is presented in Table \ref{tab:system}. We list sources of systematics, the assumed percentual variation, whether it acts as an increase or a decrease and the impact on iron share and yield.

Let us start with stellar mass estimates. We identify the following key points in the calculation that goes from the magnitudes of individual galaxies to the estimate of the iron share and yield: 1) the stellar luminosities of individual galaxies are converted into stellar masses and then summed up to derive the cluster stellar mass, $M_{\rm star}$; 2) the stellar mass is converted into iron locked in stars, $M^{\rm star}_{\rm Fe,500}$, by assuming a certain value of the metallicity and 3) an estimate of the stellar mass loss is used to compute the mass of gas that went into stars, $M_{\rm star}(0)$.
Systematic errors in the stellar mass estimates is discussed extensively in Sect. \ref{sec:scale_mstar}. Although we lack a clear understanding for the origin of  different estimates of $M_{\rm star}$, our comparison shows that we may be underestimating  stellar masses by as much as 60\%.  If we correct for this systematic, we derive a reduction of the mean iron share by the same percentage and of the mean effective iron yield by about 50\% (see also Table \ref{tab:system}).
The factor $r_o$, introduced to account for the stellar mass loss (see Sect. \ref{sec:iron-yield}), is a further source of systematic uncertainty. The value we have adopted, $r_o = 1/0.58$, has been derived for a top heavy IMF, for a Salpeter IMF it would go down to 1/0.70, implying that we may be overestimating  $r_o$ by as much as 20\%.
Correcting for this systematic leads to a reduction of the mean effective iron yield by the same percentage.

Another source of uncertainty, in our estimate of the iron share and yield, is the contribution of Intra-Cluster Light (hereafter ICL) to the total cluster light. There have been several measurements of ICL on individual systems and a few on low redshift samples. We shall consider two  studies based on stacking of large samples of Survey data, SDSS and DES respectively. In the first, \citet{Zibetti:2005} found that the ICL contribution to the total optical emission in a cluster is about 10\%, while in the second, \citet{Zhang_ICL_2019} derived a much larger value, about 50\%
\footnote{Actually, \citet{Zhang_ICL_2019} provide a measure of the ICL plus Central Galaxy (CG) light. The estimate we quote has been derived by subtracting, in an approximate and conservative fashion, the contribution of the CG.}.
An evaluation of which of these two measures is more robust, is well beyond the scope of this paper. We adopt a conservative approach and assume the larger contribution of
50\%, this results in a reduction of the mean iron share by the same percentage  and of the mean effective iron yield by about 40\% (see also Table \ref{tab:system}).

Let us now turn to uncertainties on ICM Fe mass estimates.
As evident from Eq. \ref{eq:mfe}, the estimate on the Fe mass is based on two quantities: the gas mass and the Fe abundance. The former is one of the better measured in X-rays, with systematics less than a few \%, see discussion in \citet{Eckert_non_th_XCOP:2019},
while the latter is the most challenging \citep[e.g.,][]{Molendi:2016}. Thus, uncertainties on
the ICM Fe mass will be dominated by uncertainties on the Fe abundance. An estimate of
the systematic uncertainty on the Fe abundance has been presented in Sect. \ref{sec:systx}.
Here we take the 15\% uncertainty, conservatively assume that it takes the form of a reduction on the iron mass, and apply it directly to our ICM Fe mass estimates \footnote{This is a conservative approach because: 1) systematics have been estimated for the abundance in the outskirts, where they are larger, and we apply them to the integrated quantity 2) the systematic on the abundance profile is itself conservative, see discussion in Sect. \ref{sec:systx}.}. 
We find a reduction by the same amount, i.e. 
15\% of the iron share and of $\simeq$ 10\% of the effective iron yield (see also Table \ref{tab:system}).

 \begin{table}
 \centering
 \caption{Systematic uncertainties on iron share $\Im_{500}$ and effective iron yield $\mathcal{Y}_{\rm Fe,\odot}$. The down facing arrows placed next to $\Im_{500}$ and $\mathcal{Y}_{\rm Fe,\odot}$ indicate that all corrections have the effect of reducing estimates for these two variables.}
 \renewcommand\arraystretch{1.5}
 \begin{tabular}{|c|c|c|c|c|}
 \hline
 \hline
 Variable & Sys. on var. & direction & $\downarrow $ $\Im_{500}$ & $\downarrow $ $\mathcal{Y}_{\rm Fe, \odot}$ \\
 \hline
 $M_{\rm star}$  &  60\%  &  $\uparrow   $ & 60\% &  50\% \\
 $r_o $          &  20\%  &  $\downarrow $ &  -   &  20\% \\
 ICL             &  50\%  &  $\uparrow   $ & 50\% &  40\% \\ 
 \hline
 $M_{\rm Fe,500}$&  15\%   &  $\uparrow   $ & 15\% &  10\%\\ 
 \hline
$ \mathcal{R}$   &  50(20)\%  &  $\uparrow  $  & 50(20)\%  &  40(15)\% \\
 \hline
 \hline
 \end{tabular}
 \renewcommand\arraystretch{1}
 
 \label{tab:system}
 \end{table}

Finally, let us consider systematics associated to the combination of our ICM and stellar mass estimates. Our measurements are integrated out to $R_{500}$ both for stellar and ICM masses. When computing iron shares and effective yields from this data, the underlying assumption is that the radial distribution of stellar and ICM masses in cluster outskirts do not differ substantially. This is  an approximation as the ICM is more compactly distributed than the stellar matter. From our data we estimate that the ratio of mass within $R_{200}$ over mass within $R_{500}$ is respectively $\sim 1.5$ for the ICM and $\sim 1.8$ for the stellar component, leading to a value of $ \mathcal{R}$ defined as, $ \mathcal{R} \equiv (M_{\rm star,200} / M_{\rm star,500})/(M_{\rm gas,200} / M_{\rm gas,500})$ of 1.2.
Thus, under the assumption that metal abundances in the ICM continue to remain constant beyond $R_{500}$,
iron shares and effective yields would decrease by about $\sim 1.2$ and 1.15 respectively, if we were to extend our measures out to $R_{200}$. No data is available to correct masses beyond  $R_{200}$,  we consider two different possibilities.
If the stellar mass profile follows the dark matter profile, then, given  that $f_{\rm gas} =  M_{\rm gas} / M_{\rm tot}$  predicted by simulations at large radii is either constant or increasing with radius \citep{Kravtsov:2005,Planelles:2013}, the numbers derived above apply.
If, conversely,  we adopt a more conservative approach and extrapolate linearly the ratios for $M_{\rm star}$ and $M_{\rm gas}$ discussed above out to a very large radius of 1.5$R_{200}$,  we come up with a value of $ \mathcal{R}$ of about 1.5, leading to a drop in iron shares and effective yields of $\sim 1.5$ and 1.4 respectively. In Table \ref{tab:system} we report both  estimates of the systematics, with the less conservative in parenthesis. 

Having evaluated the various systematics affecting our estimates of iron shares and effective yields, we now step back and finish this section with some general comments.
Of all our systematics, those affecting ICM iron mass estimates are the smallest. This is easily understandable: albeit challenging, X-ray measures are fairly straightforward and
have been carried out on a set of observations specifically designed for cluster outskirts.
Conversely, stellar masses are characterized by the largest systematics. This is not surprising, $M_{\rm star}$ estimates are observationally challenging  (e.g. intracluster light) and 
at the same time rely on sophisticated and non-unique modeling.
The difference in radial profiles  between stellar mass and the ICM, may also lead to substantial systematics, particularly if the stellar mass does not follow the dark matter distribution. 
Inspection of Table \ref{tab:system} shows that individual sources of errors are insufficient to reconcile our measures of the iron yield with expectations based on SN rates. However,
if we combine all systematics affecting stellar mass estimates we end  up with a value of $\sim$ 3.8 for
$\mathcal{Y}_{\rm Fe, \odot}$, which is relatively close to the one expected if we assume the SN$_{\rm Ia}$ rate proposed by \citet{Maoz:2017}, see also Fig. \ref{fig:yfe}.
Clearly, combining the three systematics on stellar mass with others, particularly those on $ \mathcal{R}$, will reconcile the mean measured yield with the expected one.

\section{Discussion}
\label{sec:disc}
We have analyzed the first representative sample of massive clusters for 
which iron abundance has been measured out to $R_{500}$.
Building on significant work presented in previous X-COP papers \citep{Ghirardini:2018,Ghirardini:2019,Ettori:XCOP2019,Eckert_non_th_XCOP:2019}, we have taken a closer look at the method commonly employed to measure Fe abundances,  identifying  and correcting a major systematic error plaguing  measurements in cluster outskirts. 

Our analysis sheds light on the complicated interplay between instrumental background, effective area calibration, the Fe K$\alpha$ line and the Fe L-shell emission, from which iron abundance measurements emerge. As expounded in Sect. \ref{sec:externalbins},  Fe K$\alpha$ measures are far less prone to systematic uncertainties than L-shell ones, which implies they are amenable to statistical treatments leading to the estimation of robust sample properties such as mean and scatter. Furthermore, as discussed in Sect. \ref{sec:systx}, working with hot systems ensures that, fitting with a one temperature and one abundance model spectra that, to some extent, must be multi-temperature and multi-abundance, yields only modest systematic errors.

Our abundance measurements are unique when compared to those extracted from previous samples because: 
1)	they are based on a representative sample;
2)	they are unaffected by a systematic error that has either limited \citep{Leccardi_metal:2008} or likely biased \citep{Mernier:2017, Lovisari:2019} other measurements;
3) according to our conservative estimates, residual systematics must be smaller than 15\% and 
4)	they extend, for virtually all systems, to $R_{500}$.

Because of these properties, several important implications follow from our measurements, let us go through them. 
Our profiles flatten out at large radii, suggesting early enrichment and significant feedback, admittedly not a new result \citetext{e.g. \citealp{Fabjan:2010} and \citealp{Planelles:2014}}, however the radial range and representative nature of our sample extends its import well beyond previous findings.
We find no evidence of segregation between cool-core and non-cool-core systems beyond $\sim 0.3 R_{500}$. This shows that, as was found for thermodynamic  properties \citep{Ghirardini:2019}, the physical state of the core does not affect global cluster properties.  
Our mean abundance within $R_{500}$ shows a very modest scatter $< $15\%, suggesting the enrichment process must be quite similar in all these  massive systems.  This is a new finding and has significant implications on  feedback processes.  Together with results  from  thermodynamic properties, i.e. renormalized entropy profiles  \citep[see Fig. 6 of][]{Ghirardini:2019},  it affords a coherent picture where  feedback effects
do not vary significantly from one system to another. 
 Another way of looking at the low scatter in mean abundance is through the $M_{\rm gas,500}$ vs $M_{\rm Fe,500}$ relation. The tight nature of the correlation implies that  $M_{\rm Fe,500}$ can be estimated with good accuracy from $M_{\rm gas,500}$. It will be interesting to see if and how these properties extend to less massive systems although, as pointed out earlier in this section, deriving iron abundances for such objects may only be possible with the advent of high spectra resolution instruments such as those on board XRISM \citep{Tashiro_XRISM:2018,GT:2018} and ATHENA \citep{Nandra_Athena:2013}. 

Before moving on to discussing optical measurements, we would like to address two questions on  size namely: 1) how can a sample of only 12 systems prove so powerful in providing constraints on the cluster population? and 2) will increasing the sample by, say, two or tenfold  lead to more stringent measurements? The answer to these questions goes as follows. Measurements are limited either by statistical or systematic errors. In the case at hand, although statistical errors on individual systems are larger than systematic ones, errors on the mean are not. For example, see Sect. \ref{sec:systx}, the statistical error on the mean abundance is actually smaller than the systematic one. This implies that future measurements on larger samples will provide more stringent measurements only if the ensuing reduction on statistical errors will be accompanied by a comparable decrement on systematic uncertainties.
For this and other reasons, several of the authors of this paper are engaged in ensuring that the instrumental background on ATHENA  feature the smallest systematics of any imaging X-ray mission ever flown.  
 
For a subsample of 7 of our 12 systems, we have secured stellar masses.
We have found that $M_{\rm star,500}$ and $M_{\rm gas,500}$ correlate  well, the scatter, within the limits of the available data, is comparable with the one between $M_{\rm Fe,500}$ and $M_{\rm gas,500}$, reinforcing the concept that enrichment must be quite similar in all  our systems. By combing stellar with ICM measurements, we have been able to take an inventory of Fe in clusters. 
We find that the amount diffused in the ICM  with respect to that locked in stars, the so called iron share, is very high, about 10 times, and features a moderate scatter around the mean value. The implication is that the bulk of Fe produced by stars is expelled in the ICM and that this process proceeds essentially at the same pace in all systems. As  the iron share, the effective yield features a modest scatter of $\sim$ 13\% around the mean value. This has significant implications for the approach taken when computing the iron  masses expected from SN rates. Indeed, a description of the enrichment process through a simple equation (i.e. Eq. \ref{eq:iron-yield}), based on average properties such as  mean SN rates and mean iron mass per SN, could not be justified if the distribution of measured iron yields in the cluster population were to be characterized by a large scatter or a multi-modal distribution.  Thus, while the actual numbers we adopt could be incorrect, the procedure is most likely sound.

By comparing the measured effective iron yield with the expected one, we find that the efficiency for Fe production in cluster galaxies must be higher than predicted. Similar claims have been made before \citep[e.g.][]{RA14}, however previous measurements were based on non-representative samples and extended to significantly smaller radii. As pointed out in \citet{Molendi:2016}, this made such claims at the very least premature. 
The thorough analysis of systematic uncertainties conducted on X-ray (see Sect. \ref{sec:systx}) shows that ICM iron masses cannot explain the discrepancy between measured and predicted effective yields. The only possible alternatives are that, either SN yields or stellar masses, or both, have been significantly underestimated. 

As discussed in Sect. \ref{sec:iron-yield}, recent estimates have revised upwards the SN Ia Fe yield in galaxy clusters, however the ensuing increase in predicted effective yield is insufficient to  bring it into agreement with the measured one.  The remaining option is that stellar masses must be underestimated. In Sect. \ref{sec:scale_mstar} and \ref{sec:sys_comb} we have seen this may be happening for a number of reasons. Amongst them some, such as the conversion of multi-band optical/infrared data to stellar masses, are related to  technical issues and others, such as a substantial contribution from ICL or from a stellar population lying  in the outskirts, are of a more astrophysical and perhaps appealing nature.
However, whatever the explanation for the discrepancy will turn out to be, since the scatter in the correlations found in all our scaling relations is much smaller than the ratio between measured and predicted iron share, the correction factor will have to be very similar in all clusters. This could either imply that we need to correct the conversion  adopted to go from  multi-band optical/infrared data to stellar masses by a constant factor or, more intriguingly, that a sizeable but roughly constant fraction of ICL or stellar mass is missing from our inventory.

To summarize, although current estimates suggest, with some strength, that the measured iron mass in clusters is well in excess of the predicted one, systematic errors prevent us from making a definitive statement. Further advancements will only be possible when systematic uncertainties, principally those associated to the estimate of stellar masses, both within and beyond $R_{500}$, can be reduced. 


\section{Summary}
\label{sec:summary}
We have measured iron abundance for the X-COP sample of massive clusters, these are our main findings. 
\begin{itemize}
\item Our measurements are unprecedented for 4 reasons: 1)	they are based on a representative sample; 2)	they are unaffected by a systematic error that has plagued previous measurements;
3) they feature residual systematics smaller than 15\% and
4)	they extend, for virtually all systems, to $R_{500}$.

\item Our profiles flatten out at large radii suggesting early enrichment and significant feedback, admittedly not a new result \citetext{e.g. \citealp{Fabjan:2010} and \citealp{Planelles:2014}}, however the radial range and representative nature of our sample extends its import well beyond previous findings.
\item We find no evidence of segregation between cool-core and non-cool-core systems beyond $\sim 0.3 R_{500}$. This shows that, as was found for thermodynamic  properties \citep{Ghirardini:2019}, the physical state of the core does not affect global cluster properties.  

\item Our mean abundance within $R_{500}$ shows a very modest scatter, $< $15\%, suggesting the enrichment process must be quite similar in all these  massive systems.  This is a new finding and has significant implications on  feedback processes.  Together with results  from  thermodynamic properties, i.e. renormalized entropy profiles  \citep[see Fig. 6 of][]{Ghirardini:2019},  it affords a coherent picture where feedback effects do not vary significantly from one system to another.

\item Another way of looking at the low scatter in mean abundance is through the $M_{\rm gas,500}$ vs $M_{\rm Fe,500}$ relation. The tight nature of the correlation, $\sigma < $15\%, implies that  $M_{\rm gas,500}$ can be used as a robust proxy for $M_{\rm Fe,500}$.

\end{itemize}
For a subsample of 7 of our 12 systems we have secured stellar masses. By combing stellar with ICM measurements, we have derived the following results.
\begin{itemize}
\item The amount of Fe diffused in the ICM  with respect to that locked in stars, the so called iron share, is very high, about 10 times, and features a moderate scatter around the mean value. The implication is that the bulk of Fe produced by stars is expelled in the ICM and that this process proceeds  at a similar pace in all systems.
\item Although current estimates suggest, with some strength, that the measured iron mass in clusters is well in excess of the predicted one, systematic errors prevent us from making a definitive statement. Further advancements will only be possible when systematic uncertainties, principally those associated to the estimate of stellar masses, both within and beyond $R_{500}$, can be reduced. 

\end{itemize}

\begin{acknowledgements}
We warmly thank F.Mernier, L.Lovisari and D.Maoz for providing feedback on an early version of this paper.
This research is based on observations obtained with {\it XMM-Newton}, an ESA science mission with instruments and contributions directly funded by ESA Member States and the USA (NASA).
\end{acknowledgements}

\bibliography{biblio}
\newpage
\onecolumn

\appendix
\section{Abundance profiles for X-COP clusters}
\label{sec:app_zprofs}

\begin{figure*}[h]
\centerline{
\subfloat{\includegraphics[width=0.32\textwidth, keepaspectratio]{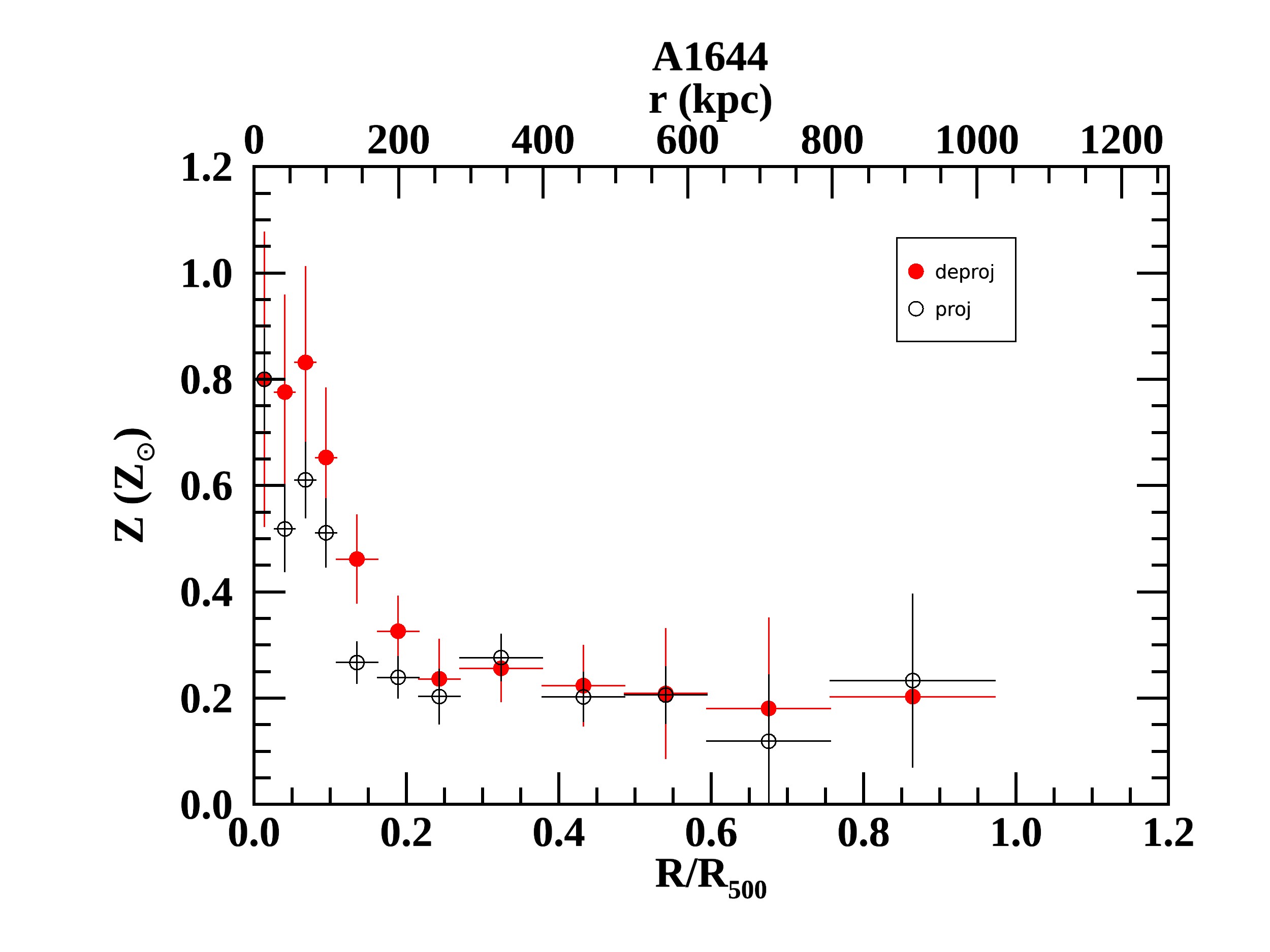}} \hfill \subfloat{
\includegraphics[width=0.32\textwidth, keepaspectratio]{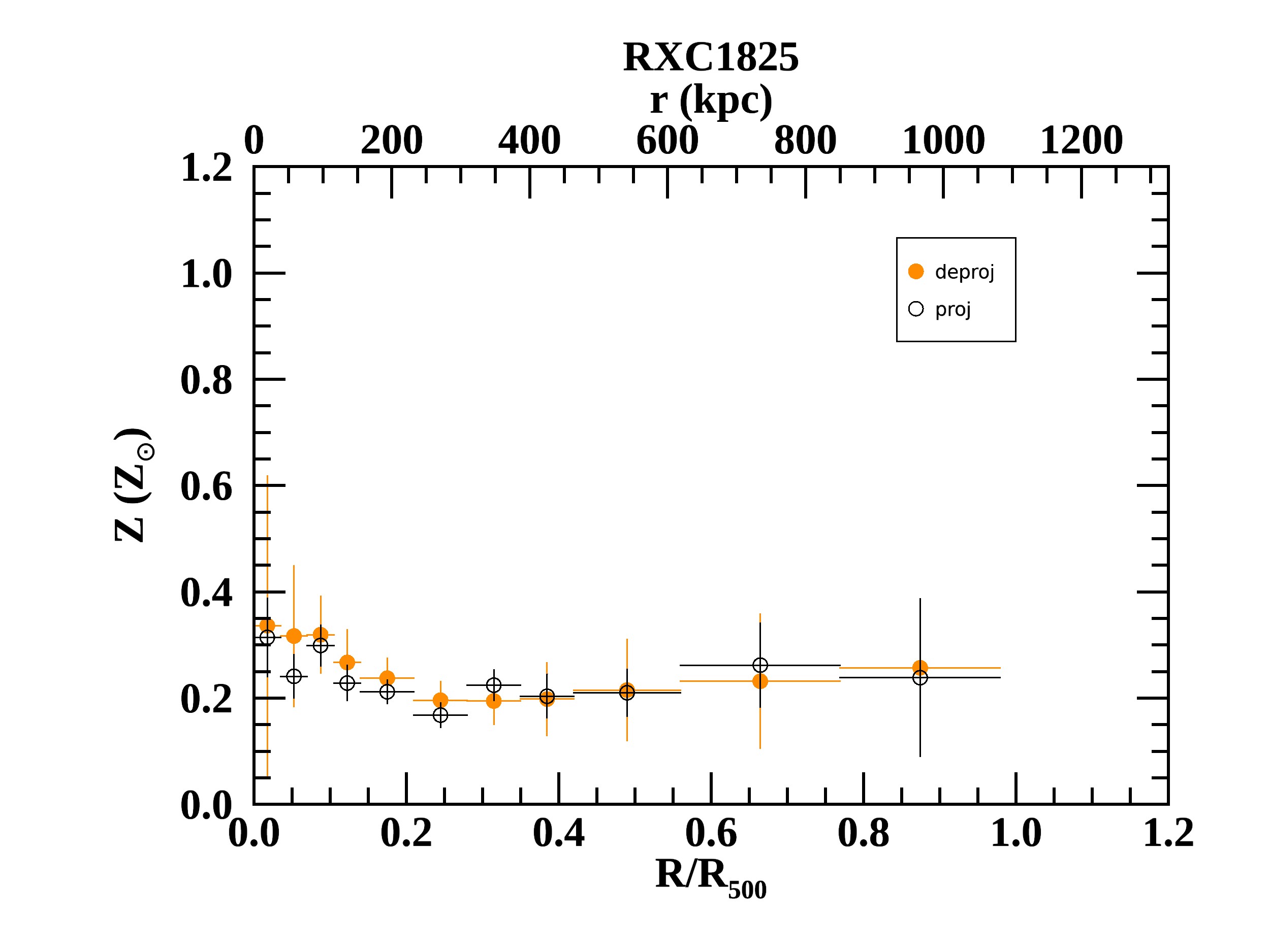}}
\hfill \subfloat{
\includegraphics[width=0.32\textwidth, keepaspectratio]{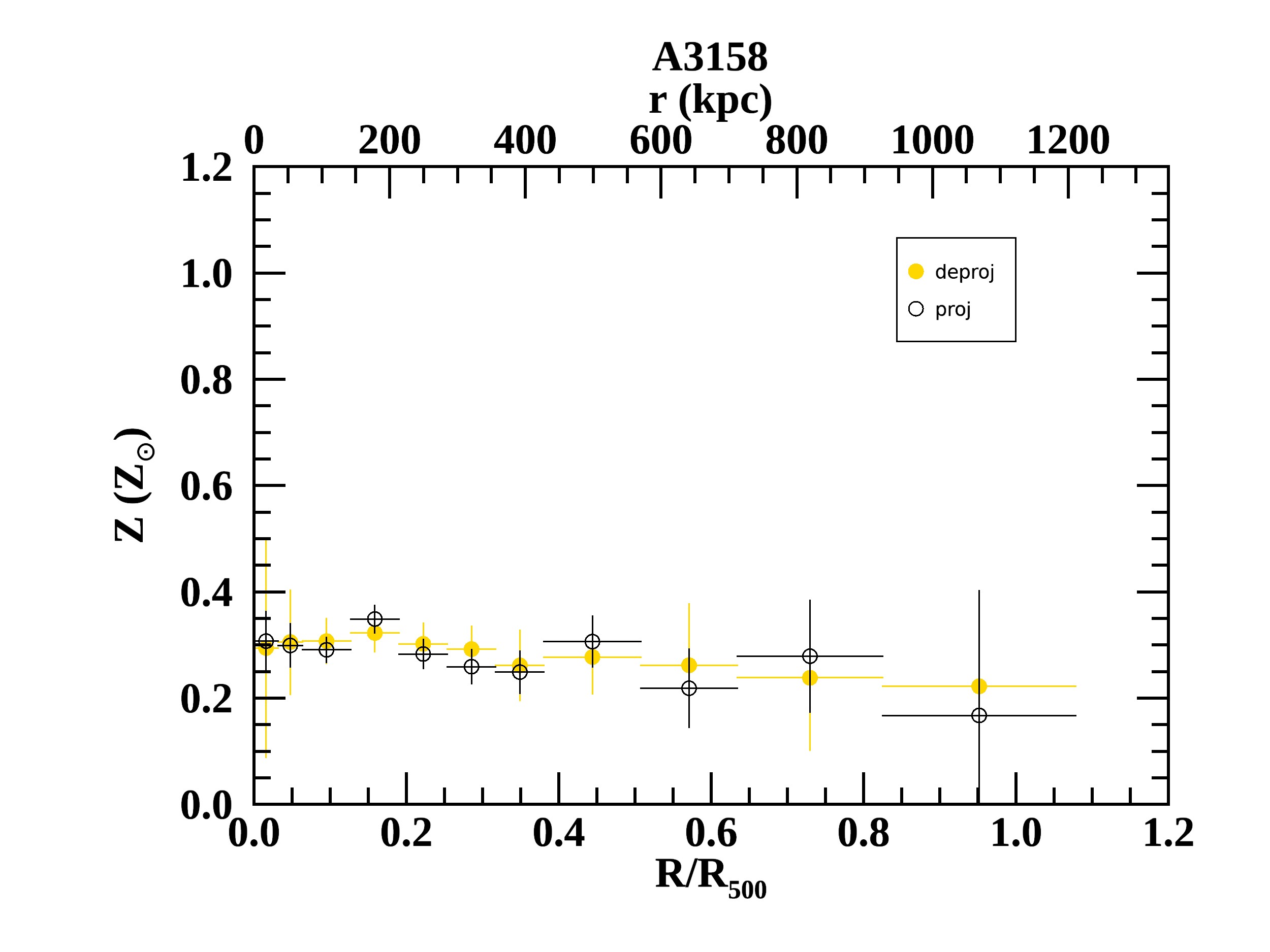}}
} 
\centerline{
\subfloat{\includegraphics[width=0.32\textwidth, keepaspectratio]{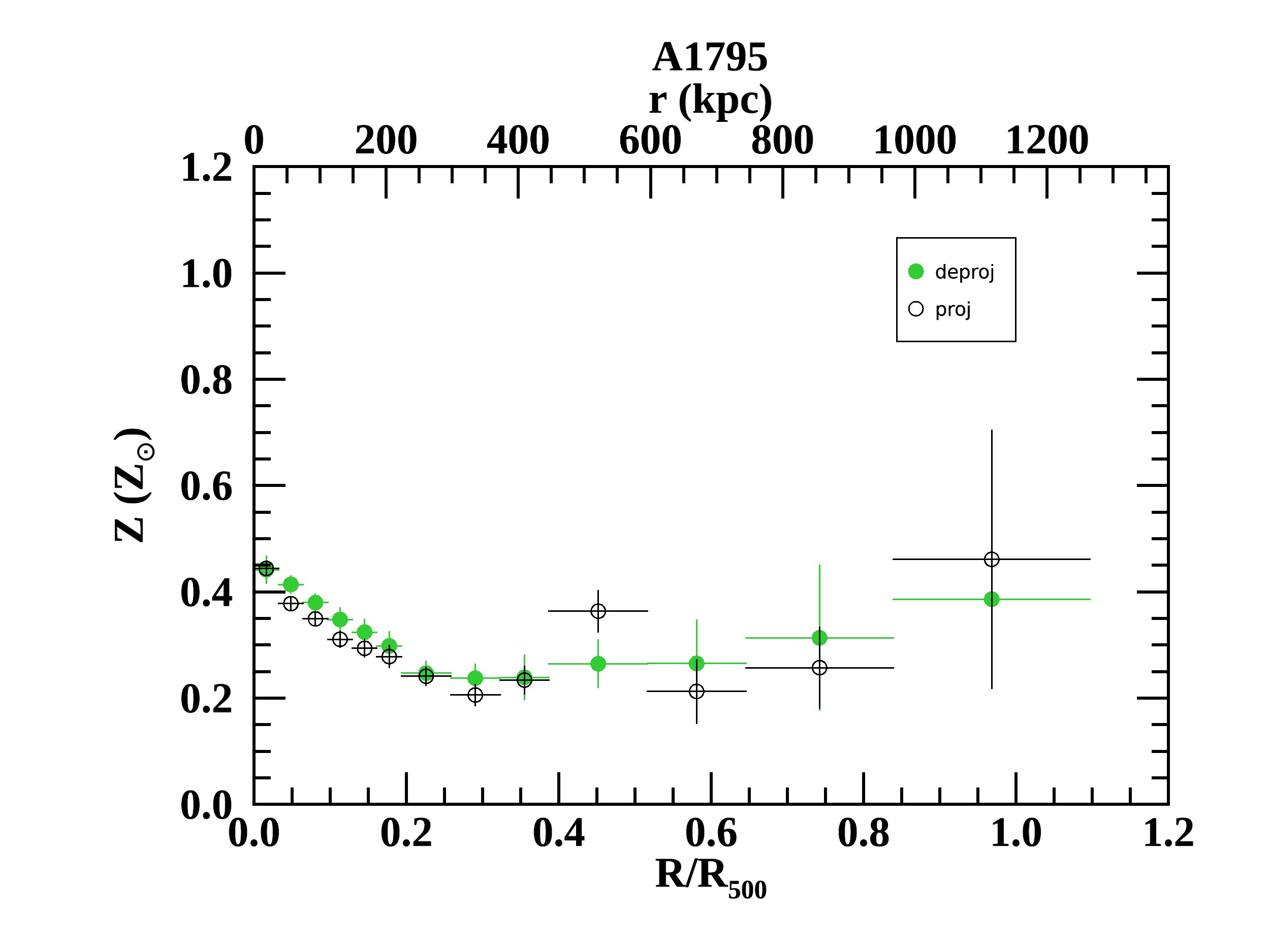}} \hfill \subfloat{
\includegraphics[width=0.32\textwidth, keepaspectratio]{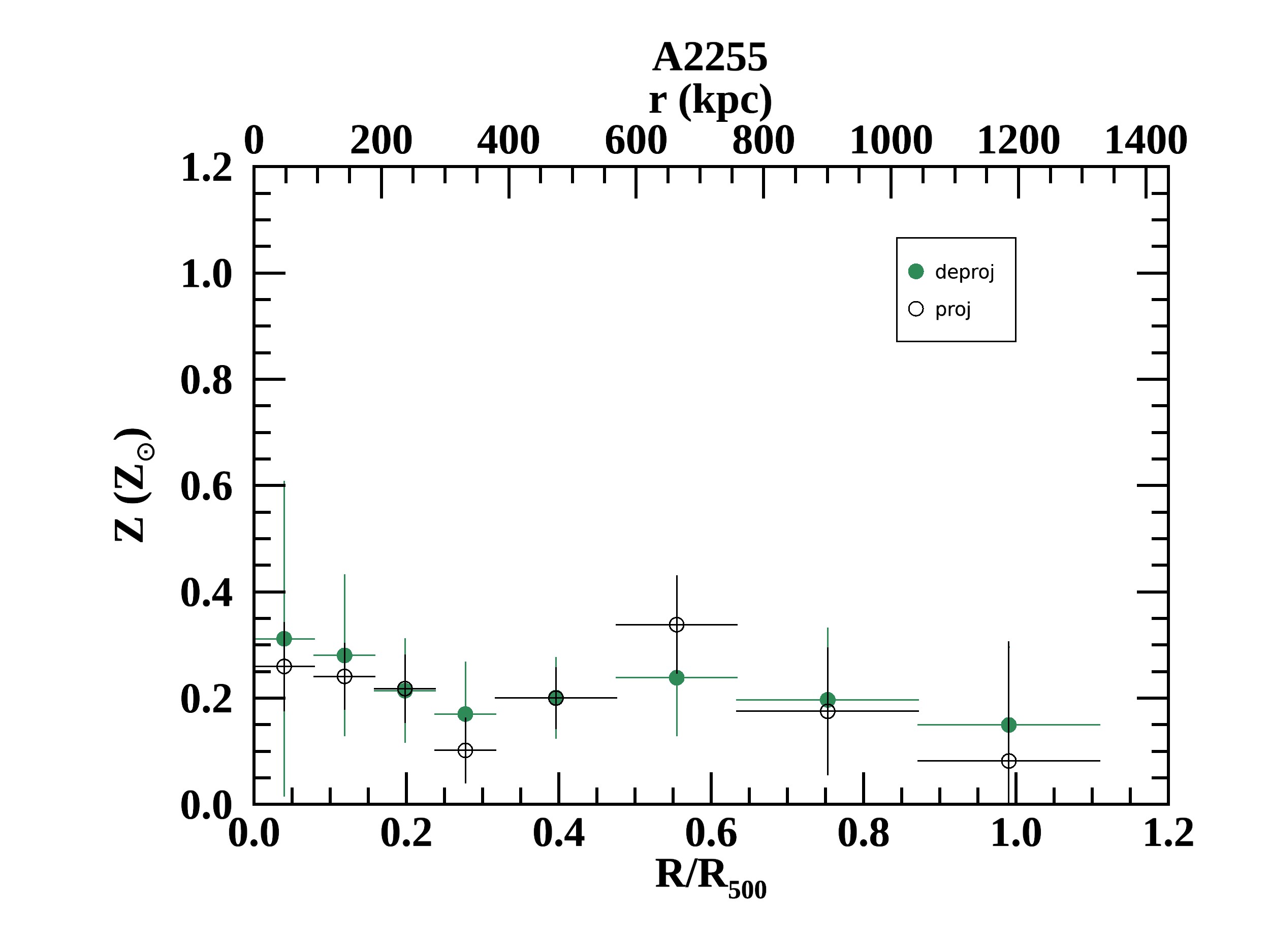}}
\hfill \subfloat{
\includegraphics[width=0.32\textwidth, keepaspectratio]{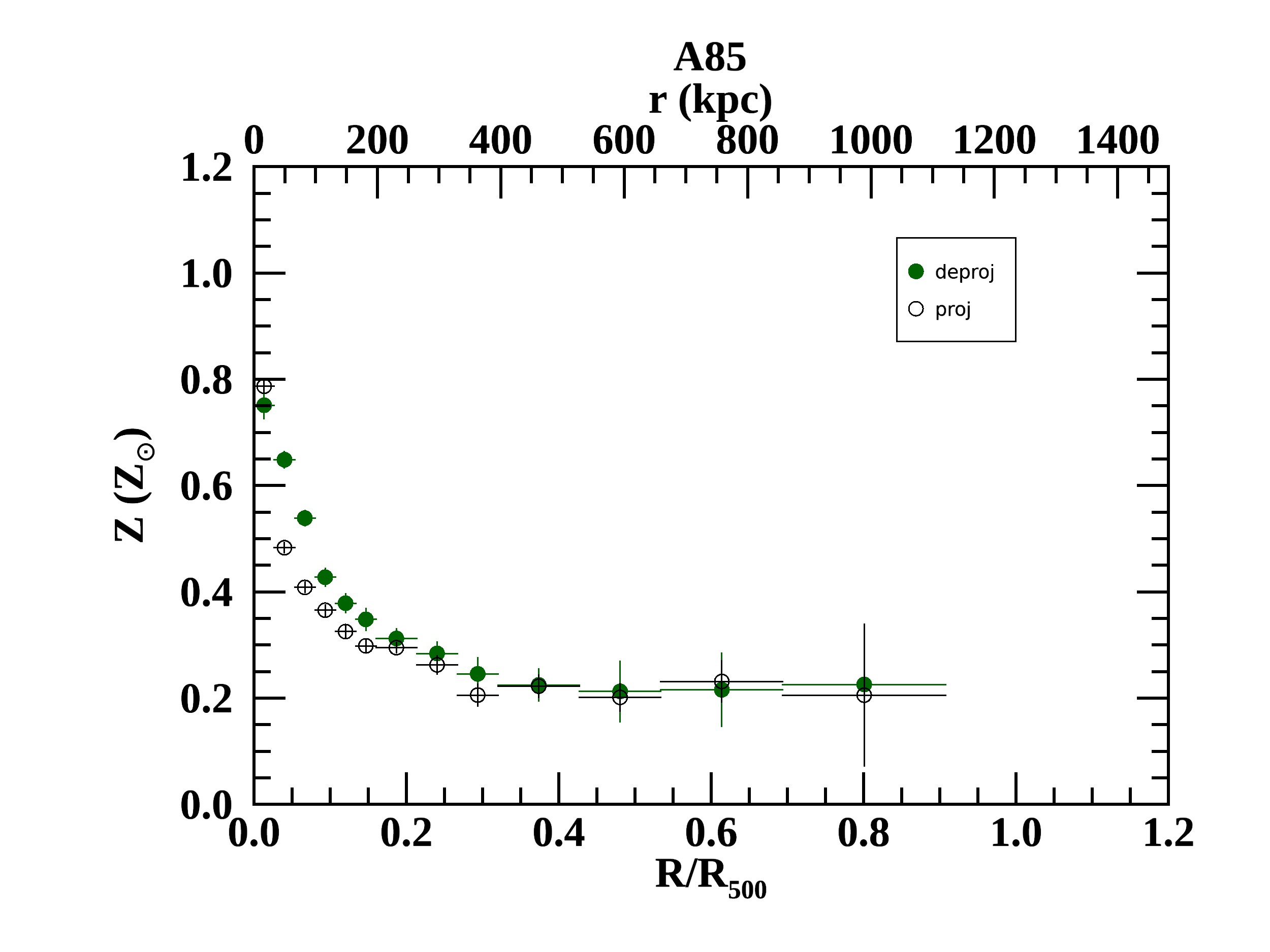}}
}
\centerline{
\subfloat{\includegraphics[width=0.32\textwidth, keepaspectratio]{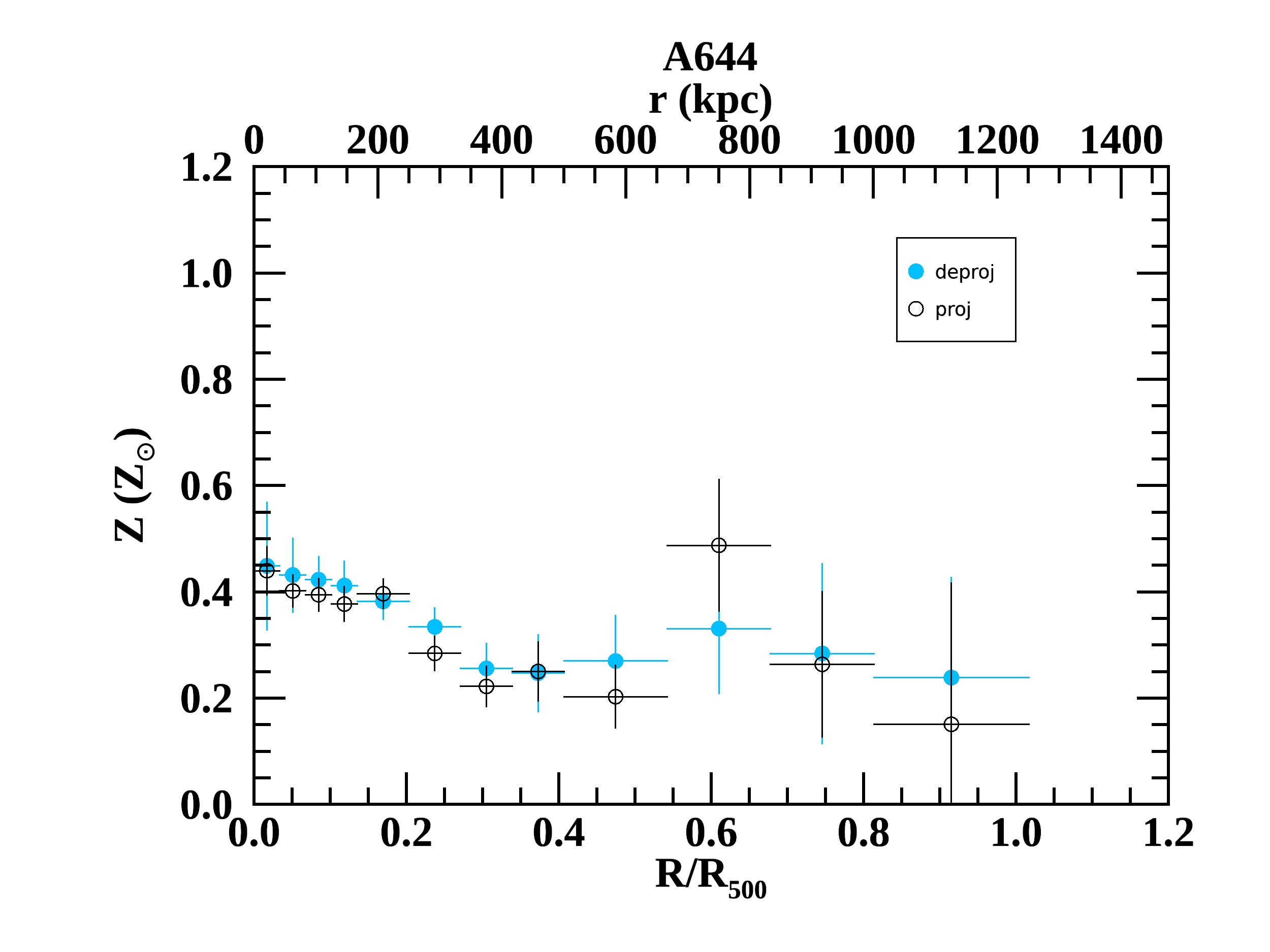}} \hfill \subfloat{
\includegraphics[width=0.32\textwidth, keepaspectratio]{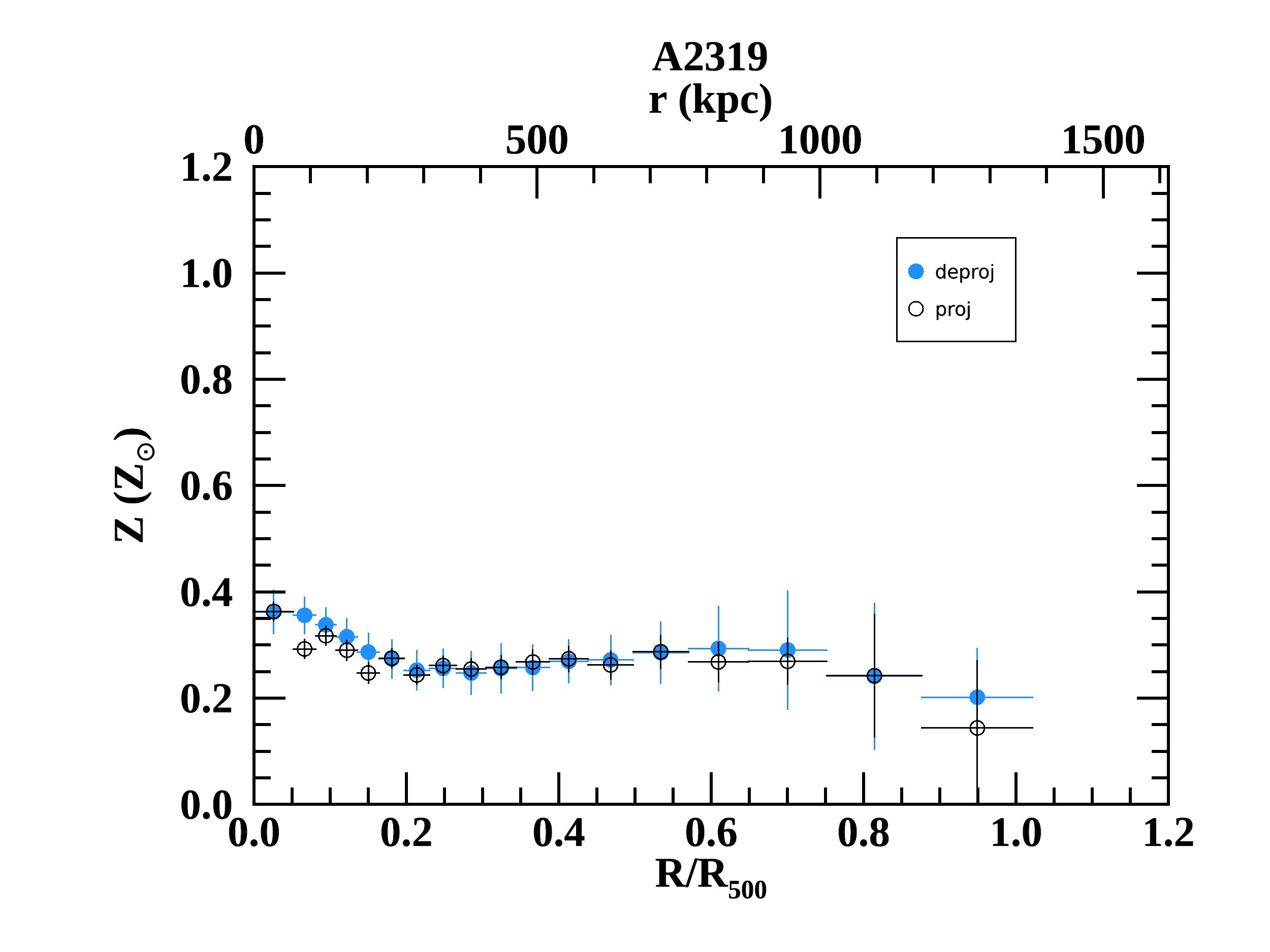}}
\hfill \subfloat{
\includegraphics[width=0.32\textwidth, keepaspectratio]{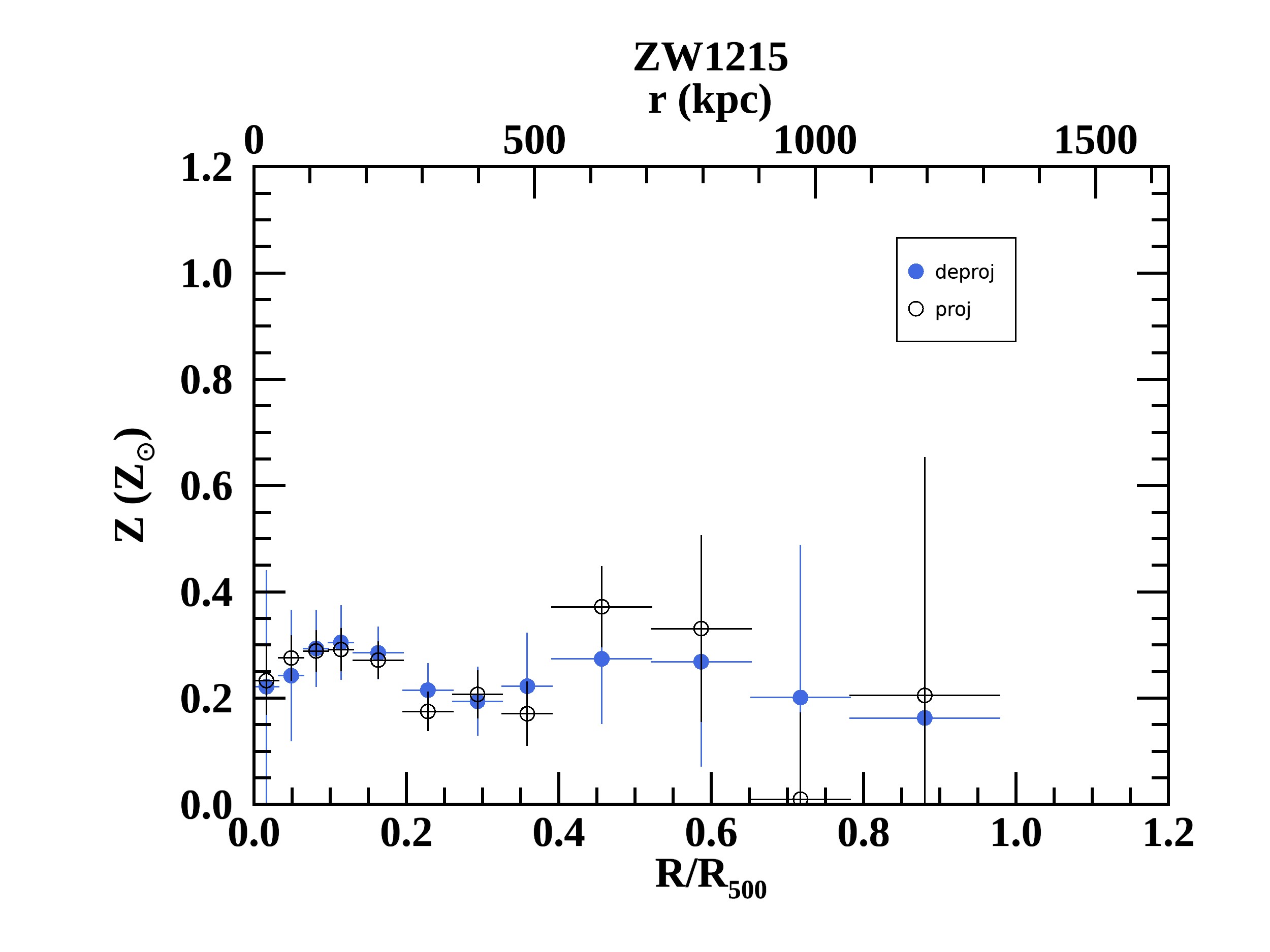}}
}
\centerline{
\subfloat{\includegraphics[width=0.32\textwidth, keepaspectratio]{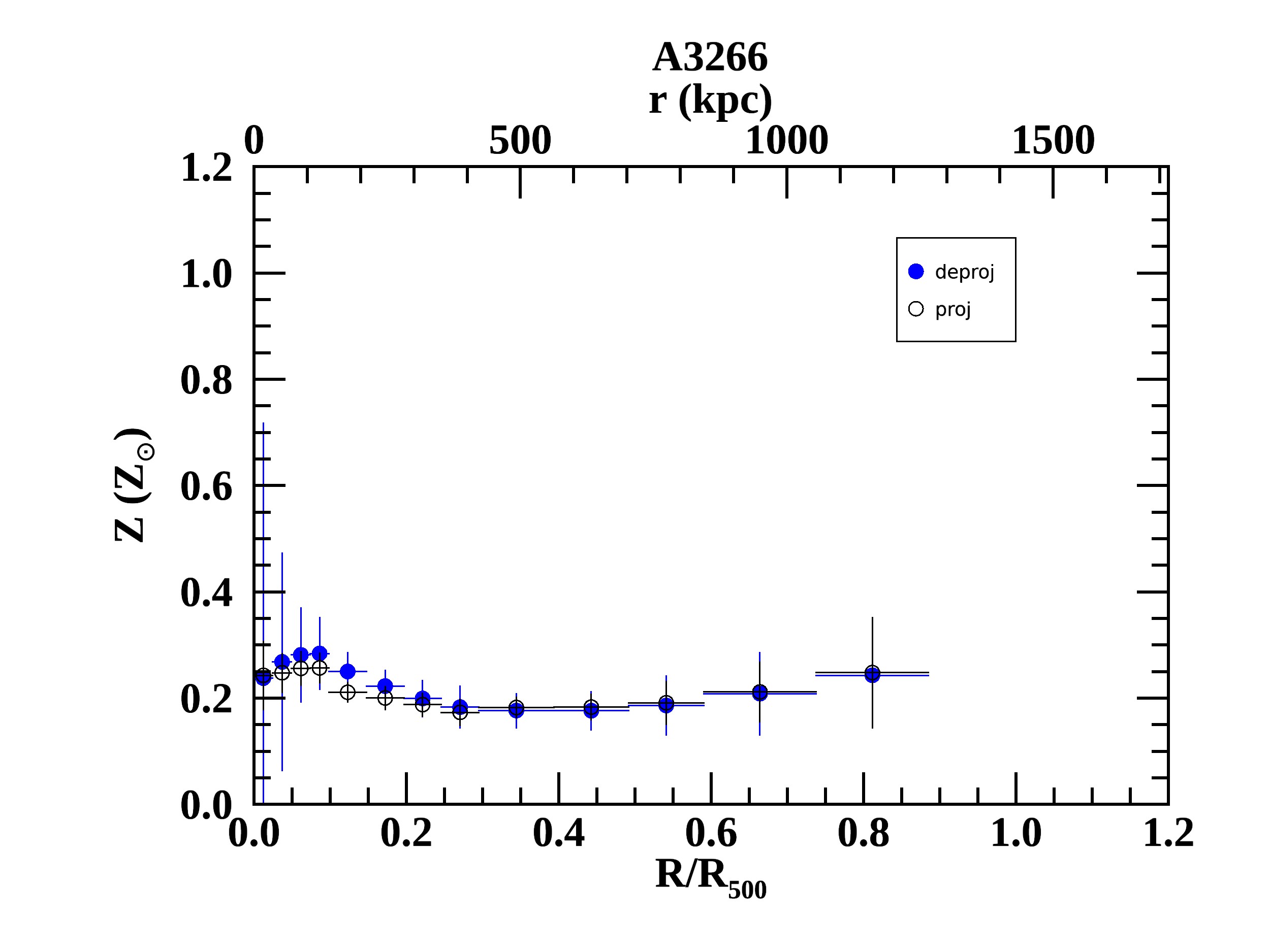}} \hfill \subfloat{
\includegraphics[width=0.32\textwidth, keepaspectratio]{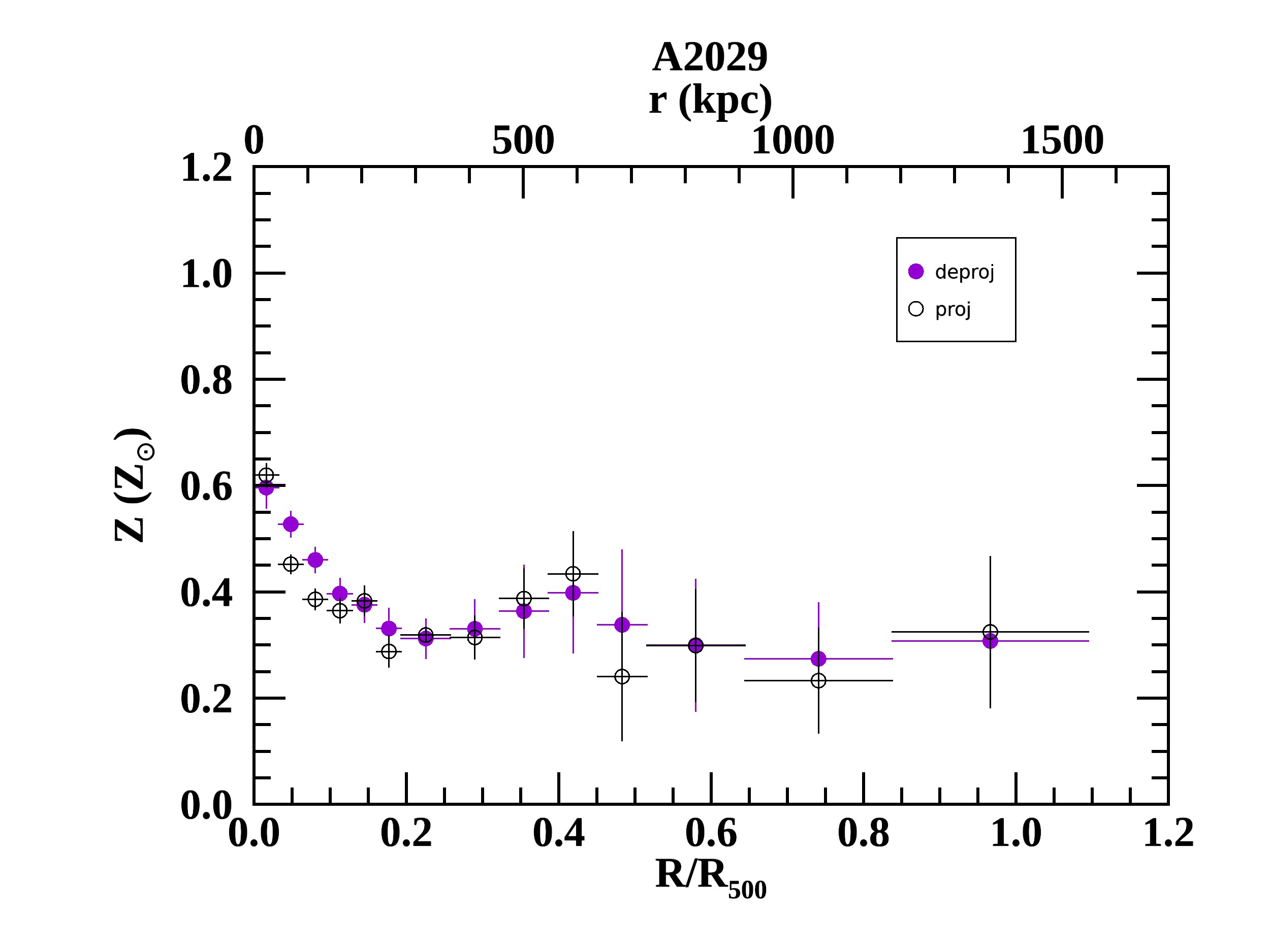}}
\hfill \subfloat{
\includegraphics[width=0.32\textwidth, keepaspectratio]{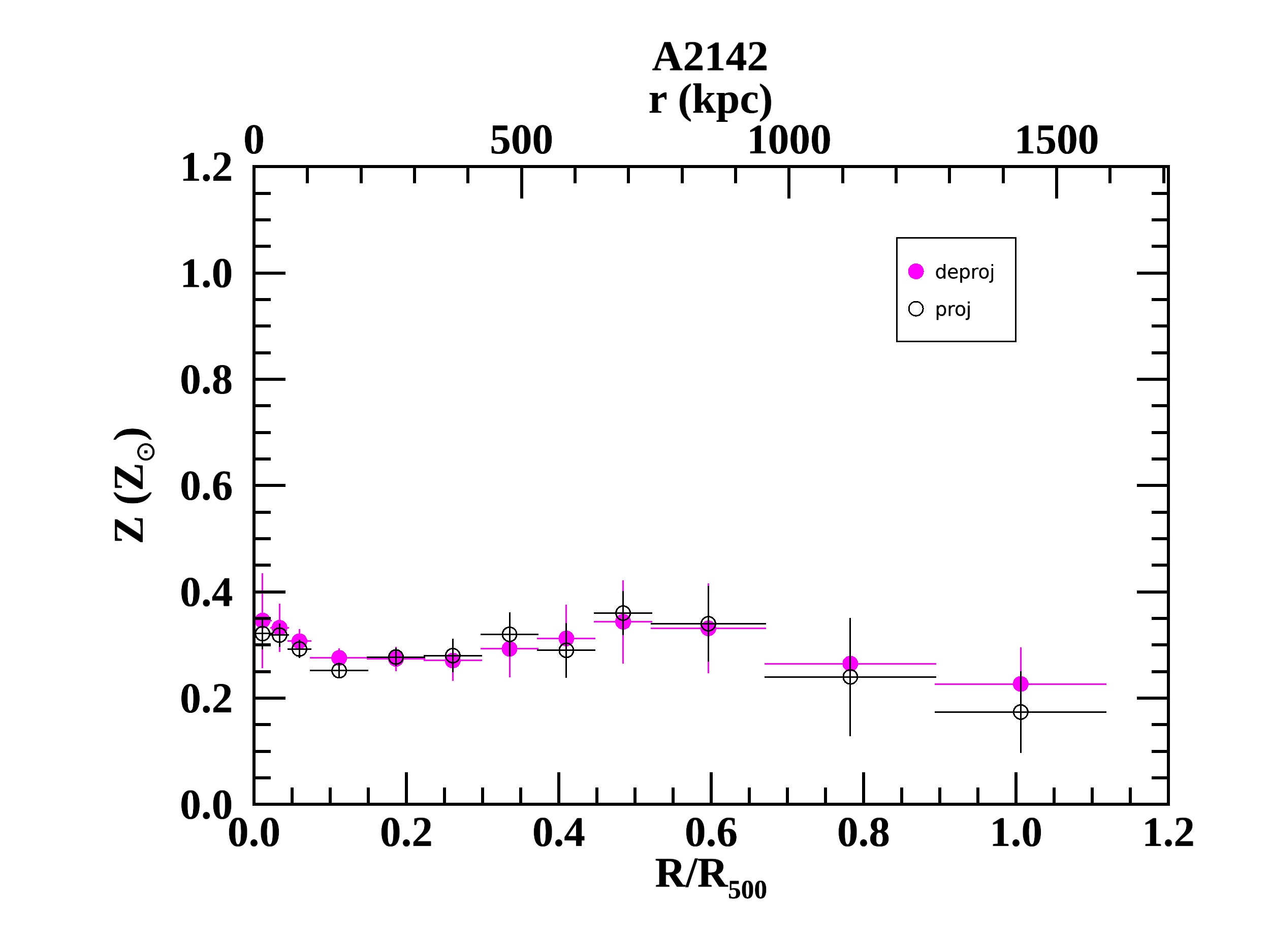}}
}
\caption{Projected (open black circles) and deprojected (coloured closed circles) abundance profiles for all  X-COP clusters.}
\label{fig:Z_XCOP_profs}
\end{figure*}

\twocolumn

\section{Biases in average metallicity measurements}
\label{appendix:bias}

The mean metallicity profile presented in Sect. \ref{sec:fe_icm} was computed by performing a weighted mean of the individual iron abundance measurements. This procedure assumes Gaussian posterior distributions for the individual measurements, which is not necessarily verified in the case of the metal abundance, since it is a positive-definite quantity. Especially in the case of low abundance and large uncertainty, we expect the posterior distribution to be strongly skewed towards high values, and the mean of the posterior to be biased high. 

To verify whether our measurements can be affected by this bias, especially in the outermost regions, we performed Monte Carlo simulations of spectra for various input metallicities. We simulated data based on the observed spectra in the outermost radial bin of a typical X-COP cluster (A3158) and fitted the simulated spectra using the X-COP procedure. For a grid of metal abundances spanning the range $[0.01-0.4]Z_\odot$, we generated 100 realizations of the spectra and studied the distribution of fitted metal abundance values. As expected, we find that, when the uncertainties are large, the distributions are highly non-Gaussian and skewed towards high values. Then, from the generated metal abundance measurements, we randomly picked a sample of 12 observed values and computed the weighted mean to mimic the X-COP sample selection. We repeated the selection 100 times to study the distribution of expected mean values for the sample. 

\begin{figure}
    \resizebox{\hsize}{!}{
    \includegraphics[width=\textwidth]{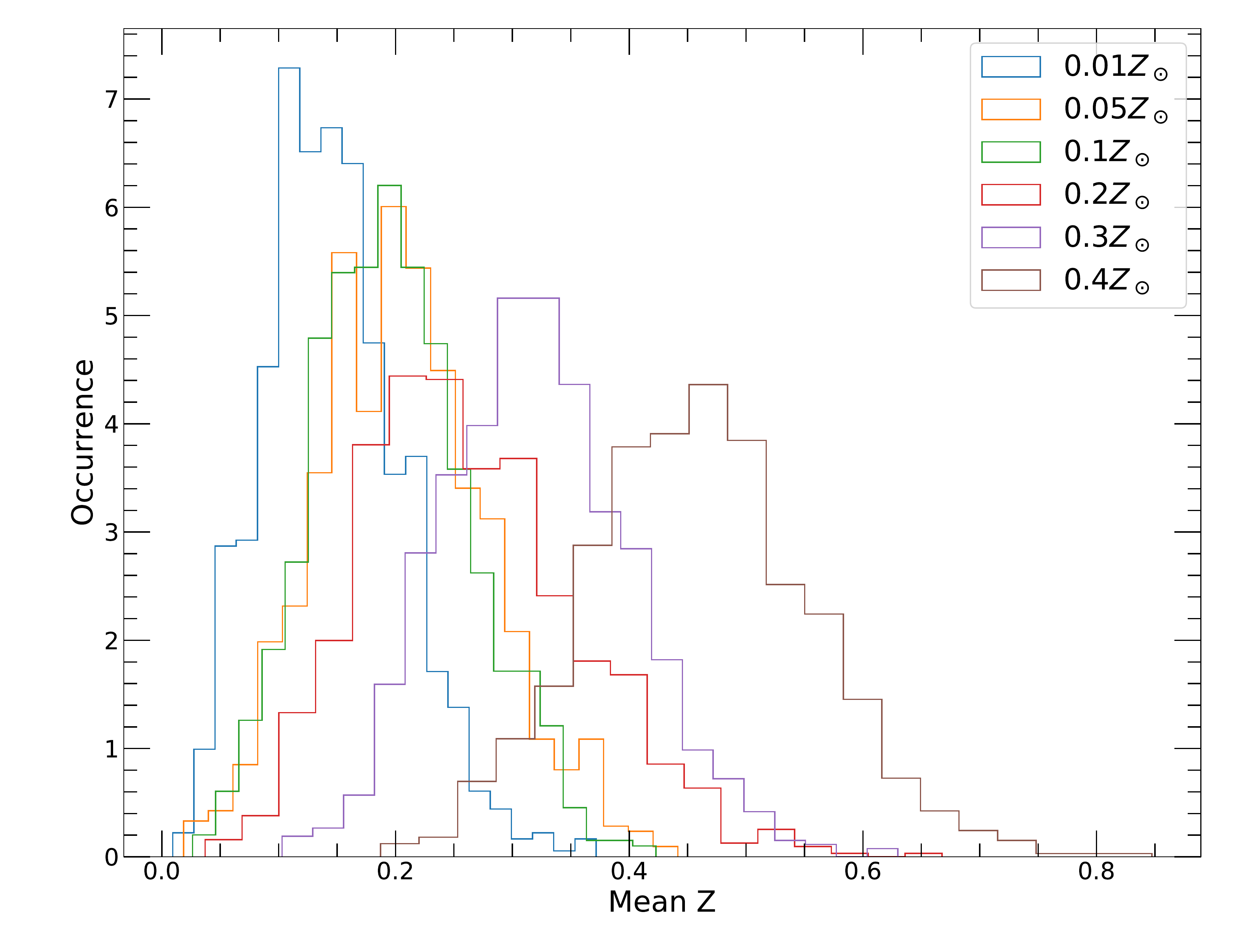}}
    \caption{Distributions of simulated sample means in the outermost radial bin for various input metallicities and for 100 X-COP-like samples. The input metallicities range from 0.01 (blue) to 0.4$Z_\odot$ (brown). }
    \label{fig:sim_sample_mean}
\end{figure}

\begin{figure}
\resizebox{\hsize}{!}{    \includegraphics{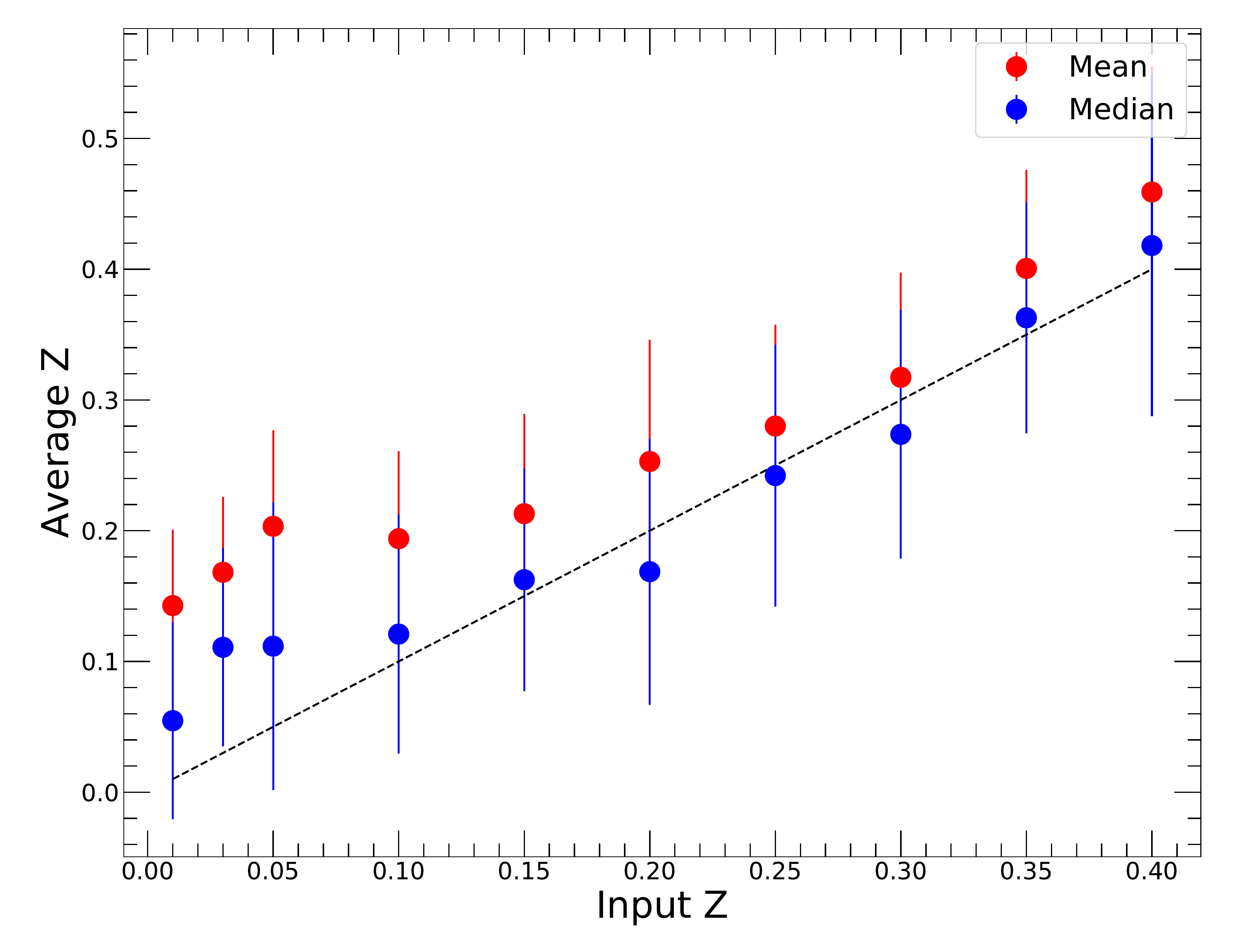}}
    \caption{Most probable average sample values (filled circles) and standard deviation of the expected sample averages (error bars) as a function of input metallicity for sets of 100 simulated spectra. The data points show the expected weighted mean (red) and median (blue). The black dashed line is the one-to-one relation. }
    \label{fig:ab_mean_vs_median}
\end{figure}

In Fig. \ref{fig:sim_sample_mean} we show the distributions of expected mean values for various input metallicities. We can see that for very low metallicity ($Z=0.01Z_\odot$) the most probable mean value is $\sim0.12$ and recovering the true value is very unlikely. For higher metal abundance values the peak of the distribution gets progressively closer to the true value, although it remains biased, even for $Z=0.4Z_\odot$. However, when performing the same exercise with the median of the distribution instead of the weighted mean, we find values that are much closer to the true value. In Fig. \ref{fig:ab_mean_vs_median} we show the most probable sample values as a function of metallicity for the weighted mean and the median. The error bars show the standard deviation of the distribution of expected mean values. While at very low metallicity ($Z<0.1Z_\odot$) the median is biased as well, above this threshold the median accurately recovers the true value, whereas the weighted mean remains biased. Therefore, the comparison between the mean and the median value can tell us whether our measurement is affected by a strong bias. 

\begin{figure}
\centerline{\includegraphics[angle=0,width=9.8cm]{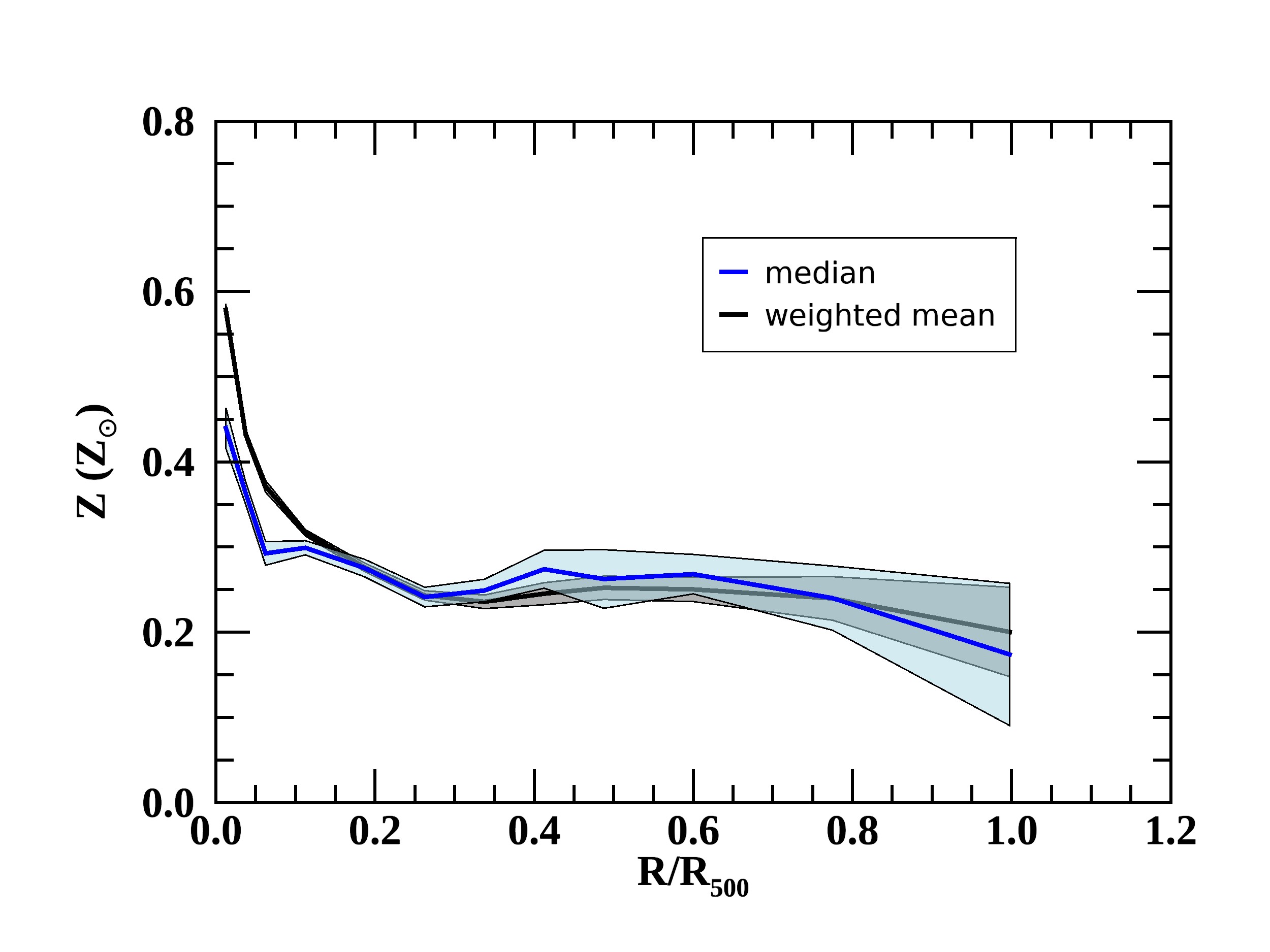}}
\caption{Comparison between the weighted - mean abundance profile (black line), as derived in Sect. \ref{sec:projz}  and median abundance profile (blue line) for the X-COP sample. The  shaded areas represent the statistical errors of the profiles. The two profiles exhibit an excellent agreement.}
\label{fig:cfr_median_wave_prof}
\end{figure}

To this aim, we computed the median profile of X-COP metal abundance measurements and compared it with the weighted mean (see Fig. \ref{fig:cfr_median_wave_prof}).  The median profile is consistent with the mean profile at each radius, implying that our measurements are robust. However, we caution that, in other contexts, averaging metal abundance values can introduce important biases, when the non-Gaussianity of posterior metal abundance values is ignored. 

\end{document}